\documentclass[11pt]{cambrian}

\usepackage{xspace}
\usepackage{multirow}
\usepackage{subcaption}
\usepackage{float}
\usepackage{wrapfig}
\usepackage{tikz}
\usetikzlibrary{arrows.meta,calc,fit}
\usepackage{placeins}  
\usepackage[most]{tcolorbox}

\newif\ifshowtodos
\showtodostrue
\ifshowtodos
    \newcommand{\todotxt}[1]{#1}
\else
    \newcommand{\todotxt}[1]{}
\fi

\makeatletter
\DeclareRobustCommand\onedot{\futurelet\@let@token\@onedot}
\def\@onedot{\ifx\@let@token.\else.\null\fi\xspace}

\def\eg{\emph{e.g}\onedot}

\makeatother

\newcommand{\paintemoji}{\raisebox{-0.15em}{\includegraphics[height=1em]{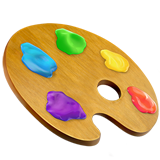}}}

\newcommand{\bench}{\textsc{PaintBench}\xspace}
\newcommand{\tinygrafixbench}{\textsc{TinyGrafixBench}\xspace}

\newcommand{\nanobanana}{\textsc{Nano-Banana-2}\xspace}
\newcommand{\nanobananaone}{\textsc{Nano-Banana-1}\xspace}
\newcommand{\gptitwo}{\textsc{GPT-Image-2}\xspace}
\newcommand{\qwenedit}{\textsc{Qwen-Image-Edit-2511}\xspace}
\newcommand{\fluxdev}{\textsc{FLUX.2-dev}\xspace}
\newcommand{\fluxkontext}{\textsc{FLUX.1-Kontext-dev}\xspace}
\newcommand{\fluxklein}{\textsc{FLUX.2-klein-9B}\xspace}
\newcommand{\longcat}{\textsc{LongCat-Image-Edit}\xspace}
\newcommand{\iptop}{\textsc{InstructPix2Pix}\xspace}
\newcommand{\bagel}{\textsc{BAGEL}\xspace}
\newcommand{\hunyuanimage}{\textsc{HunyuanImage-3.0}\xspace}
\newcommand{\gemini}{\textsc{Gemini 3.1 Thinking}\xspace}

\newcommand{\nanobananaS}{\textsc{NB-2}\xspace}
\newcommand{\nanobananaoneS}{\textsc{NB-1}\xspace}
\newcommand{\gptitwoS}{\textsc{GPT-I2}\xspace}
\newcommand{\qweneditS}{\textsc{Qwen-IE}\xspace}
\newcommand{\fluxdevS}{\textsc{FLUX.2-D}\xspace}

\newcommand{\longcatS}{\textsc{LCat-IE}\xspace}
\newcommand{\iptopS}{\textsc{IP2P}\xspace}
\newcommand{\bagelS}{\textsc{BAGEL}\xspace}
\newcommand{\hunyuanimageS}{\textsc{HY-3}\xspace}

\definecolor{linkblue}{rgb}{0.12,0.49,0.85}
\hypersetup{
    colorlinks=true,
    linkcolor=linkblue,
    citecolor=linkblue,
    urlcolor=linkblue,
}

\definecolor{rowheader}{RGB}{240,240,240}
\definecolor{oursrowheader}{RGB}{200,220,255}
\definecolor{ours}{RGB}{220,240,255}
\definecolor{pos}{RGB}{0,128,0}
\definecolor{neg}{RGB}{200,0,0}
\definecolor{cigrey}{RGB}{130,130,130}
\definecolor{olsteal}{RGB}{27,110,123}  

\newcommand{\dE}{\Delta E^*}

\newcommand{\iouat}[1]{\text{IoU}@{#1}}
\newcommand{\miou}{\text{mIoU}\xspace}

\definecolor{ceColor}{HTML}{2CA02C}
\definecolor{ieColor}{HTML}{FF7F0E}
\definecolor{cpColor}{HTML}{1F77B4}
\definecolor{ipColor}{HTML}{D62728}

\newcommand{\colCE}{{\color{ceColor}\text{CE}}}
\newcommand{\colIE}{{\color{ieColor}\text{IE}}}
\newcommand{\colCP}{{\color{cpColor}\text{CP}}}
\newcommand{\colIP}{{\color{ipColor}\text{IP}}}

\newcommand{\colCEtxt}[1]{\textcolor{ceColor}{#1}}
\newcommand{\colIEtxt}[1]{\textcolor{ieColor}{#1}}
\newcommand{\colCPtxt}[1]{\textcolor{cpColor}{#1}}
\newcommand{\colIPtxt}[1]{\textcolor{ipColor}{#1}}

\definecolor{editColor}{HTML}{EC4899}
\definecolor{presColor}{HTML}{06B6D4}

\newcommand{\colE}{{\color{editColor}\mathcal{E}}}
\newcommand{\colP}{{\color{presColor}\mathcal{P}}}

\newcommand{\colEtxt}[1]{\textcolor{editColor}{#1}}
\newcommand{\colPtxt}[1]{\textcolor{presColor}{#1}}

\newcommand{\finding}[2]{
    \vspace{1em}
    \vspace{-0.1cm}
    \begin{tcolorbox}[
        enhanced,
        colback=white!90!gray,
        colframe=magenta!50!black,
        arc=5pt,
        boxsep=5pt,
        left=7pt, right=7pt, top=2pt, bottom=2pt,
        boxrule=0.8pt
    ]
    \vspace{-0.1cm}
        \paintemoji~~#2
    \vspace{-0.1cm}
    \end{tcolorbox}
    \vspace{-0.3cm}
    \vspace{1em}
}

\usepackage{fontawesome}
\newcommand{\huggingface}{\raisebox{-1.5pt}{\includegraphics[height=1.05em]{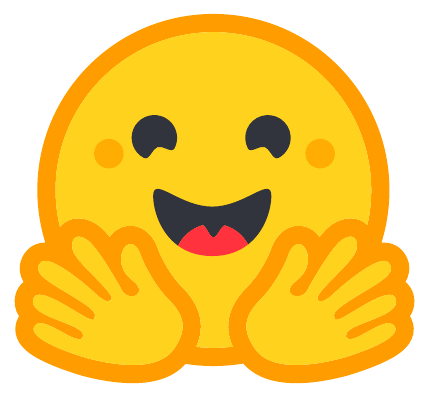}}\xspace}
\newcommand{\github}{\raisebox{-1.5pt}{\includegraphics[height=1.05em]{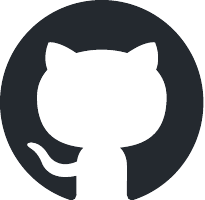}}\xspace}
\newcommand{\worldwideweb}{\raisebox{-1.5pt}{\includegraphics[height=1.05em]{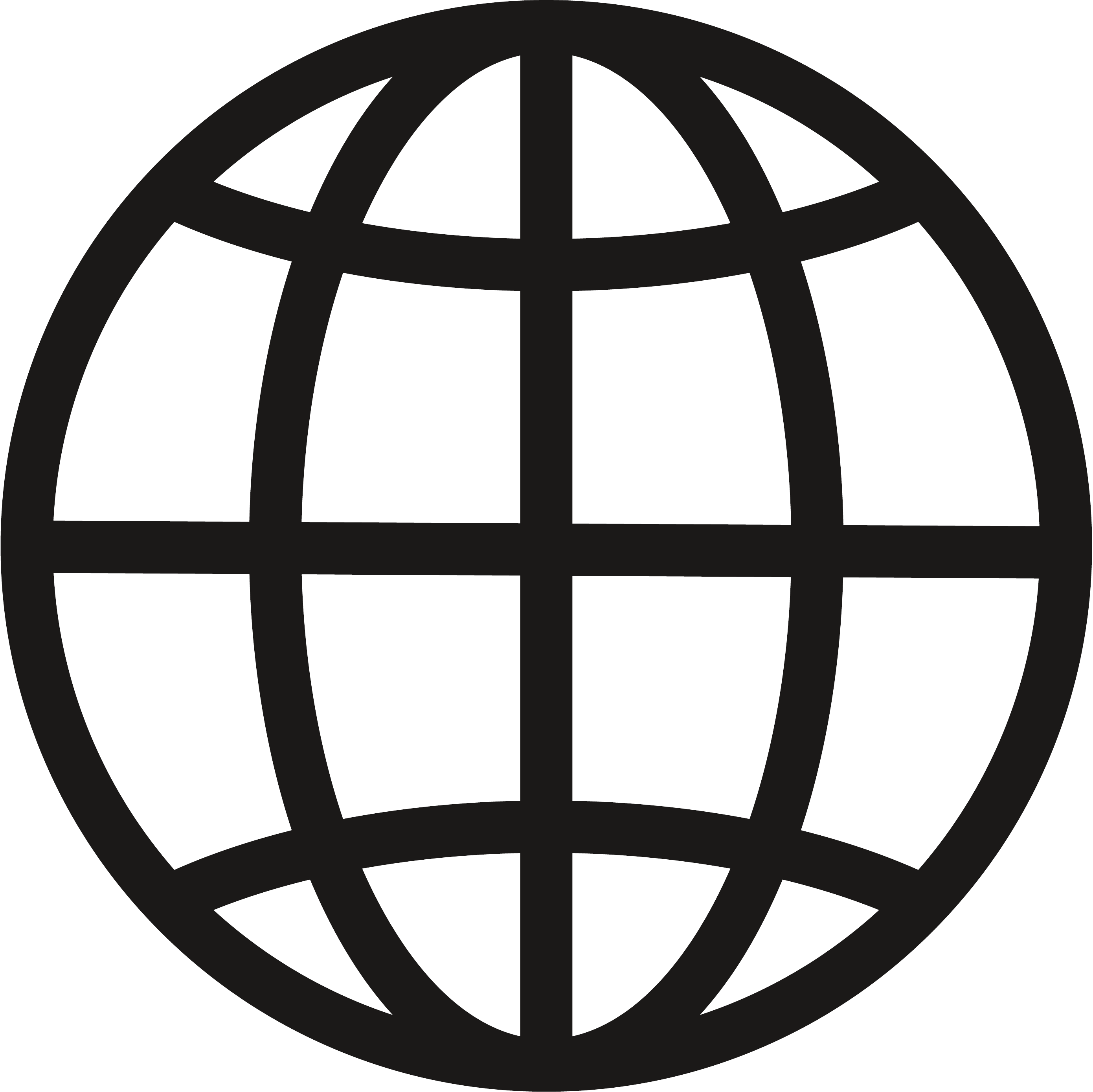}}\xspace}

\begin{document}
\pagestyle{fancy}

\title{
    \center 
    \paintemoji~\bench\\ Deterministic Evaluation of Precise Visual Editing
}

\author{
    Kai~Xu$^{*}$ \quad
    Ellis~Brown$^{*}$ \quad
    Shrikar~Madhu \quad
    Rob~Fergus \quad
    He~He \quad
    Saining~Xie\\
    New~York~University
}

\begin{abstract}
While current multimodal models are proficient at open-ended visual editing, executing precise single-answer edits remains an important obstacle.
To probe this challenge, we introduce \bench, a dynamically scalable benchmark targeting 20 fundamental precise visual editing operations across four categories: geometric transformation, structural manipulation, color change, and symbolic reasoning.
Procedural generation with configurable complexity enables an effectively infinite, contamination-resistant evaluation suite, and deterministic pixel-level evaluation eliminates reliance on bias-prone judge models.
Across 11 image editing models, we find overall low performance, with the current highest-performing industry leader scoring only 17.1\% (mIoU).
Task decomposition reveals especially challenging operation types (geometric transformation, most structural manipulation, formula-based color change) and model-specific specializations.
Fine-grained benchmark diagnostics further show performance degradations induced by scene variations in object count, background complexity, color scheme, and edit-region size.
To test generalization of \bench scores to applied task performance, we create a procedural, deterministic evaluation for data visualization editing (\tinygrafixbench) and find strong linear correlation with \bench scores ($R^2 = 0.91$, $p < 0.001$).
Altogether, \bench provides a rigorous foundation for measuring and driving progress in precise multimodal visual editing.
\end{abstract}

\renewcommand{\and}{, }

\maketitle

\begingroup
    \renewcommand{\thefootnote}{\fnsymbol{footnote}}
    \footnotetext[1]{Project lead}
\endgroup

\vspace{0.5em}
\begin{center}
    \renewcommand{\arraystretch}{1.75}
    \begin{tabular}{rll}
        \worldwideweb{} & \textbf{Website} & \url{https://PaintBench.github.io}\\
        \github{} & \textbf{Code} & \url{https://github.com/PaintBench/PaintBench}\\
        \huggingface{} & \textbf{Benchmarks} & \url{https://hf.co/datasets/PaintBench/PaintBench} \\
    \end{tabular}
\end{center}

\clearpage
{
    \small
    \hypersetup{linkcolor=black}
    \tableofcontents
}
\clearpage

\section{Introduction}
\label{sec:intro}

Generative image editing has
advanced rapidly through the development of architectures such as instruction-tuned diffusion models~\cite{brooks2023instructpix2pix},
flow-matching editors~\cite{flux2025kontext}, and native multimodal generators~\cite{openai2026gptimage2,gemini31flashimage2026}.
Yet evaluating whether these models can execute precise single-answer edits remains an open challenge.
Consider a simple task: recolor the olive-colored triangle to cyan (\textcolor[HTML]{0FE1DF}{\texttt{\#0FE1DF}}).
A model may produce an output that looks correct at first glance, but the generated cyan may
deviate substantially from the target color, and the background may be slightly altered.
Reducing such errors is crucial for a myriad of applications that require exact precision.

\begin{figure}[t]
    \centering
    \includegraphics[width=\textwidth]{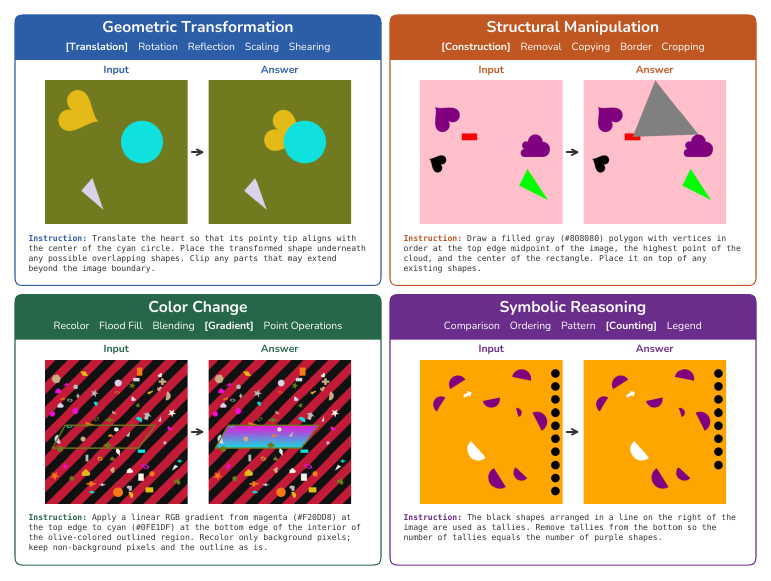}
    \vspace{-2em}
    \caption{
        \textbf{\bench\ spans 20 fundamental visual editing operations in four categories.}
        Each panel shows an example seed-generated problem from one operation (highlighted as \(\boldsymbol{[*]}\)), with an input image, instruction, and ground-truth answer.
        The deterministic, single-answer design enables automated, pixel-level evaluation without bias-prone judge models.
    }
    \label{fig:teaser}
\end{figure}

Existing image editing benchmarks~\cite{zhang2023magicbrush,basu2023editval,ma2024i2ebench}
largely rely on human judgment or model-based evaluation via vision-language models and perceptual
metrics (\eg, SSIM~\cite{wang2004ssim}, LPIPS~\cite{zhang2018lpips}, FID~\cite{heusel2017fid}).
These are natural choices for many natural-image editing tasks without a single correct answer:
``make the sky more blue'' admits infinitely many valid outputs, so evaluation is closer to
preference scoring than correctness verification.
Recent benchmarks targeting complex editing~\cite{han2025unireditbench,yang2025complexedit} continue
this tradition of using powerful learned judge models.

However, many editing tasks \emph{do} have unique correct answers, in which case evaluation can
be performed deterministically without potential biases of model-based scoring or human judgment.
Examples of such tasks include moving a shape 50 pixels to the right, flood-filling a region to a target color, or removing the smallest instance of a shape.
Motivated by this observation, we create \bench, a benchmark of such deterministic tasks.
The name reflects the simplicity of its constituent operations: these tasks can be completed easily in a raster image editor, yet they are far from easy for current multimodal models.

\bench generates problems procedurally: a random seed and parameter configuration produces an
input image, natural-language instruction, and unique correct answer
(see \cref{sec:method} for design principles).
In the same way that HumanEval~\cite{chen2021humaneval} evaluates code generation by executing
programs against test cases rather than judging surface plausibility, \bench evaluates visual
editing by direct pixel comparison against ground truth across a range of color precision
thresholds, without relying on bias-prone judge models.
Because problems are generated from seeds with configurable scene parameters, \bench also functions as a flexible diagnostic instrument: practitioners can generate task sets targeting specific operations or visual conditions of interest, going beyond a static snapshot of capabilities.
In line with the broader vision-first push in multimodal research~\cite{wang2024emu3,tong2024cambrian,tong2026beyond,zhao2025unifiedmm,visionbanana2026}, we focus our evaluation on native pixel-space image editing models, rather than agentic or programmatic approaches that delegate the edit to a raster engine.

Across 11 models spanning native multimodal generators, diffusion editors, and flow-matching editors, we find that even the best-performing model reaches only 17.1\% mIoU.
Task difficulty tracks the underlying fundamental operation: geometric transformation, formula-based color change, and most structural manipulation tasks are consistently hard across all models (\cref{sec:experiments}), while removal and single-color operations are more tractable, though still far from solved.
Beyond overall rankings, models exhibit notable task-specific specializations on individual operations that diverge from their overall standing.
By varying \bench's procedural scene parameters, we find that striped backgrounds, high object counts, nonstandard color palettes, and small edit-regions all substantially degrade performance.
To test whether \bench scores predict performance on \emph{applied tasks} built upon the same fundamental operations, we introduce \tinygrafixbench, applying our procedurally generated, deterministically evaluated philosophy to data visualization editing.
Model scores between the two are strongly correlated ($R^2 = 0.91$, $p < 0.001$), suggesting that \bench captures capabilities that generalize beyond its synthetic-shape scope.

\noindent Our contributions include:
\begin{enumerate}[leftmargin=*,itemsep=0pt,topsep=0pt]
    \item \bench: a seed-generated deterministic benchmark of 20 fundamental precise visual editing operations across four categories, comprising a 1{,}920-problem test set.
    \item A pixel-level evaluation protocol that factors in both edit- and preservation quality using perceptual color similarity, with no reliance on bias-prone judge models.
    \item A systematic evaluation of 11 models showing that even the best-performing model reaches only 17.1\% mIoU, with consistent challenges across geometric transformation, formula-based color change, and structural manipulation tasks.
    \item \bench's configurable procedural generator enables controlled scene-variation analysis: striped backgrounds, high object counts, nonstandard color palettes, and small edit-regions all substantially degrade performance.
    \item \tinygrafixbench: a companion benchmark applying the same procedural, deterministic principles to data visualization editing. Model scores between the two are strongly correlated, suggesting that \bench captures capabilities that generalize to applied visual editing.
\end{enumerate}

\section{Related Work}
\label{sec:related}

We situate \bench against prior image editing benchmarks, broader generative model evaluation, synthetic benchmarking methodologies, and image editing model capabilities.

\paragraph{Image Editing Benchmarks.}
Several benchmarks evaluate instruction-following image editors using real-image edit triplets and subjective evaluation.
MagicBrush~\cite{zhang2023magicbrush}, EditVal~\cite{basu2023editval}, I2EBench~\cite{ma2024i2ebench}, and ImgEdit~\cite{cai2025imgedit} curate such triplets and rely on human judgment, vision-language model scoring, or perceptual metrics.
Complex-Edit~\cite{yang2025complexedit} probes how performance degrades with instruction complexity, and UniREditBench~\cite{han2025unireditbench} adds a reasoning-heavy dual-reference variant.
Recent work proposes fine-grained MLLM judges~\cite{liu2026mlmmjudges} for editing evaluation, but the community has not converged on reliable methods, and judge models still yield highly uncertain absolute scores~\cite{kumar2026vlm}.
\bench departs from these designs: by focusing on edits with deterministic ground truth, pixel-level comparison against a known answer eliminates the need for judge models entirely.
Benchmarks for reasoning-heavy edits with inherently subjective outputs (\eg, RISEBench~\cite{zhao2025envisioning}, KRIS-Bench~\cite{wu2025kris}) address a complementary problem.

\paragraph{Evaluation of Generative Models.}
GenEval~\cite{ghosh2023geneval} establishes object-focused text-to-image evaluation using detection models as proxy judges.
T2I-CompBench~\cite{huang2023t2icompbench} evaluates compositional generation via BLIP-based scoring.
GenEval~2~\cite{ghosh2025geneval2} identifies \emph{benchmark drift} (static benchmarks becoming misaligned as models improve) and proposes harder evaluation sets.
This motivates \bench's dynamic generation: fresh problems can be generated at will, preventing saturation and contamination.
A Very Big Video Reasoning Suite~\cite{wang2026vbvr} proposes rule-based, human-aligned scoring for video reasoning benchmarks as an alternative to model-based judging.

\paragraph{Synthetic Visual Reasoning Benchmarks.}
CLEVR~\cite{johnson2017clevr} established programmatic visual scene generation for diagnostic evaluation; ARC~\cite{chollet2019arc}, RAVEN~\cite{zhang2019raven}, and Bongard-Logo~\cite{nie2020bongard} extend this tradition to abstract visual reasoning with deterministic answers.
In the language domain, HumanEval~\cite{chen2021humaneval} measures functional correctness of code generation by executing programs against test cases, philosophically aligned with \bench's pixel-level verification.
We extend this tradition of synthetic, deterministic evaluation to generative image editing.
Concurrent work Gen-ViRe~\cite{liu2025genvire} evaluates generative visual reasoning for video generation, but targets world simulators and relies on VLM-assisted evaluation rather than deterministic comparison.

\paragraph{Image Editing Models.}
Modern image editing models span three architectural families:
instruction-tuned diffusion editors~\cite{brooks2023instructpix2pix,wu2025qwen,team2025longcat},
flow-matching editors~\cite{flux2025kontext,bfl2025flux2},
and native multimodal generators that produce text and images from a single backbone~\cite{wang2024emu3,deng2025bagel,cao2025hunyuanimage,google2025gemini25flashimage,gemini31flashimage2026,openai2026gptimage2}.
Complementary approaches may write code or call external tools to perform edits.
Such approaches lie outside our scope: \bench is a proxy benchmark designed to evaluate models that natively \emph{output} in pixel-space.
The native pixel-space class is increasingly the product of a research thread on \emph{unified multimodal models} aimed at combining image understanding and generation in one backbone~\cite{teamchameleon2024,zhou2024transfusion,tong2024metamorph,zhao2025unifiedmm,tong2026beyond}, further motivated by evidence that image generators have become competent generalist vision systems on their own~\cite{visionbanana2026,tong2024cambrian}.
\bench tests the pixel-level precision side of that unified claim: whether models that can \emph{describe} a scene can also \emph{edit} it exactly.

\begin{figure}[t]
    \centering
    \includegraphics[width=\linewidth]{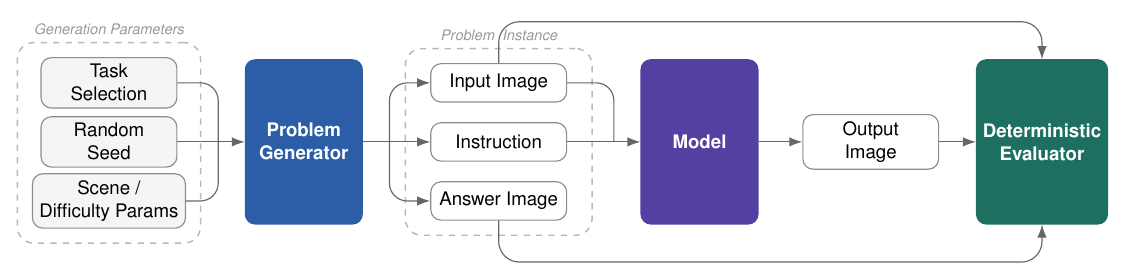}
    \vspace{-2em}
    \caption{
        \textbf{Procedurally generated problems are evaluated against pixel-exact ground truth.}
        A random seed, task mode, and scene parameters (number of shapes, color palette, image dimensions, background style) produce a problem consisting of an input image, instruction, and answer image.
        The model output is then deterministically compared pixel-wise to the answer and input images, without relying on bias-prone judge models.
    }\label{fig:generation-pipeline}
\end{figure}

\section{Benchmarking Precise Visual Editing}
\label{sec:method}

\bench is a procedurally-generated deterministic evaluation of multimodal model capabilities on fundamental operations in precise single-answer visual editing.
Here, we describe the design principles (\cref{sec:design}), task categories (\cref{sec:tasks}), and visual conditions (\cref{sec:conditions}).

Each problem consists of an (input image, instruction, answer image) triple generated from a seed.
Scenes consist of geometric shapes of varying types and colors rendered on solid-color or striped backgrounds.
For each of the 20 tasks, we generate 12 problems for each of the 8 visual conditions,
yielding 1{,}920 problems in total (20 tasks $\times$ 8 conditions $\times$ 12 problems; see \cref{sec:appx:bench-config}).

\subsection{Design Principles}
\label{sec:design}

Four principles guide the design of \bench.
(1) \textbf{Determinism}: every problem has exactly one correct output image, produced by a
deterministic transformation $A = f(I, t)$ from the input image $I$ and instruction $t$;
evaluation reduces to a pixel-level comparison of the model's output $\hat{A}$ against $A$ and $I$,
with no judge models, no perceptual proxies, and no ambiguity.
(2) \textbf{Dynamic Generation}: problems are produced procedurally from random seeds, so fresh
problem sets can be generated at will to prevent memorization or contamination.
(3) \textbf{Controlled Difficulty}: tasks expose explicit parameters (canvas dimensions, object
count, background texture, color palette) that vary scene conditions, enabling
precise ablations (\cref{sec:exp:ablations}).
(4) \textbf{Atomic Operations}: tasks target fundamental visual editing operations that serve
as the building blocks of complex workflows.

\subsection{Task Categories}
\label{sec:tasks}

\bench comprises 20 task types organized into four categories (\cref{fig:teaser}; full
taxonomy in \cref{tab:appx:taxonomy} and per-task descriptions in
\cref{sec:appx:tasks}).
\textit{Geometric Transformation} (translation, rotation, reflection, scaling, shearing)
tests affine transformation of shapes.
\textit{Structural Manipulation} (construction, removal, copying, border, cropping) tests
addition and removal of elements, and modification of scene composition.
\textit{Color Change} (recolor, flood fill, blending, gradient, point operations) tests manipulation of pixel color values.
\textit{Symbolic Reasoning} (comparison, ordering, pattern, counting, legend) tests edits
that require spatial or numerical inference before execution.
Generation details (including modes that vary the operation within each task) appear in \cref{sec:appx:construction}.

\subsection{Visual Conditions}
\label{sec:conditions}

\begin{figure}[t]
    \centering
    \includegraphics[width=0.95\linewidth]{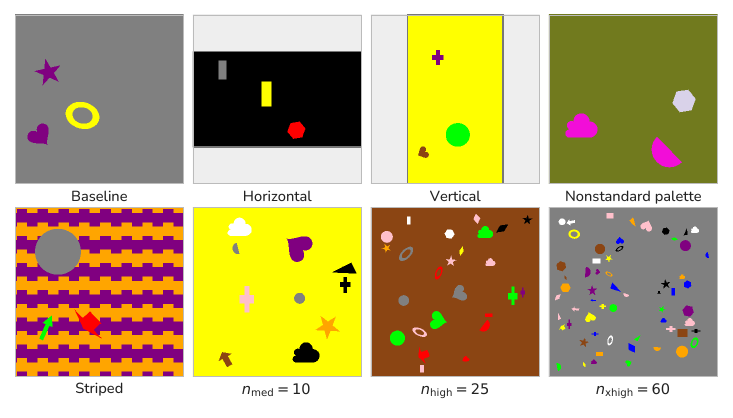}
    \vspace{-1.5em}
    \caption{
        \textbf{Visual conditions isolate one scene parameter at a time.}
        Each panel shows an input image for the removal task.
        Starting from the baseline ($n=3$, $1024 \times 1024$, solid background, standard palette), we vary aspect ratio, palette, background texture, or object count ($n\in\{10, 25, 60\}$).
    }
    \label{fig:ablation-examples}
\end{figure}

Every task is evaluated across eight visual conditions, each changing exactly one scene parameter relative to the baseline condition ($1024 \times 1024$, $n=3$ shapes, standard color palette, single-color background; \cref{fig:ablation-examples}, full enumeration in \cref{tab:appx:ablation-variants}).
This breakdown reveals model sensitivity to scene variations and ensures overall benchmark scores reflect a diverse mix of conditions.

\section{Pixel-Level Evaluation}
\label{sec:metrics}

\begin{wrapfigure}{r}{156pt}
    \vspace{-5em}
    \centering
    \includegraphics[width=\linewidth]{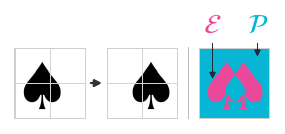}
    \vspace{-2.25em}
    \caption{
        \small
        Edit ($\colE$) and preservation ($\colP$) regions for a simple translation.
    }
    \label{fig:edit-preservation}
    \vspace{-3em}
\end{wrapfigure}

Given an input image $I$ and ground-truth answer image $A$, we define the \colEtxt{\emph{edit-region}} $\colE$ as pixels that differ between them, and the \colPtxt{\emph{preservation-region}} $\colP$ as pixels that are identical (\cref{fig:edit-preservation}).

\begin{figure}
    \centering
    \includegraphics[width=\linewidth]{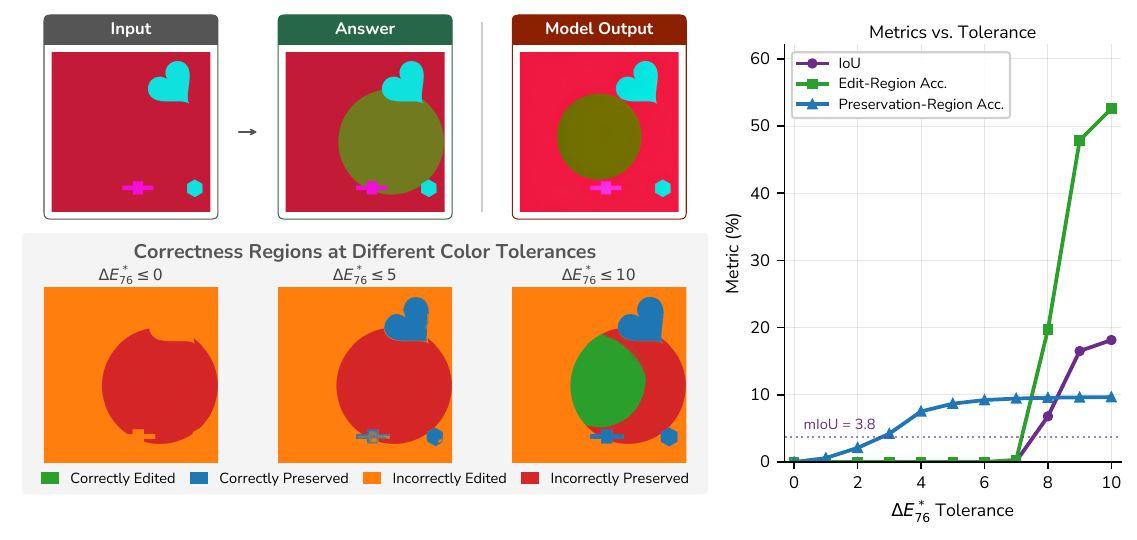}
    \vspace{-2em}
    \caption{
        \textbf{Deterministic evaluation measures geometric and color accuracy at the pixel-level.}
        \textit{Top left:} The model output is graded against the input and answer images.
        \textit{Bottom left:} At a color tolerance $t$ (defined by the $\dE_{76}$ convention; the lower the stricter), each pixel is classified as
            \colCEtxt{\textbf{\textit{correctly} edited}},
            \colCPtxt{\textbf{\textit{correctly} preserved}},
            \colIEtxt{\textbf{\textit{incorrectly} edited}}, or
            \colIPtxt{\textbf{\textit{incorrectly} preserved}}.
        \textit{Right:} \textbf{IoU@t} increases with rising $\dE_{76}$ tolerance $t$ due to increasing \textit{edit-} and \textit{preservation}-region accuracies (\S\ref{sec:iou});
        averaging over $t \in \{0, \ldots, 10\}$ gives an overall score \textbf{\miou} for one problem (\S\ref{sec:miou}).
    }
    \label{fig:eval-breakdown}
\end{figure}

\subsection{Color Distance}
\label{sec:color_metrics}

We measure per-pixel color difference using $\dE_{76}$~\cite{cie1976lab}, the Euclidean
distance in CIE~L*a*b* color space:
$\dE_{76}(o_i, a_i) = \sqrt{(L_o - L_a)^2 + (a_o - a_a)^2 + (b_o - b_a)^2}$,
where $L$, $a$, $b$ are the lightness and chromaticity coordinates of output pixel $o_i$ and
answer pixel $a_i$.
Under a commonly used rule of thumb, $\dE_{76} \leq 1$ is
imperceptible, $1$--$2$ is perceptible under close observation, $2$--$10$ is perceptible
at a glance, and ${>}10$ indicates distinct colors~\cite{minaker2021optimizing,schuessler2016deltae}.
A pixel is \emph{correct} if $\dE_{76}(o_i, a_i) \leq t$ for a chosen color tolerance $t$;
we evaluate at integer tolerances $t \in \{0, 1, \ldots, 10\}$, spanning exact pixel
match ($t = 0$) to a lenient tolerance ($t = 10$).

\subsection{Per-Problem Score: $\iouat{t}$}
\label{sec:iou}

Each problem's pixels fall into four disjoint sets at tolerance $t$:
\colCEtxt{correctly edited} ($\colCE$) and \colIEtxt{incorrectly edited} ($\colIE$) pixels in the edit-region $\colE$, partitioned by whether $\dE_{76} \leq t$;
and \colCPtxt{correctly preserved} ($\colCP$) and \colIPtxt{incorrectly preserved} ($\colIP$) pixels in the preservation-region $\colP$, similarly partitioned.
We score each problem with the Jaccard index applied to pixel correctness:
\begin{equation}
    \iouat{t} = \frac{|\colCE|}{|\colCE| + |\colIE| + |\colIP|}
    \label{eq:iou}
\end{equation}
This formulation jointly penalizes failure to execute the requested edit (increasing $|\colIE|$) and corruption of the preservation-region (increasing $|\colIP|$).
$\iouat{t}$ is robust to edit-region size, naturally handling the common case where the preservation-region is much larger than the edit-region.

\subsection{Summary Metric: mIoU}
\label{sec:miou}

To capture model performance across the full spectrum of color tolerances (and avoid arbitrarily committing to a single one), we summarize $\iouat{t}$ by sweeping $t$ across $\{0, 1, \ldots, 10\}$.
This is analogous to COCO's Average Precision metric~\cite{lin2014coco}, which sweeps IoU thresholds rather than reporting one.
The resulting summary metric \textbf{\miou} (mean IoU) averages $\iouat{t}$ over these 11 tolerances and all $N$ problems:
\begin{equation}
    \miou = \frac{1}{N} \sum_{i=1}^{N} \frac{1}{11} \sum_{t=0}^{10} \iouat{t}(i)
    \label{eq:ep}
\end{equation}
We use \miou throughout; tolerance-specific values are written as $\iou@t$.

\definecolor{cigrey}{RGB}{136,136,136}
\definecolor{benchbg}{RGB}{240,240,240}
\definecolor{catgeometrictransformationlight}{RGB}{235,242,255}
\definecolor{catgeometrictransformationdark}{RGB}{43,94,167}
\definecolor{catstructuralmanipulationlight}{RGB}{255,243,235}
\definecolor{catstructuralmanipulationdark}{RGB}{192,86,33}
\definecolor{catcolorchangelight}{RGB}{234,250,240}
\definecolor{catcolorchangedark}{RGB}{39,103,73}
\definecolor{catsymbolicreasoninglight}{RGB}{245,238,255}
\definecolor{catsymbolicreasoningdark}{RGB}{107,45,139}

\begin{table}[t]
\centering
\small
\caption{\textbf{\bench\ \miou (\%) per task.} Best score per task in bold. \textit{Category Avg.}\ rows are macro-averages over the 5 tasks in the category; \textit{Benchmark Avg.}\ is the macro-average over 20 tasks with 95\% bootstrap CIs ($\pm$; full intervals in \cref{tab:appx:bootstrap_ci}).}
\vspace{-0.5em}
\label{tab:paintbench_metrics}
\resizebox{\linewidth}{!}{%
\begin{tabular}{lrrrrrrrrrrr}
\toprule
\textbf{Task} & \textbf{NB-2} & \textbf{GPT-I2} & \textbf{NB-1} & \textbf{Qwen-IE} & \textbf{BAGEL} & \textbf{FLUX.2-D} & \textbf{FLUX.1-Kt} & \textbf{LCat-IE} & \textbf{FLUX.2-Kl} & \textbf{HY-3} & \textbf{IP2P} \\
\midrule
\rowcolor{catgeometrictransformationdark}
\multicolumn{12}{l}{\textbf{\color{white}Geometric Transformation}\strut} \\
\rowcolor{catgeometrictransformationlight!50!white}
 & \cellcolor[RGB]{255,236,196}6.1 & \cellcolor[RGB]{254,221,147}\textbf{11.1} & \cellcolor[RGB]{255,236,195}6.2 & \cellcolor[RGB]{255,245,222}3.4 & \cellcolor[RGB]{255,248,232}2.4 & \cellcolor[RGB]{255,245,223}3.3 & \cellcolor[RGB]{255,248,232}2.4 & \cellcolor[RGB]{255,248,233}2.2 & \cellcolor[RGB]{255,251,241}1.4 & \cellcolor[RGB]{255,255,254}0.1 & \cellcolor[RGB]{255,255,255}0.0 \\[-2pt]
\rowcolor{catgeometrictransformationlight!50!white}
\multirow{-2}{*}{\textit{\quad Category Avg.}} & \cellcolor[RGB]{255,236,196}{\color{black}\scriptsize $\pm0.7$} & \cellcolor[RGB]{254,221,147}{\color{black}\scriptsize $\pm0.9$} & \cellcolor[RGB]{255,236,195}{\color{black}\scriptsize $\pm0.8$} & \cellcolor[RGB]{255,245,222}{\color{black}\scriptsize $\pm0.5$} & \cellcolor[RGB]{255,248,232}{\color{black}\scriptsize $\pm0.4$} & \cellcolor[RGB]{255,245,223}{\color{black}\scriptsize $\pm0.5$} & \cellcolor[RGB]{255,248,232}{\color{black}\scriptsize $\pm0.3$} & \cellcolor[RGB]{255,248,233}{\color{black}\scriptsize $\pm0.4$} & \cellcolor[RGB]{255,251,241}{\color{black}\scriptsize $\pm0.3$} & \cellcolor[RGB]{255,255,254}{\color{black}\scriptsize $\pm0.1$} & \cellcolor[RGB]{255,255,255}{\color{black}\scriptsize $\pm0.0$} \\
Translation & \cellcolor[RGB]{254,217,135}12.3 & \cellcolor[RGB]{254,201,90}\textbf{17.5} & \cellcolor[RGB]{254,226,161}9.6 & \cellcolor[RGB]{255,239,204}5.3 & \cellcolor[RGB]{255,244,220}3.6 & \cellcolor[RGB]{255,241,211}4.5 & \cellcolor[RGB]{255,247,229}2.7 & \cellcolor[RGB]{255,243,217}3.9 & \cellcolor[RGB]{255,246,226}3.0 & \cellcolor[RGB]{255,255,254}0.1 & \cellcolor[RGB]{255,255,255}0.0 \\
Rotation & \cellcolor[RGB]{255,232,181}7.6 & \cellcolor[RGB]{254,215,126}\textbf{13.2} & \cellcolor[RGB]{255,233,186}7.1 & \cellcolor[RGB]{255,238,201}5.5 & \cellcolor[RGB]{255,243,216}4.0 & \cellcolor[RGB]{255,238,200}5.6 & \cellcolor[RGB]{255,246,226}3.0 & \cellcolor[RGB]{255,246,227}2.9 & \cellcolor[RGB]{255,249,237}1.9 & \cellcolor[RGB]{255,255,254}0.1 & \cellcolor[RGB]{255,255,254}0.1 \\
Reflection & \cellcolor[RGB]{255,242,212}4.4 & \cellcolor[RGB]{254,227,167}\textbf{9.1} & \cellcolor[RGB]{255,239,205}5.1 & \cellcolor[RGB]{255,246,226}3.0 & \cellcolor[RGB]{255,250,238}1.8 & \cellcolor[RGB]{255,243,216}4.0 & \cellcolor[RGB]{255,247,229}2.7 & \cellcolor[RGB]{255,247,231}2.5 & \cellcolor[RGB]{255,252,245}1.1 & \cellcolor[RGB]{255,255,254}0.1 & \cellcolor[RGB]{255,255,255}0.0 \\
Scaling & \cellcolor[RGB]{255,246,225}3.0 & \cellcolor[RGB]{255,231,179}\textbf{7.8} & \cellcolor[RGB]{255,241,210}4.6 & \cellcolor[RGB]{255,249,237}1.9 & \cellcolor[RGB]{255,252,245}1.0 & \cellcolor[RGB]{255,252,247}0.8 & \cellcolor[RGB]{255,250,238}1.8 & \cellcolor[RGB]{255,253,248}0.7 & \cellcolor[RGB]{255,253,248}0.8 & \cellcolor[RGB]{255,254,253}0.2 & \cellcolor[RGB]{255,255,255}0.0 \\
Shearing & \cellcolor[RGB]{255,246,225}3.1 & \cellcolor[RGB]{255,231,178}\textbf{7.8} & \cellcolor[RGB]{255,242,212}4.4 & \cellcolor[RGB]{255,251,242}1.4 & \cellcolor[RGB]{255,251,241}1.4 & \cellcolor[RGB]{255,250,239}1.6 & \cellcolor[RGB]{255,250,238}1.8 & \cellcolor[RGB]{255,251,244}1.2 & \cellcolor[RGB]{255,253,250}0.5 & \cellcolor[RGB]{255,255,255}0.0 & \cellcolor[RGB]{255,255,255}0.0 \\
\rowcolor{catstructuralmanipulationdark}
\multicolumn{12}{l}{\textbf{\color{white}Structural Manipulation}\strut} \\
\rowcolor{catstructuralmanipulationlight!50!white}
 & \cellcolor[RGB]{254,182,81}22.7 & \cellcolor[RGB]{254,175,77}\textbf{24.5} & \cellcolor[RGB]{254,212,119}14.0 & \cellcolor[RGB]{254,224,155}10.3 & \cellcolor[RGB]{254,225,158}10.0 & \cellcolor[RGB]{255,230,175}8.2 & \cellcolor[RGB]{255,231,177}7.9 & \cellcolor[RGB]{255,233,186}7.1 & \cellcolor[RGB]{255,233,185}7.2 & \cellcolor[RGB]{255,252,247}0.8 & \cellcolor[RGB]{255,252,246}0.9 \\[-2pt]
\rowcolor{catstructuralmanipulationlight!50!white}
\multirow{-2}{*}{\textit{\quad Category Avg.}} & \cellcolor[RGB]{254,182,81}{\color{black}\scriptsize $\pm1.7$} & \cellcolor[RGB]{254,175,77}{\color{black}\scriptsize $\pm1.5$} & \cellcolor[RGB]{254,212,119}{\color{black}\scriptsize $\pm1.5$} & \cellcolor[RGB]{254,224,155}{\color{black}\scriptsize $\pm1.3$} & \cellcolor[RGB]{254,225,158}{\color{black}\scriptsize $\pm1.4$} & \cellcolor[RGB]{255,230,175}{\color{black}\scriptsize $\pm1.1$} & \cellcolor[RGB]{255,231,177}{\color{black}\scriptsize $\pm1.1$} & \cellcolor[RGB]{255,233,186}{\color{black}\scriptsize $\pm1.2$} & \cellcolor[RGB]{255,233,185}{\color{black}\scriptsize $\pm1.0$} & \cellcolor[RGB]{255,252,247}{\color{black}\scriptsize $\pm0.3$} & \cellcolor[RGB]{255,252,246}{\color{black}\scriptsize $\pm0.4$} \\
Construction & \cellcolor[RGB]{254,207,102}\textbf{15.7} & \cellcolor[RGB]{254,211,115}14.3 & \cellcolor[RGB]{255,240,208}4.8 & \cellcolor[RGB]{255,239,203}5.3 & \cellcolor[RGB]{255,252,246}1.0 & \cellcolor[RGB]{255,245,222}3.4 & \cellcolor[RGB]{255,239,204}5.3 & \cellcolor[RGB]{255,248,232}2.4 & \cellcolor[RGB]{255,245,224}3.1 & \cellcolor[RGB]{255,253,249}0.6 & \cellcolor[RGB]{255,251,243}1.2 \\
Removal & \cellcolor[RGB]{207,40,44}{\color{white}45.8} & \cellcolor[RGB]{189,0,38}{\color{white}\textbf{50.6}} & \cellcolor[RGB]{237,105,54}{\color{white}38.2} & \cellcolor[RGB]{253,149,64}31.7 & \cellcolor[RGB]{253,165,72}27.5 & \cellcolor[RGB]{254,188,84}21.1 & \cellcolor[RGB]{254,177,78}24.1 & \cellcolor[RGB]{254,197,88}18.7 & \cellcolor[RGB]{254,182,81}22.6 & \cellcolor[RGB]{255,248,233}2.3 & \cellcolor[RGB]{255,245,224}3.1 \\
Copying & \cellcolor[RGB]{254,212,118}\textbf{14.0} & \cellcolor[RGB]{254,212,119}13.9 & \cellcolor[RGB]{254,216,130}12.8 & \cellcolor[RGB]{255,237,196}6.0 & \cellcolor[RGB]{255,240,208}4.9 & \cellcolor[RGB]{255,251,241}1.4 & \cellcolor[RGB]{255,252,247}0.8 & \cellcolor[RGB]{255,245,224}3.2 & \cellcolor[RGB]{255,241,211}4.5 & \cellcolor[RGB]{255,255,254}0.1 & \cellcolor[RGB]{255,255,255}0.0 \\
Border & \cellcolor[RGB]{254,196,88}\textbf{18.9} & \cellcolor[RGB]{254,208,106}15.2 & \cellcolor[RGB]{255,241,211}4.6 & \cellcolor[RGB]{255,253,249}0.6 & \cellcolor[RGB]{255,255,254}0.1 & \cellcolor[RGB]{255,254,251}0.4 & \cellcolor[RGB]{255,254,252}0.3 & \cellcolor[RGB]{255,255,254}0.1 & \cellcolor[RGB]{255,255,254}0.1 & \cellcolor[RGB]{255,255,254}0.1 & \cellcolor[RGB]{255,255,254}0.1 \\
Cropping & \cellcolor[RGB]{254,195,88}19.1 & \cellcolor[RGB]{253,161,70}\textbf{28.5} & \cellcolor[RGB]{254,226,163}9.5 & \cellcolor[RGB]{255,231,179}7.8 & \cellcolor[RGB]{254,205,94}16.5 & \cellcolor[RGB]{254,210,111}14.7 & \cellcolor[RGB]{254,227,165}9.2 & \cellcolor[RGB]{254,222,149}10.9 & \cellcolor[RGB]{255,239,203}5.4 & \cellcolor[RGB]{255,252,244}1.1 & \cellcolor[RGB]{255,255,254}0.1 \\
\rowcolor{catcolorchangedark}
\multicolumn{12}{l}{\textbf{\color{white}Color Change}\strut} \\
\rowcolor{catcolorchangelight!50!white}
 & \cellcolor[RGB]{254,202,91}\textbf{17.2} & \cellcolor[RGB]{254,213,120}13.8 & \cellcolor[RGB]{255,236,193}6.4 & \cellcolor[RGB]{255,238,202}5.4 & \cellcolor[RGB]{255,247,229}2.6 & \cellcolor[RGB]{255,247,229}2.7 & \cellcolor[RGB]{255,250,240}1.6 & \cellcolor[RGB]{255,250,238}1.8 & \cellcolor[RGB]{255,248,232}2.4 & \cellcolor[RGB]{255,254,253}0.2 & \cellcolor[RGB]{255,254,253}0.2 \\[-2pt]
\rowcolor{catcolorchangelight!50!white}
\multirow{-2}{*}{\textit{\quad Category Avg.}} & \cellcolor[RGB]{254,202,91}{\color{black}\scriptsize $\pm1.6$} & \cellcolor[RGB]{254,213,120}{\color{black}\scriptsize $\pm1.5$} & \cellcolor[RGB]{255,236,193}{\color{black}\scriptsize $\pm1.2$} & \cellcolor[RGB]{255,238,202}{\color{black}\scriptsize $\pm1.1$} & \cellcolor[RGB]{255,247,229}{\color{black}\scriptsize $\pm0.8$} & \cellcolor[RGB]{255,247,229}{\color{black}\scriptsize $\pm0.8$} & \cellcolor[RGB]{255,250,240}{\color{black}\scriptsize $\pm0.5$} & \cellcolor[RGB]{255,250,238}{\color{black}\scriptsize $\pm0.6$} & \cellcolor[RGB]{255,248,232}{\color{black}\scriptsize $\pm0.7$} & \cellcolor[RGB]{255,254,253}{\color{black}\scriptsize $\pm0.1$} & \cellcolor[RGB]{255,254,253}{\color{black}\scriptsize $\pm0.1$} \\
Recolor & \cellcolor[RGB]{253,154,67}\textbf{30.4} & \cellcolor[RGB]{253,159,69}29.0 & \cellcolor[RGB]{255,231,179}7.8 & \cellcolor[RGB]{255,234,188}6.8 & \cellcolor[RGB]{254,228,169}8.8 & \cellcolor[RGB]{255,240,207}5.0 & \cellcolor[RGB]{255,249,235}2.0 & \cellcolor[RGB]{255,248,234}2.2 & \cellcolor[RGB]{255,242,214}4.2 & \cellcolor[RGB]{255,254,252}0.3 & \cellcolor[RGB]{255,254,252}0.3 \\
Flood Fill & \cellcolor[RGB]{254,174,77}24.8 & \cellcolor[RGB]{253,166,73}\textbf{27.1} & \cellcolor[RGB]{254,221,146}11.2 & \cellcolor[RGB]{254,205,96}16.3 & \cellcolor[RGB]{255,248,232}2.3 & \cellcolor[RGB]{255,240,207}4.9 & \cellcolor[RGB]{255,247,229}2.6 & \cellcolor[RGB]{255,241,209}4.7 & \cellcolor[RGB]{255,243,218}3.8 & \cellcolor[RGB]{255,254,252}0.3 & \cellcolor[RGB]{255,254,253}0.2 \\
Blending & \cellcolor[RGB]{255,239,203}5.3 & \cellcolor[RGB]{255,236,193}\textbf{6.4} & \cellcolor[RGB]{255,247,230}2.6 & \cellcolor[RGB]{255,251,243}1.2 & \cellcolor[RGB]{255,252,244}1.1 & \cellcolor[RGB]{255,253,248}0.7 & \cellcolor[RGB]{255,250,239}1.7 & \cellcolor[RGB]{255,252,246}0.9 & \cellcolor[RGB]{255,251,243}1.2 & \cellcolor[RGB]{255,255,254}0.1 & \cellcolor[RGB]{255,255,254}0.1 \\
Gradient & \cellcolor[RGB]{254,215,128}\textbf{13.0} & \cellcolor[RGB]{255,251,241}1.4 & \cellcolor[RGB]{255,246,226}2.9 & \cellcolor[RGB]{255,253,248}0.7 & \cellcolor[RGB]{255,254,253}0.2 & \cellcolor[RGB]{255,253,247}0.8 & \cellcolor[RGB]{255,253,250}0.5 & \cellcolor[RGB]{255,255,254}0.1 & \cellcolor[RGB]{255,252,245}1.1 & \cellcolor[RGB]{255,255,254}0.1 & \cellcolor[RGB]{255,255,255}0.0 \\
Point Operations & \cellcolor[RGB]{254,217,135}\textbf{12.3} & \cellcolor[RGB]{255,238,202}5.4 & \cellcolor[RGB]{255,232,183}7.4 & \cellcolor[RGB]{255,248,234}2.2 & \cellcolor[RGB]{255,253,249}0.7 & \cellcolor[RGB]{255,249,234}2.1 & \cellcolor[RGB]{255,252,245}1.1 & \cellcolor[RGB]{255,252,246}0.9 & \cellcolor[RGB]{255,250,239}1.7 & \cellcolor[RGB]{255,254,252}0.3 & \cellcolor[RGB]{255,253,249}0.6 \\
\rowcolor{catsymbolicreasoningdark}
\multicolumn{12}{l}{\textbf{\color{white}Symbolic Reasoning}\strut} \\
\rowcolor{catsymbolicreasoninglight!50!white}
 & \cellcolor[RGB]{254,182,81}\textbf{22.6} & \cellcolor[RGB]{254,206,100}15.9 & \cellcolor[RGB]{254,199,90}18.0 & \cellcolor[RGB]{255,231,180}7.7 & \cellcolor[RGB]{255,240,206}5.1 & \cellcolor[RGB]{255,242,215}4.1 & \cellcolor[RGB]{255,244,220}3.6 & \cellcolor[RGB]{255,244,221}3.5 & \cellcolor[RGB]{255,245,224}3.2 & \cellcolor[RGB]{255,254,252}0.3 & \cellcolor[RGB]{255,255,254}0.1 \\[-2pt]
\rowcolor{catsymbolicreasoninglight!50!white}
\multirow{-2}{*}{\textit{\quad Category Avg.}} & \cellcolor[RGB]{254,182,81}{\color{black}\scriptsize $\pm1.6$} & \cellcolor[RGB]{254,206,100}{\color{black}\scriptsize $\pm1.4$} & \cellcolor[RGB]{254,199,90}{\color{black}\scriptsize $\pm1.4$} & \cellcolor[RGB]{255,231,180}{\color{black}\scriptsize $\pm1.1$} & \cellcolor[RGB]{255,240,206}{\color{black}\scriptsize $\pm0.9$} & \cellcolor[RGB]{255,242,215}{\color{black}\scriptsize $\pm0.6$} & \cellcolor[RGB]{255,244,220}{\color{black}\scriptsize $\pm0.6$} & \cellcolor[RGB]{255,244,221}{\color{black}\scriptsize $\pm0.6$} & \cellcolor[RGB]{255,245,224}{\color{black}\scriptsize $\pm0.5$} & \cellcolor[RGB]{255,254,252}{\color{black}\scriptsize $\pm0.1$} & \cellcolor[RGB]{255,255,254}{\color{black}\scriptsize $\pm0.1$} \\
Comparison & \cellcolor[RGB]{254,206,98}\textbf{16.1} & \cellcolor[RGB]{254,222,150}10.7 & \cellcolor[RGB]{254,211,116}14.3 & \cellcolor[RGB]{254,216,129}12.9 & \cellcolor[RGB]{255,242,214}4.2 & \cellcolor[RGB]{255,236,195}6.2 & \cellcolor[RGB]{255,237,196}6.0 & \cellcolor[RGB]{255,232,181}7.5 & \cellcolor[RGB]{255,230,175}8.2 & \cellcolor[RGB]{255,254,251}0.4 & \cellcolor[RGB]{255,254,250}0.5 \\
Ordering & \cellcolor[RGB]{254,192,86}20.0 & \cellcolor[RGB]{254,188,84}\textbf{21.0} & \cellcolor[RGB]{254,198,89}18.2 & \cellcolor[RGB]{255,231,177}8.0 & \cellcolor[RGB]{255,240,206}5.1 & \cellcolor[RGB]{255,236,194}6.2 & \cellcolor[RGB]{255,243,217}3.9 & \cellcolor[RGB]{255,241,211}4.5 & \cellcolor[RGB]{255,250,238}1.8 & \cellcolor[RGB]{255,254,253}0.2 & \cellcolor[RGB]{255,255,254}0.1 \\
Pattern & \cellcolor[RGB]{254,214,124}13.4 & \cellcolor[RGB]{254,213,122}\textbf{13.7} & \cellcolor[RGB]{254,229,171}8.7 & \cellcolor[RGB]{255,230,177}8.0 & \cellcolor[RGB]{255,241,212}4.4 & \cellcolor[RGB]{255,239,203}5.4 & \cellcolor[RGB]{255,248,233}2.3 & \cellcolor[RGB]{255,252,247}0.9 & \cellcolor[RGB]{255,245,223}3.2 & \cellcolor[RGB]{255,254,251}0.4 & \cellcolor[RGB]{255,255,255}0.0 \\
Counting & \cellcolor[RGB]{254,205,96}\textbf{16.3} & \cellcolor[RGB]{254,210,110}14.9 & \cellcolor[RGB]{254,210,111}14.8 & \cellcolor[RGB]{254,229,173}8.4 & \cellcolor[RGB]{255,238,201}5.5 & \cellcolor[RGB]{255,250,239}1.6 & \cellcolor[RGB]{255,249,236}1.9 & \cellcolor[RGB]{255,248,234}2.2 & \cellcolor[RGB]{255,248,233}2.3 & \cellcolor[RGB]{255,254,252}0.3 & \cellcolor[RGB]{255,255,255}0.0 \\
Legend & \cellcolor[RGB]{202,30,43}{\color{white}\textbf{47.1}} & \cellcolor[RGB]{254,194,87}19.4 & \cellcolor[RGB]{252,139,60}34.2 & \cellcolor[RGB]{255,252,244}1.1 & \cellcolor[RGB]{255,236,196}6.1 & \cellcolor[RGB]{255,252,245}1.1 & \cellcolor[RGB]{255,243,217}3.9 & \cellcolor[RGB]{255,248,231}2.4 & \cellcolor[RGB]{255,254,252}0.3 & \cellcolor[RGB]{255,255,254}0.1 & \cellcolor[RGB]{255,255,255}0.0 \\
\midrule
\rowcolor{benchbg}
 & \cellcolor[RGB]{254,202,91}\textbf{17.1} & \cellcolor[RGB]{254,205,95}16.3 & \cellcolor[RGB]{254,221,146}11.1 & \cellcolor[RGB]{255,235,190}6.7 & \cellcolor[RGB]{255,240,206}5.0 & \cellcolor[RGB]{255,241,210}4.6 & \cellcolor[RGB]{255,243,217}3.9 & \cellcolor[RGB]{255,244,220}3.6 & \cellcolor[RGB]{255,244,220}3.5 & \cellcolor[RGB]{255,254,251}0.4 & \cellcolor[RGB]{255,254,252}0.3 \\[-2pt]
\rowcolor{benchbg}
\multirow{-2}{*}{\textbf{Benchmark Avg.}} & \cellcolor[RGB]{254,202,91}{\color{black}\scriptsize $\pm0.7$} & \cellcolor[RGB]{254,205,95}{\color{black}\scriptsize $\pm0.7$} & \cellcolor[RGB]{254,221,146}{\color{black}\scriptsize $\pm0.6$} & \cellcolor[RGB]{255,235,190}{\color{black}\scriptsize $\pm0.5$} & \cellcolor[RGB]{255,240,206}{\color{black}\scriptsize $\pm0.5$} & \cellcolor[RGB]{255,241,210}{\color{black}\scriptsize $\pm0.4$} & \cellcolor[RGB]{255,243,217}{\color{black}\scriptsize $\pm0.3$} & \cellcolor[RGB]{255,244,220}{\color{black}\scriptsize $\pm0.4$} & \cellcolor[RGB]{255,244,220}{\color{black}\scriptsize $\pm0.3$} & \cellcolor[RGB]{255,254,251}{\color{black}\scriptsize $\pm0.1$} & \cellcolor[RGB]{255,254,252}{\color{black}\scriptsize $\pm0.1$} \\
\bottomrule
\end{tabular}%
}
\end{table}

\section{Leading Models Fail to Execute Precise Edits}
\label{sec:experiments}

We evaluate 11 models on \bench, organizing our analyses around four findings: operation difficulty and model specialization (\cref{sec:exp:tasks}); common failure modes (\cref{sec:exp:failure-modes}); sensitivity to scene variations (\cref{sec:exp:ablations}); and a pervasive over-editing tendency (\cref{sec:exp:spillover}).

\subsection{Setup}
\label{sec:exp:setup}

We evaluate 11 models spanning distinct architectural families.
\nanobanana~\cite{gemini31flashimage2026}, \nanobananaone~\cite{google2025gemini25flashimage}, and
\gptitwo~\cite{openai2026gptimage2} are closed-weights native multimodal generators; the remaining eight are open-weights.
\qwenedit~\cite{wu2025qwen} is a 20B instruction-tuned diffusion editor.
\fluxdev~\cite{bfl2025flux2} is a 32B flow-matching generator, and \fluxklein~\cite{bfl2025flux2} is a step-distilled variant from the same family targeting sub-second inference.
\fluxkontext~\cite{flux2025kontext} is a 12B rectified flow transformer specialized for instruction-based image editing.
\bagel~\cite{deng2025bagel} is a 7B mixture-of-transformer-experts model.
\hunyuanimage~\cite{cao2025hunyuanimage} is an 80B mixture-of-experts instruction-following image generator.
\longcat~\cite{team2025longcat} is a 6B long-context diffusion editor.
\iptop~\cite{brooks2023instructpix2pix} is a 1B diffusion editor frequently cited as an older baseline.
We generate one output per problem per model (see \cref{sec:appx:details}).

\subsection{Operation difficulty and model specialization}
\label{sec:exp:tasks}

\Cref{tab:paintbench_metrics} reports \miou per task and model across all 1{,}920 \bench problems.
The strongest model (\nanobananaS) reaches only 17.1\% \miou; \gptitwoS follows at 16.3\%, \nanobananaoneS at 11.1\%, and open-weights models range from 6.7\% (\qweneditS) down to below 1\% (\hunyuanimageS, \iptopS).
Remarkably, despite being the largest open-weights model we evaluate (80B), \hunyuanimageS scores near zero on almost all tasks.
Aggregate rankings tell only part of the story; the table reveals a clear difficulty gradient across operation types, and per-task scores expose pronounced model specializations.

\paragraph{Geometric and structural operations are consistently hard.}
The entire geometric transformation category is consistently difficult.
No model exceeds 17.5\% \miou on any geometric task, with shearing and scaling especially difficult across the board ($\leq$ 7.8\% for both).
Most structural manipulation tasks and formula-based color changes are also challenging: unlike single-color operations (recolor and flood fill), tasks such as gradient, blending, and point operations require different pixels to be colored differently, and scores are correspondingly low across all models.

\paragraph{Removal and single-color operations are more tractable.}
Removal reaches 50.6\% (\gptitwoS), flood fill 27.1\% (\gptitwoS), and recolor 30.4\% (\nanobananaS).
Open-weights models are strongest on removal, reaching 31.7\% for \qweneditS and 27.5\% for \bagel.
These tasks all involve filling a connected region with a single color or removing content entirely, which is structurally simpler than the per-pixel computation that gradient, blending, or point operations demand.

\paragraph{Symbolic reasoning presents a mixed picture.}
Pattern is consistently hard, reaching only 13.7\% (\gptitwoS).
Comparison (16.1\% for \nanobananaS) and counting (16.3\% for \nanobananaS) are also fairly challenging.
Ordering is moderately difficult, reaching 21.0\% (\gptitwoS).
Legend is relatively easier for closed-weights models (up to 47.1\% for \nanobananaS) but not for open-weights models (up to 6.1\% for \bagel).

\finding{}{Overall performance is low (best model at 17.1\% \miou), with difficulty tracking the underlying primitive: geometric transformation, formula-based color change, and most structural manipulation tasks are consistently hard, while removal and single-color operations are more tractable, though still far from solved.}

\paragraph{\nanobanana and \gptitwo specialize in complementary categories.}
The two leading models split the four \bench categories down the middle: \gptitwoS leads geometric transformation (11.1\% vs.\ 6.1\%) and structural manipulation (24.5\% vs.\ 22.7\%), while \nanobananaS leads color change (17.2\% vs.\ 13.8\%) and symbolic reasoning (22.6\% vs.\ 15.9\%).
\gptitwoS's geometric transformation advantage holds across every task in the category: translation (17.5\% vs.\ 12.3\%), rotation (13.2\% vs.\ 7.6\%), reflection (9.1\% vs.\ 4.4\%), scaling (7.8\% vs.\ 3.0\%), and shearing (7.8\% vs.\ 3.1\%).
Within structural manipulation, \gptitwoS leads on removal (50.6\% vs.\ 45.8\%) and cropping (28.5\% vs.\ 19.1\%).
Meanwhile, \nanobananaS scores substantially higher than \gptitwoS on gradient (13.0\% vs.\ 1.4\%) and point operations (12.3\% vs.\ 5.4\%).
A notable outlier is legend, where \nanobananaS scores 47.1\% vs.\ \gptitwoS's 19.4\%, a 28-point spread despite comparable overall scores.
We hypothesize this reflects different training-data composition or fine-tuning emphases between the models.

\begin{figure}[t]
    \centering
    \includegraphics[width=0.95\linewidth]{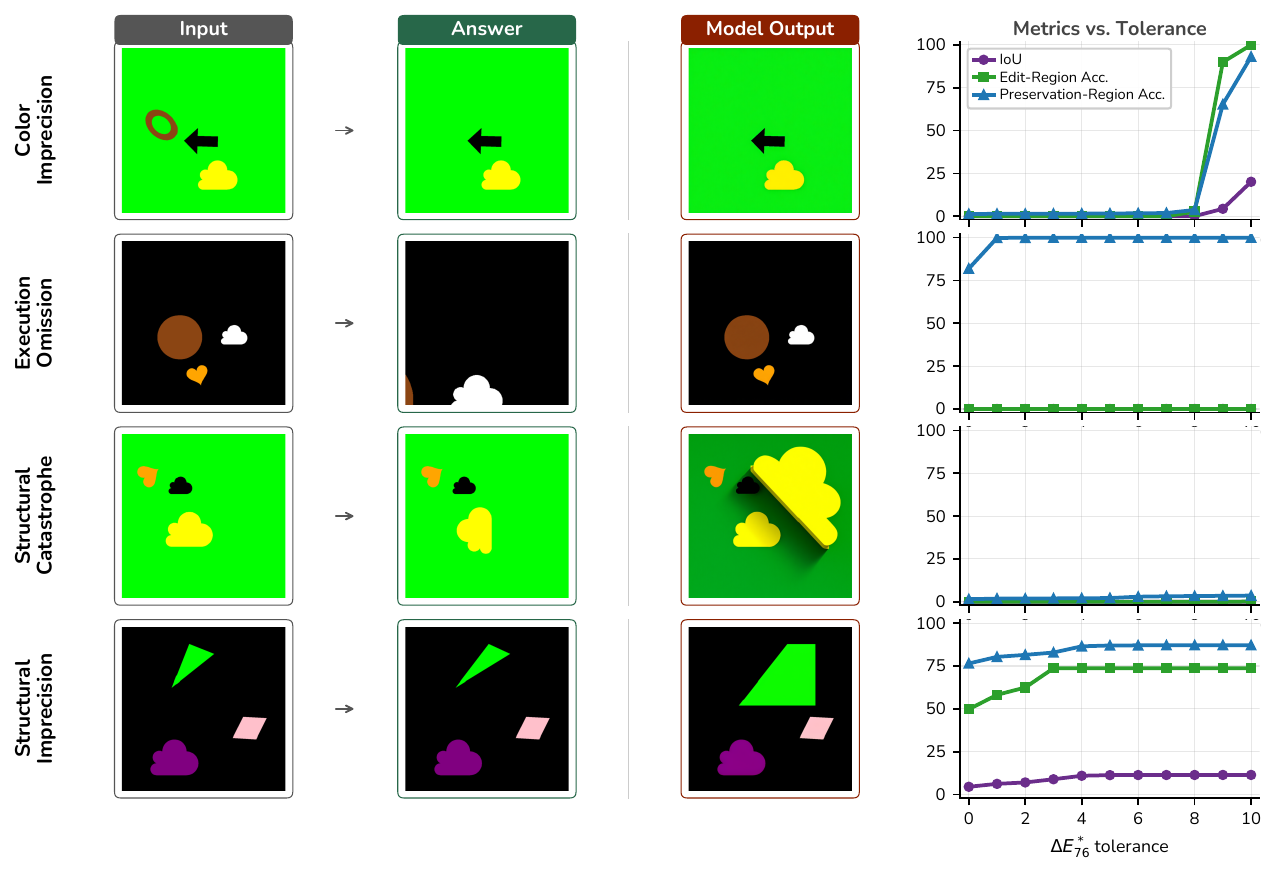}
    \vspace{-1em}
    \caption{
        \textbf{Four common failure modes diagnosed by color-tolerance metric curves.}
        Each row shows an input, answer, and model output, along with the metric curves across $\dE_{76}$ tolerances.
        Curve shape pinpoints the failure mode: \emph{color imprecision} reaches high accuracy only at high tolerances; \emph{execution omission} keeps edit-region accuracy near zero; \emph{structural catastrophe} collapses all three metrics to near-zero; \emph{structural imprecision} plateaus at moderate edit-region accuracy.
    }
    \label{fig:failure-modes}
    \vspace{-1em}
\end{figure}

\paragraph{Individual models depart from their own average on specific tasks.}
Despite leading in geometric transformation and structural manipulation, \gptitwoS scores only 1.4\% on gradient, far below its 16.3\% average and \nanobananaS's 13.0\%.
Among open-weights models, \qweneditS is especially proficient at flood fill (16.3\%, outperforming the closed-weights \nanobananaoneS; 11.4 points ahead of the next open-weights model) and comparison (12.9\%).
\bagel and \fluxdevS are also especially proficient at cropping (16.5\% and 14.7\%, respectively, outperforming the closed-weights \nanobananaoneS).
The border task has a sharp capability cliff: only \nanobananaS (18.9\%), \gptitwoS (15.2\%), and \nanobananaoneS (4.6\%) score substantially; all other models score near zero.

\finding{}{\bench diagnoses task-specific profiles that benchmark averages conceal: \gptitwoS\ and \nanobananaS\ specialize in complementary categories, and individual models exhibit pronounced strengths and weaknesses on specific tasks that their overall score does not predict.}

\subsection{Four common failure modes}
\label{sec:exp:failure-modes}

Beyond aggregate scores, the per-tolerance metric curves diagnose \emph{how} a model fails on individual problems (\cref{fig:failure-modes}).
\textbf{Color imprecision} appears when the edit is structurally correct but the model's output colors deviate from the target: edit-region accuracy is near zero at $\dE_{76} = 0$ but climbs steeply with tolerance, reaching high values only at lenient thresholds.
\textbf{Execution omission} appears when the model fails to attempt the edit at all: edit-region accuracy stays near zero across the entire tolerance range, while preservation accuracy stays high.
\textbf{Structural catastrophe} appears when the model's output bears little resemblance to either the input or the answer: all three metrics collapse to near-zero across all tolerances.
\textbf{Structural imprecision} appears when the model attempts the right edit in roughly the right place but misaligns the affected region: edit-region accuracy plateaus at a moderate value that does not improve with looser color tolerance.
These curve shapes provide a compact diagnostic that pinpoints failure mode from a single per-problem evaluation, surfacing patterns that aggregate \miou scores obscure.

\begin{table}[t]
\centering
\small
\caption{\textbf{Striped backgrounds and high object counts cause the largest \miou drops.} All values are averaged over all 20 \bench tasks. The baseline row shows absolute \miou (\%); all other rows report $\Delta$\miou (\%) relative to baseline ($n=3$ for most tasks, 1024\texttimes{}1024, solid background, standard palette). Cell shading: \colorbox{pos!18}{green} = above baseline, \colorbox{neg!18}{red} = below baseline, white = no change. Per-task breakdowns in \cref{tab:ablations_full_iou}.}
\label{tab:ablations}
\resizebox{\linewidth}{!}{%
\begin{tabular}{lrrrrrrrrrrr}
\toprule
\textbf{Condition} & \textbf{NB-2} & \textbf{GPT-I2} & \textbf{NB-1} & \textbf{Qwen-IE} & \textbf{BAGEL} & \textbf{FLUX.2-D} & \textbf{FLUX.1-Kt} & \textbf{LCat-IE} & \textbf{FLUX.2-Kl} & \textbf{HY-3} & \textbf{IP2P} \\
\midrule
Baseline & \textbf{21.9} & 20.9 & 13.4 & 7.2 & 6.4 & 5.2 & 4.3 & 3.7 & 3.9 & 0.4 & 0.4 \\
\midrule
\multicolumn{12}{@{}l}{\cellcolor{rowheader}\textit{Aspect Ratio} {\scriptsize (baseline: square 1024\texttimes{}1024)}} \\
Horizontal (1024\texttimes{}576) & \cellcolor[RGB]{242,205,205}$-$2.7 & \cellcolor[RGB]{247,226,226}$-$1.6 & \cellcolor[RGB]{246,221,221}$-$1.9 & \cellcolor[RGB]{229,239,230}+1.5 & \cellcolor[RGB]{251,252,251}+0.2 & \cellcolor[RGB]{247,250,247}+0.5 & \cellcolor[RGB]{234,242,234}+1.2 & \cellcolor[RGB]{231,240,231}+1.4 & \cellcolor[RGB]{229,239,230}+1.5 & \cellcolor[RGB]{252,253,252}+0.2 & \cellcolor[RGB]{255,253,253}$-$0.1 \\
Vertical (576\texttimes{}1024) & \cellcolor[RGB]{251,238,238}$-$0.9 & \cellcolor[RGB]{238,191,191}$-$3.6 & \cellcolor[RGB]{255,253,253}$-$0.1 & \cellcolor[RGB]{242,247,242}+0.8 & \cellcolor[RGB]{248,230,230}$-$1.4 & \cellcolor[RGB]{234,242,235}+1.2 & \cellcolor[RGB]{252,245,245}$-$0.6 & \cellcolor[RGB]{254,251,251}$-$0.2 & \cellcolor[RGB]{223,235,224}+1.8 & \cellcolor[RGB]{253,254,253}+0.1 & \cellcolor[RGB]{254,251,251}$-$0.2 \\
\cmidrule(l){1-12}
\multicolumn{12}{@{}l}{\cellcolor{rowheader}\textit{Color Palette} {\scriptsize (baseline: standard palette)}} \\
Nonstandard & \cellcolor[RGB]{234,242,234}+1.2 & \cellcolor[RGB]{222,234,222}+1.9 & \cellcolor[RGB]{184,211,186}+4.0 & \cellcolor[RGB]{231,240,231}+1.4 & \cellcolor[RGB]{246,220,220}$-$1.9 & \cellcolor[RGB]{252,245,245}$-$0.6 & \cellcolor[RGB]{238,244,238}+1.0 & \cellcolor[RGB]{203,223,204}+3.0 & \cellcolor[RGB]{254,252,252}$-$0.2 & \cellcolor[RGB]{255,253,253}$-$0.1 & \cellcolor[RGB]{254,251,251}$-$0.2 \\
\cmidrule(l){1-12}
\multicolumn{12}{@{}l}{\cellcolor{rowheader}\textit{Background} {\scriptsize (baseline: solid)}} \\
Striped & \cellcolor[RGB]{202,55,55}{\color{white}$-$11.1} & \cellcolor[RGB]{213,97,97}{\color{white}$-$8.8} & \cellcolor[RGB]{226,146,146}$-$6.0 & \cellcolor[RGB]{242,207,207}$-$2.7 & \cellcolor[RGB]{244,212,212}$-$2.4 & \cellcolor[RGB]{253,246,246}$-$0.5 & \cellcolor[RGB]{246,221,221}$-$1.9 & \cellcolor[RGB]{245,219,219}$-$2.0 & \cellcolor[RGB]{246,221,221}$-$1.9 & \cellcolor[RGB]{255,255,255}0.0 & \cellcolor[RGB]{254,252,252}$-$0.1 \\
\cmidrule(l){1-12}
\multicolumn{12}{@{}l}{\cellcolor{rowheader}\textit{Object Count} {\scriptsize (baseline: $n=3$ for most tasks; see \cref{tab:appx:n-values})}} \\
$n_{\text{med}}$ & \cellcolor[RGB]{236,183,183}$-$4.0 & \cellcolor[RGB]{236,184,184}$-$4.0 & \cellcolor[RGB]{248,227,227}$-$1.6 & \cellcolor[RGB]{253,248,248}$-$0.4 & \cellcolor[RGB]{248,230,230}$-$1.4 & \cellcolor[RGB]{247,226,226}$-$1.6 & \cellcolor[RGB]{253,247,247}$-$0.4 & \cellcolor[RGB]{254,250,250}$-$0.3 & \cellcolor[RGB]{251,241,241}$-$0.8 & \cellcolor[RGB]{255,255,255}0.0 & \cellcolor[RGB]{252,253,252}+0.2 \\
$n_{\text{high}}$ & \cellcolor[RGB]{213,95,95}{\color{white}$-$8.9} & \cellcolor[RGB]{210,85,85}{\color{white}$-$9.4} & \cellcolor[RGB]{228,152,152}$-$5.7 & \cellcolor[RGB]{246,221,221}$-$1.9 & \cellcolor[RGB]{244,214,214}$-$2.2 & \cellcolor[RGB]{244,212,212}$-$2.4 & \cellcolor[RGB]{250,237,237}$-$1.0 & \cellcolor[RGB]{249,232,232}$-$1.3 & \cellcolor[RGB]{247,225,225}$-$1.7 & \cellcolor[RGB]{254,252,252}$-$0.2 & \cellcolor[RGB]{254,254,254}+0.1 \\
$n_{\text{xhigh}}$ & \cellcolor[RGB]{198,40,40}{\color{white}$-$11.9} & \cellcolor[RGB]{201,51,51}{\color{white}$-$11.3} & \cellcolor[RGB]{221,127,127}{\color{white}$-$7.1} & \cellcolor[RGB]{242,207,207}$-$2.6 & \cellcolor[RGB]{244,214,214}$-$2.3 & \cellcolor[RGB]{246,222,222}$-$1.8 & \cellcolor[RGB]{246,222,222}$-$1.8 & \cellcolor[RGB]{250,235,235}$-$1.1 & \cellcolor[RGB]{248,227,227}$-$1.6 & \cellcolor[RGB]{254,252,252}$-$0.2 & \cellcolor[RGB]{253,249,249}$-$0.3 \\
\bottomrule
\end{tabular}%
}
\end{table}

\subsection{Brittleness to scene variation}
\label{sec:exp:ablations}

\begin{figure}[t]
    \centering
    \includegraphics[width=\linewidth]{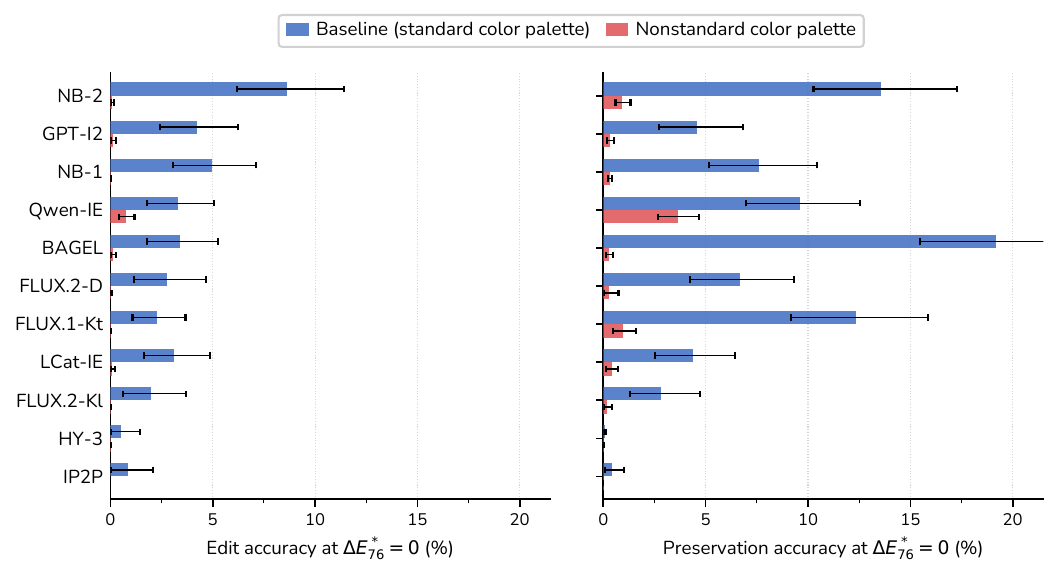}
    \caption{
        \textbf{Nonstandard colors substantially reduce exact pixel-match accuracy.}
        Edit- and preservation-region accuracies at exact pixel-match ($\dE_{76} = 0$) under baseline (standard palette) and nonstandard palettes are displayed for different models.
        The large accuracy gap between the two palettes reveals that models struggle to exactly reproduce nonstandard colors.
    }
    \label{fig:palette-comparison}
\end{figure}

Every \bench task is evaluated under eight visual conditions, enabling direct measurement of how scene variations affect model performance.
\Cref{tab:ablations} reports \miou averaged over all 20 tasks for each condition; per-task breakdowns appear in \cref{tab:ablations_full_iou}.

\paragraph{Striped backgrounds and high object counts cause the largest drops.}
Striped backgrounds knock \nanobanana down 11.1 points and \gptitwo down 8.8; all other models also decline, with the exception of \hunyuanimage, which remains near zero under both conditions.
Performance degrades similarly as object count increases: \gptitwo drops 11.3 points from baseline to $n_{\text{xhigh}}$ and \nanobanana drops 11.9, with $n_{\text{med}}$ and $n_{\text{high}}$ causing proportionally smaller but still substantial declines across almost all models.

\paragraph{Non-square aspect ratios reduce closed-weights performance modestly and inconsistently.}
The three closed-weights models all decline on non-square canvases, but with no consistent directional preference: \nanobanana falls 2.7 points on horizontal and 0.9 on vertical; \gptitwo falls 1.6 on horizontal and 3.6 on vertical (the only model with a larger vertical-axis penalty); \nanobananaone falls 1.9 and 0.1 points respectively.
Open-weights models show no consistent directional effect, with most slightly improving.

\paragraph{Models fail at exactly reproducing nonstandard colors, despite flat \miou.}
At the \miou level, the nonstandard palette appears to produce no consistent directional effect: roughly half the models slightly improve and half slightly decline.
But exact pixel-match ($\dE_{76} = 0$) accuracies tell a different story.
Both edit- and preservation-region accuracies are \textit{much} lower under the nonstandard palette for \textit{every} model (\Cref{fig:palette-comparison}).
To illustrate, \bagel's preservation accuracy plummets from 19.2\% to 0.3\%, and \nanobananaS from 13.6\% to 0.9\%.
Even \qweneditS, the most proficient at reproducing nonstandard colors, achieves only 3.6\% preservation accuracy versus 9.6\% baseline.
We hypothesize this reflects training-data composition: standard colors are far more prevalent.

\finding{}{Models are brittle to scene variation: striped backgrounds and high object counts cause large and consistent \miou drops, and models struggle at exactly reproducing nonstandard colors despite flat headline scores.}

\subsection{Models over-edit relative to the target region}
\label{sec:exp:spillover}

\begin{figure}[t]
    \centering
    \includegraphics[width=\linewidth]{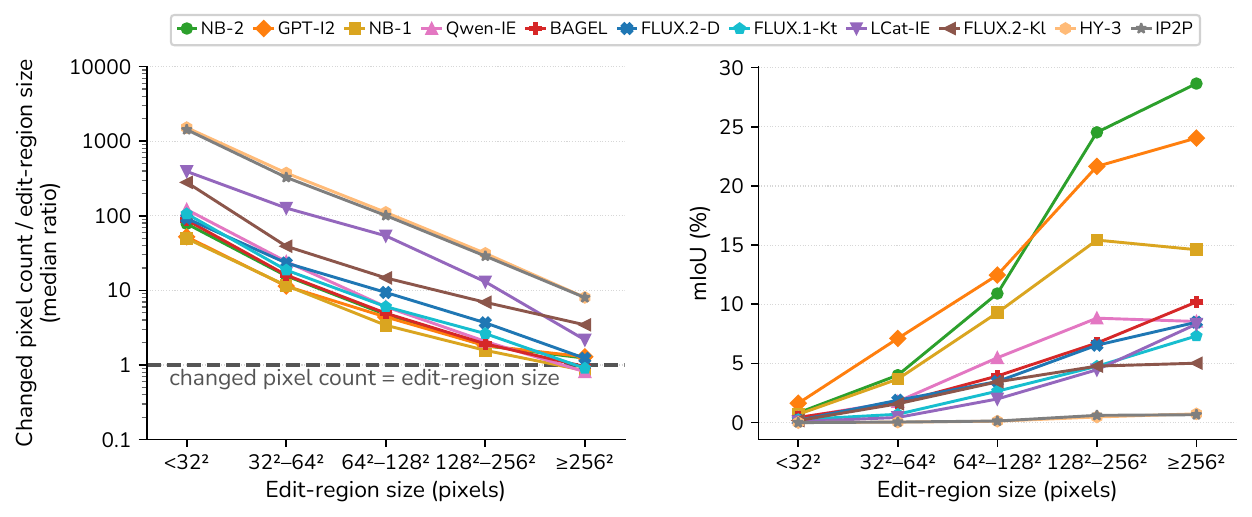}
    \caption{
        \textbf{Models \emph{over-edit} relative to edit-region area, especially for smaller regions, driving worse \miou performance.}
        Left: ratio of changed pixel count (at $\dE_{76} \leq 5$) to edit-region size (median ratio) vs.\ edit-region size.
        Right: \miou (\%) vs.\ edit-region size.
        Bins range from $<\!32^2$ to $\geq\!256^2$ pixels ($<\!0.1\%$ to $\geq\!6\%$ of the $1024^2$ canvas).
        Both metrics degrade sharply for smaller edit-regions, and the pattern holds across all models (each bin contains $\geq 80$ problems per model).
    }
    \label{fig:spillover}
\end{figure}

Beyond the scene parameters we explicitly control for, models change \textit{far} more pixels between input and output than the edit-region requires.
We quantify this as the ratio of changed pixel count (at color tolerance $\dE_{76} \leq 5$\footnote{$\dE_{76} \leq 5$ provides a principled median of changed pixel counts across all $\dE_{76}$ tolerances in $[0, 10]$.}) divided by edit-region size; a perfect edit implies a ratio of 1.
\Cref{fig:spillover} plots this ratio (left) and \miou (right) as a function of edit-region size, bucketed from $<32^2$ to $\geq 256^2$ pixels (roughly $0.1\%$ to $6\%$ of the $1024^2$ canvas).
Median ratios vary dramatically across edit-region size, ranging from $\sim$1\,--\,8$\times$ for the largest edit-regions ($\geq 256^2$ pixels) all the way to $\sim$50\,--\,1{,}400$\times$ for the smallest ($< 32^2$ pixels).
This discrepancy is starkly reflected in \miou: \nanobanana rises from a \miou of only 0.9\% at $< 32^2$ pixels to 28.7\% at $\geq 256^2$ pixels, and \gptitwo from 1.7\% to 24.1\%.
All models exhibit a pronounced increase in \miou as edit-region size increases.

\finding{}{Models over-edit by 1\,--\,8$\times$ the edit-region area for large regions and by 50\,--\,1{,}400$\times$ for the smallest, driving sharp \miou declines on problems with small edit-regions.}

\section{\tinygrafixbench: Generalization Beyond Synthetic Shapes}
\label{sec:tinygrafixbench}

\begin{figure}[t]
    \centering
    \includegraphics[width=0.8\linewidth]{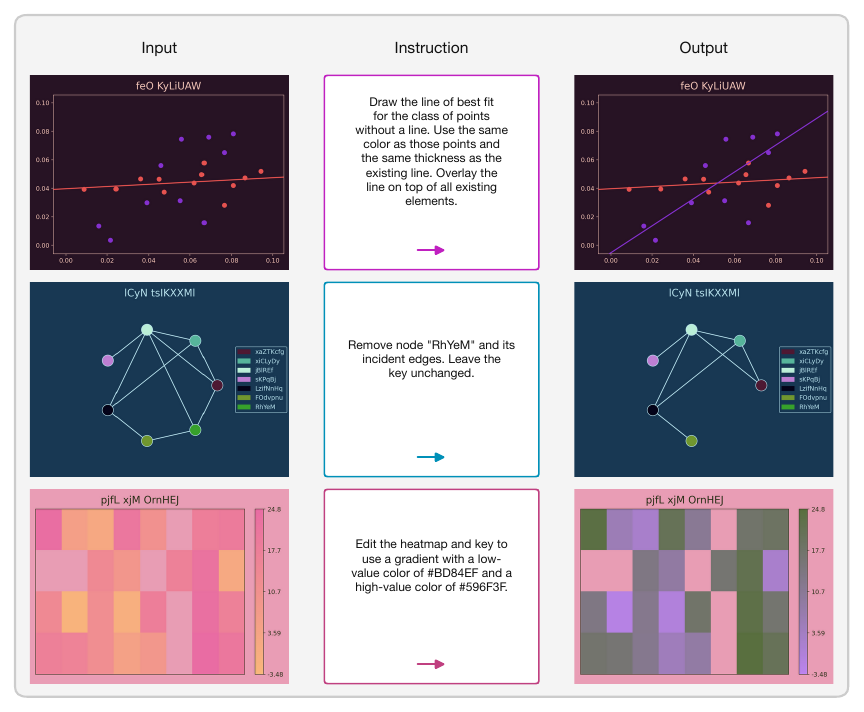}
    \caption{
        \textbf{\tinygrafixbench applies \bench's procedural, deterministic methodology to data visualization.}
        Examples from \texttt{scatter}, \texttt{network}, and \texttt{heatmap} chart types (see all in \cref{sec:appx:tgb-gallery}).
    }
    \label{fig:tinygrafixbench-examples}
\end{figure}

\tinygrafixbench applies \bench's procedurally generated, deterministically evaluated framework to data visualization editing, testing whether \bench's fundamental operations generalize to chart editing tasks.

\paragraph{Benchmark design.}
\tinygrafixbench comprises 600 problems across five chart types (bar chart, scatter plot, line chart, heatmap, network graph), with four tasks per chart type corresponding to four fundamental editing operations: construction, transformation, removal, and recoloring (\cref{fig:tinygrafixbench-examples}).
Charts are rendered with Matplotlib at $1024 \times 768$ pixels using deterministic seeds and evaluated with the same \miou protocol as \bench (see \cref{sec:appx:tgb} for details; full per-task results in \cref{tab:tinygrafixbench_full_iou}).

\paragraph{\tinygrafixbench is harder, but scores track \bench.}
Overall performances on \tinygrafixbench are slightly lower than on \bench, potentially reflecting the additional difficulty of understanding and editing chart imagery.
However, model scores between the benchmarks exhibit a \textit{strong} correlation ($R^2 = 0.91$, $p < 0.001$; \cref{fig:pb-tgb-scatter}).
\nanobanana leads on \tinygrafixbench at 15.9\% and \gptitwo follows close behind at 15.6\%, consistent with their relative standings on \bench; the other models are at 5.3\% (\nanobananaoneS) or below.

\definecolor{cigrey}{RGB}{136,136,136}
\definecolor{benchbg}{RGB}{240,240,240}
\definecolor{chartbarchartlight}{RGB}{235,242,255}
\definecolor{chartbarchartdark}{RGB}{43,94,167}
\definecolor{chartscatterplotlight}{RGB}{234,250,240}
\definecolor{chartscatterplotdark}{RGB}{39,103,73}
\definecolor{chartlinechartlight}{RGB}{255,243,235}
\definecolor{chartlinechartdark}{RGB}{192,86,33}
\definecolor{chartheatmaplight}{RGB}{245,238,255}
\definecolor{chartheatmapdark}{RGB}{107,45,139}
\definecolor{chartnetworklight}{RGB}{230,244,247}
\definecolor{chartnetworkdark}{RGB}{27,110,123}

\begin{table}[t]
\centering
\small
\caption{\textbf{\tinygrafixbench\ \miou (\%) per task.} Best score per task in bold. \textit{Plot Type Avg.}\ rows are macro-averages over the 4 tasks per chart type; \textit{\tinygrafixbench\ Avg.}\ is the macro-average over 5 chart types with 95\% bootstrap CIs ($\pm$; full intervals in \cref{tab:appx:bootstrap_ci_tgb}; task descriptions in \cref{tab:appx:tgb-tasks}).}
\label{tab:tinygrafixbench_metrics}
\resizebox{\linewidth}{!}{%
\begin{tabular}{lrrrrrrrrrrr}
\toprule
\textbf{Task} & \textbf{NB-2} & \textbf{GPT-I2} & \textbf{NB-1} & \textbf{Qwen-IE} & \textbf{BAGEL} & \textbf{FLUX.2-D} & \textbf{FLUX.1-Kt} & \textbf{LCat-IE} & \textbf{FLUX.2-Kl} & \textbf{HY-3} & \textbf{IP2P} \\
\midrule
\rowcolor{chartbarchartdark}
\multicolumn{12}{l}{\textbf{\color{white}Bar Chart}\strut} \\[-2pt]
\rowcolor{chartbarchartlight!50!white}
 & \cellcolor[RGB]{253,156,67}\textbf{38.9} & \cellcolor[RGB]{253,167,73}34.8 & \cellcolor[RGB]{254,228,169}11.5 & \cellcolor[RGB]{255,244,219}4.8 & \cellcolor[RGB]{255,249,234}2.7 & \cellcolor[RGB]{255,250,238}2.3 & \cellcolor[RGB]{255,252,245}1.3 & \cellcolor[RGB]{255,248,233}2.9 & \cellcolor[RGB]{255,248,233}2.9 & \cellcolor[RGB]{255,255,254}0.2 & \cellcolor[RGB]{255,255,254}0.1 \\[-2pt]
\rowcolor{chartbarchartlight!50!white}
\multirow{-2}{*}{\textit{\quad Plot Type Avg.}} & \cellcolor[RGB]{253,156,67}{\color{black}\scriptsize $\pm2.3$} & \cellcolor[RGB]{253,167,73}{\color{black}\scriptsize $\pm2.3$} & \cellcolor[RGB]{254,228,169}{\color{black}\scriptsize $\pm1.4$} & \cellcolor[RGB]{255,244,219}{\color{black}\scriptsize $\pm1.5$} & \cellcolor[RGB]{255,249,234}{\color{black}\scriptsize $\pm1.1$} & \cellcolor[RGB]{255,250,238}{\color{black}\scriptsize $\pm0.7$} & \cellcolor[RGB]{255,252,245}{\color{black}\scriptsize $\pm0.4$} & \cellcolor[RGB]{255,248,233}{\color{black}\scriptsize $\pm0.9$} & \cellcolor[RGB]{255,248,233}{\color{black}\scriptsize $\pm0.6$} & \cellcolor[RGB]{255,255,254}{\color{black}\scriptsize $\pm0.2$} & \cellcolor[RGB]{255,255,254}{\color{black}\scriptsize $\pm0.1$} \\
Add Bar & \cellcolor[RGB]{254,174,77}\textbf{32.3} & \cellcolor[RGB]{254,175,77}31.9 & \cellcolor[RGB]{255,254,252}0.3 & \cellcolor[RGB]{255,254,250}0.6 & \cellcolor[RGB]{255,255,255}0.0 & \cellcolor[RGB]{255,255,255}0.0 & \cellcolor[RGB]{255,255,255}0.0 & \cellcolor[RGB]{255,254,253}0.2 & \cellcolor[RGB]{255,255,255}0.0 & \cellcolor[RGB]{255,255,255}0.0 & \cellcolor[RGB]{255,255,255}0.0 \\
Sort Bars & \cellcolor[RGB]{193,9,39}{\color{white}64.4} & \cellcolor[RGB]{189,0,38}{\color{white}\textbf{65.8}} & \cellcolor[RGB]{254,177,78}31.3 & \cellcolor[RGB]{255,231,177}10.4 & \cellcolor[RGB]{255,245,224}4.1 & \cellcolor[RGB]{255,234,189}8.8 & \cellcolor[RGB]{255,247,228}3.5 & \cellcolor[RGB]{255,234,188}8.9 & \cellcolor[RGB]{254,228,168}11.6 & \cellcolor[RGB]{255,255,254}0.1 & \cellcolor[RGB]{255,254,252}0.4 \\
Remove Bar & \cellcolor[RGB]{254,181,81}\textbf{29.8} & \cellcolor[RGB]{254,224,155}13.3 & \cellcolor[RGB]{254,223,151}13.8 & \cellcolor[RGB]{255,236,195}8.0 & \cellcolor[RGB]{255,239,204}6.8 & \cellcolor[RGB]{255,255,255}0.0 & \cellcolor[RGB]{255,251,243}1.6 & \cellcolor[RGB]{255,250,240}2.1 & \cellcolor[RGB]{255,255,254}0.1 & \cellcolor[RGB]{255,253,250}0.7 & \cellcolor[RGB]{255,255,254}0.1 \\
Recolor Bar & \cellcolor[RGB]{254,183,81}\textbf{29.1} & \cellcolor[RGB]{254,186,83}28.2 & \cellcolor[RGB]{255,253,250}0.6 & \cellcolor[RGB]{255,255,255}0.1 & \cellcolor[RGB]{255,255,255}0.0 & \cellcolor[RGB]{255,254,253}0.3 & \cellcolor[RGB]{255,255,255}0.0 & \cellcolor[RGB]{255,254,251}0.5 & \cellcolor[RGB]{255,255,254}0.1 & \cellcolor[RGB]{255,255,255}0.0 & \cellcolor[RGB]{255,255,255}0.0 \\
\rowcolor{chartscatterplotdark}
\multicolumn{12}{l}{\textbf{\color{white}Scatter Plot}\strut} \\[-2pt]
\rowcolor{chartscatterplotlight!50!white}
 & \cellcolor[RGB]{255,245,224}4.2 & \cellcolor[RGB]{255,239,204}\textbf{6.8} & \cellcolor[RGB]{255,246,227}3.7 & \cellcolor[RGB]{255,252,245}1.3 & \cellcolor[RGB]{255,250,239}2.1 & \cellcolor[RGB]{255,245,223}4.2 & \cellcolor[RGB]{255,245,222}4.3 & \cellcolor[RGB]{255,253,247}1.0 & \cellcolor[RGB]{255,250,240}2.1 & \cellcolor[RGB]{255,255,254}0.1 & \cellcolor[RGB]{255,254,252}0.3 \\[-2pt]
\rowcolor{chartscatterplotlight!50!white}
\multirow{-2}{*}{\textit{\quad Plot Type Avg.}} & \cellcolor[RGB]{255,245,224}{\color{black}\scriptsize $\pm0.6$} & \cellcolor[RGB]{255,239,204}{\color{black}\scriptsize $\pm0.7$} & \cellcolor[RGB]{255,246,227}{\color{black}\scriptsize $\pm0.2$} & \cellcolor[RGB]{255,252,245}{\color{black}\scriptsize $\pm0.2$} & \cellcolor[RGB]{255,250,239}{\color{black}\scriptsize $\pm0.6$} & \cellcolor[RGB]{255,245,223}{\color{black}\scriptsize $\pm0.4$} & \cellcolor[RGB]{255,245,222}{\color{black}\scriptsize $\pm0.3$} & \cellcolor[RGB]{255,253,247}{\color{black}\scriptsize $\pm0.3$} & \cellcolor[RGB]{255,250,240}{\color{black}\scriptsize $\pm0.4$} & \cellcolor[RGB]{255,255,254}{\color{black}\scriptsize $\pm0.0$} & \cellcolor[RGB]{255,254,252}{\color{black}\scriptsize $\pm0.2$} \\
Draw Best Fit Line & \cellcolor[RGB]{255,253,249}0.8 & \cellcolor[RGB]{255,252,247}\textbf{1.1} & \cellcolor[RGB]{255,254,251}0.5 & \cellcolor[RGB]{255,253,250}0.7 & \cellcolor[RGB]{255,253,249}0.8 & \cellcolor[RGB]{255,253,248}0.9 & \cellcolor[RGB]{255,254,252}0.5 & \cellcolor[RGB]{255,255,253}0.2 & \cellcolor[RGB]{255,253,248}0.9 & \cellcolor[RGB]{255,255,255}0.0 & \cellcolor[RGB]{255,255,254}0.1 \\
Swap Axes & \cellcolor[RGB]{255,234,187}9.0 & \cellcolor[RGB]{254,217,135}\textbf{16.0} & \cellcolor[RGB]{254,224,156}13.1 & \cellcolor[RGB]{255,248,231}3.1 & \cellcolor[RGB]{255,239,203}6.9 & \cellcolor[RGB]{254,220,145}14.7 & \cellcolor[RGB]{254,218,137}15.7 & \cellcolor[RGB]{255,247,229}3.4 & \cellcolor[RGB]{255,242,215}5.4 & \cellcolor[RGB]{255,254,253}0.3 & \cellcolor[RGB]{255,253,248}0.9 \\
Remove Outlier & \cellcolor[RGB]{255,254,252}0.4 & \cellcolor[RGB]{255,254,253}0.3 & \cellcolor[RGB]{255,254,253}0.3 & \cellcolor[RGB]{255,254,251}0.5 & \cellcolor[RGB]{255,254,251}0.5 & \cellcolor[RGB]{255,253,248}\textbf{0.9} & \cellcolor[RGB]{255,253,250}0.7 & \cellcolor[RGB]{255,254,253}0.3 & \cellcolor[RGB]{255,253,249}0.8 & \cellcolor[RGB]{255,255,255}0.0 & \cellcolor[RGB]{255,254,253}0.2 \\
Recolor Class & \cellcolor[RGB]{255,239,205}6.6 & \cellcolor[RGB]{255,232,183}\textbf{9.6} & \cellcolor[RGB]{255,253,248}0.9 & \cellcolor[RGB]{255,253,250}0.7 & \cellcolor[RGB]{255,255,254}0.1 & \cellcolor[RGB]{255,254,252}0.4 & \cellcolor[RGB]{255,254,251}0.5 & \cellcolor[RGB]{255,255,254}0.1 & \cellcolor[RGB]{255,252,246}1.2 & \cellcolor[RGB]{255,255,255}0.0 & \cellcolor[RGB]{255,255,254}0.1 \\
\rowcolor{chartlinechartdark}
\multicolumn{12}{l}{\textbf{\color{white}Line Chart}\strut} \\[-2pt]
\rowcolor{chartlinechartlight!50!white}
 & \cellcolor[RGB]{254,228,169}11.5 & \cellcolor[RGB]{254,217,135}\textbf{16.0} & \cellcolor[RGB]{255,244,220}4.7 & \cellcolor[RGB]{255,245,222}4.4 & \cellcolor[RGB]{255,243,218}4.9 & \cellcolor[RGB]{255,246,227}3.8 & \cellcolor[RGB]{255,244,221}4.5 & \cellcolor[RGB]{255,248,231}3.2 & \cellcolor[RGB]{255,243,218}4.9 & \cellcolor[RGB]{255,255,255}0.0 & \cellcolor[RGB]{255,255,254}0.1 \\[-2pt]
\rowcolor{chartlinechartlight!50!white}
\multirow{-2}{*}{\textit{\quad Plot Type Avg.}} & \cellcolor[RGB]{254,228,169}{\color{black}\scriptsize $\pm1.6$} & \cellcolor[RGB]{254,217,135}{\color{black}\scriptsize $\pm1.5$} & \cellcolor[RGB]{255,244,220}{\color{black}\scriptsize $\pm0.4$} & \cellcolor[RGB]{255,245,222}{\color{black}\scriptsize $\pm0.7$} & \cellcolor[RGB]{255,243,218}{\color{black}\scriptsize $\pm0.8$} & \cellcolor[RGB]{255,246,227}{\color{black}\scriptsize $\pm0.7$} & \cellcolor[RGB]{255,244,221}{\color{black}\scriptsize $\pm0.3$} & \cellcolor[RGB]{255,248,231}{\color{black}\scriptsize $\pm0.4$} & \cellcolor[RGB]{255,243,218}{\color{black}\scriptsize $\pm0.9$} & \cellcolor[RGB]{255,255,255}{\color{black}\scriptsize $\pm0.0$} & \cellcolor[RGB]{255,255,254}{\color{black}\scriptsize $\pm0.0$} \\
Draw Segments & \cellcolor[RGB]{255,251,243}\textbf{1.6} & \cellcolor[RGB]{255,253,247}1.0 & \cellcolor[RGB]{255,254,252}0.4 & \cellcolor[RGB]{255,253,247}1.0 & \cellcolor[RGB]{255,253,250}0.6 & \cellcolor[RGB]{255,253,249}0.9 & \cellcolor[RGB]{255,254,252}0.4 & \cellcolor[RGB]{255,254,253}0.3 & \cellcolor[RGB]{255,254,250}0.6 & \cellcolor[RGB]{255,255,255}0.0 & \cellcolor[RGB]{255,255,254}0.1 \\
Normalize Series & \cellcolor[RGB]{255,241,209}6.2 & \cellcolor[RGB]{254,219,139}\textbf{15.5} & \cellcolor[RGB]{254,229,172}11.0 & \cellcolor[RGB]{254,228,168}11.5 & \cellcolor[RGB]{255,236,193}8.3 & \cellcolor[RGB]{255,234,188}8.9 & \cellcolor[RGB]{255,230,174}10.8 & \cellcolor[RGB]{255,235,190}8.6 & \cellcolor[RGB]{255,232,180}10.0 & \cellcolor[RGB]{255,255,255}0.0 & \cellcolor[RGB]{255,255,255}0.1 \\
Filter Series & \cellcolor[RGB]{255,232,181}9.8 & \cellcolor[RGB]{255,232,182}9.7 & \cellcolor[RGB]{255,238,202}7.1 & \cellcolor[RGB]{255,251,243}1.6 & \cellcolor[RGB]{255,231,178}\textbf{10.2} & \cellcolor[RGB]{255,244,220}4.7 & \cellcolor[RGB]{255,239,203}6.9 & \cellcolor[RGB]{255,247,231}3.2 & \cellcolor[RGB]{255,238,201}7.2 & \cellcolor[RGB]{255,255,254}0.1 & \cellcolor[RGB]{255,254,253}0.2 \\
Shade Interval & \cellcolor[RGB]{254,185,82}28.4 & \cellcolor[RGB]{253,158,69}\textbf{37.9} & \cellcolor[RGB]{255,254,253}0.3 & \cellcolor[RGB]{255,247,229}3.4 & \cellcolor[RGB]{255,254,251}0.5 & \cellcolor[RGB]{255,254,251}0.6 & \cellcolor[RGB]{255,255,255}0.0 & \cellcolor[RGB]{255,254,252}0.5 & \cellcolor[RGB]{255,251,241}1.9 & \cellcolor[RGB]{255,255,255}0.0 & \cellcolor[RGB]{255,255,255}0.0 \\
\rowcolor{chartheatmapdark}
\multicolumn{12}{l}{\textbf{\color{white}Heatmap}\strut} \\[-2pt]
\rowcolor{chartheatmaplight!50!white}
 & \cellcolor[RGB]{254,207,103}\textbf{20.2} & \cellcolor[RGB]{254,219,138}15.5 & \cellcolor[RGB]{255,246,225}4.0 & \cellcolor[RGB]{255,243,218}4.9 & \cellcolor[RGB]{255,250,240}2.0 & \cellcolor[RGB]{255,247,231}3.2 & \cellcolor[RGB]{255,247,229}3.5 & \cellcolor[RGB]{255,236,196}7.9 & \cellcolor[RGB]{255,244,219}4.8 & \cellcolor[RGB]{255,253,248}0.9 & \cellcolor[RGB]{255,255,255}0.0 \\[-2pt]
\rowcolor{chartheatmaplight!50!white}
\multirow{-2}{*}{\textit{\quad Plot Type Avg.}} & \cellcolor[RGB]{254,207,103}{\color{black}\scriptsize $\pm2.8$} & \cellcolor[RGB]{254,219,138}{\color{black}\scriptsize $\pm2.5$} & \cellcolor[RGB]{255,246,225}{\color{black}\scriptsize $\pm0.7$} & \cellcolor[RGB]{255,243,218}{\color{black}\scriptsize $\pm1.3$} & \cellcolor[RGB]{255,250,240}{\color{black}\scriptsize $\pm0.5$} & \cellcolor[RGB]{255,247,231}{\color{black}\scriptsize $\pm1.2$} & \cellcolor[RGB]{255,247,229}{\color{black}\scriptsize $\pm0.6$} & \cellcolor[RGB]{255,236,196}{\color{black}\scriptsize $\pm1.4$} & \cellcolor[RGB]{255,244,219}{\color{black}\scriptsize $\pm0.8$} & \cellcolor[RGB]{255,253,248}{\color{black}\scriptsize $\pm0.3$} & \cellcolor[RGB]{255,255,255}{\color{black}\scriptsize $\pm0.0$} \\
Add Cell & \cellcolor[RGB]{255,249,235}2.6 & \cellcolor[RGB]{255,237,198}\textbf{7.6} & \cellcolor[RGB]{255,254,252}0.4 & \cellcolor[RGB]{255,254,250}0.6 & \cellcolor[RGB]{255,255,255}0.0 & \cellcolor[RGB]{255,253,248}1.0 & \cellcolor[RGB]{255,254,252}0.4 & \cellcolor[RGB]{255,254,253}0.2 & \cellcolor[RGB]{255,252,247}1.1 & \cellcolor[RGB]{255,255,255}0.0 & \cellcolor[RGB]{255,255,255}0.0 \\
Shift Heatmap & \cellcolor[RGB]{253,149,64}\textbf{41.1} & \cellcolor[RGB]{254,179,79}30.5 & \cellcolor[RGB]{254,223,153}13.5 & \cellcolor[RGB]{254,222,150}14.0 & \cellcolor[RGB]{255,238,200}7.4 & \cellcolor[RGB]{255,239,204}6.8 & \cellcolor[RGB]{255,231,179}10.1 & \cellcolor[RGB]{254,213,120}18.0 & \cellcolor[RGB]{254,216,131}16.5 & \cellcolor[RGB]{255,250,239}2.1 & \cellcolor[RGB]{255,255,254}0.1 \\
Mask Cells & \cellcolor[RGB]{254,197,89}\textbf{24.1} & \cellcolor[RGB]{254,213,122}17.8 & \cellcolor[RGB]{255,252,246}1.2 & \cellcolor[RGB]{255,244,220}4.6 & \cellcolor[RGB]{255,255,254}0.1 & \cellcolor[RGB]{255,244,219}4.7 & \cellcolor[RGB]{255,247,229}3.4 & \cellcolor[RGB]{254,224,155}13.3 & \cellcolor[RGB]{255,255,253}0.2 & \cellcolor[RGB]{255,252,246}1.2 & \cellcolor[RGB]{255,255,255}0.0 \\
Change Colormap & \cellcolor[RGB]{254,224,157}\textbf{13.1} & \cellcolor[RGB]{255,240,208}6.2 & \cellcolor[RGB]{255,253,249}0.8 & \cellcolor[RGB]{255,254,252}0.4 & \cellcolor[RGB]{255,254,252}0.4 & \cellcolor[RGB]{255,254,252}0.3 & \cellcolor[RGB]{255,255,255}0.0 & \cellcolor[RGB]{255,255,255}0.0 & \cellcolor[RGB]{255,252,244}1.5 & \cellcolor[RGB]{255,255,254}0.2 & \cellcolor[RGB]{255,255,255}0.0 \\
\rowcolor{chartnetworkdark}
\multicolumn{12}{l}{\textbf{\color{white}Network}\strut} \\[-2pt]
\rowcolor{chartnetworklight!50!white}
 & \cellcolor[RGB]{255,244,219}\textbf{4.8} & \cellcolor[RGB]{255,244,219}\textbf{4.8} & \cellcolor[RGB]{255,249,235}2.7 & \cellcolor[RGB]{255,251,243}1.6 & \cellcolor[RGB]{255,251,242}1.8 & \cellcolor[RGB]{255,251,241}1.9 & \cellcolor[RGB]{255,250,241}1.9 & \cellcolor[RGB]{255,253,249}0.8 & \cellcolor[RGB]{255,249,237}2.4 & \cellcolor[RGB]{255,255,254}0.1 & \cellcolor[RGB]{255,255,254}0.1 \\[-2pt]
\rowcolor{chartnetworklight!50!white}
\multirow{-2}{*}{\textit{\quad Plot Type Avg.}} & \cellcolor[RGB]{255,244,219}{\color{black}\scriptsize $\pm0.6$} & \cellcolor[RGB]{255,244,219}{\color{black}\scriptsize $\pm0.8$} & \cellcolor[RGB]{255,249,235}{\color{black}\scriptsize $\pm0.4$} & \cellcolor[RGB]{255,251,243}{\color{black}\scriptsize $\pm0.4$} & \cellcolor[RGB]{255,251,242}{\color{black}\scriptsize $\pm0.5$} & \cellcolor[RGB]{255,251,241}{\color{black}\scriptsize $\pm0.4$} & \cellcolor[RGB]{255,250,241}{\color{black}\scriptsize $\pm0.2$} & \cellcolor[RGB]{255,253,249}{\color{black}\scriptsize $\pm0.3$} & \cellcolor[RGB]{255,249,237}{\color{black}\scriptsize $\pm0.6$} & \cellcolor[RGB]{255,255,254}{\color{black}\scriptsize $\pm0.0$} & \cellcolor[RGB]{255,255,254}{\color{black}\scriptsize $\pm0.1$} \\
Add Node & \cellcolor[RGB]{255,253,249}0.7 & \cellcolor[RGB]{255,252,246}\textbf{1.2} & \cellcolor[RGB]{255,254,251}0.5 & \cellcolor[RGB]{255,254,251}0.6 & \cellcolor[RGB]{255,253,248}1.0 & \cellcolor[RGB]{255,254,253}0.3 & \cellcolor[RGB]{255,254,253}0.3 & \cellcolor[RGB]{255,255,254}0.1 & \cellcolor[RGB]{255,254,251}0.6 & \cellcolor[RGB]{255,255,255}0.0 & \cellcolor[RGB]{255,255,254}0.1 \\
Swap Nodes & \cellcolor[RGB]{255,243,217}\textbf{5.0} & \cellcolor[RGB]{255,243,217}\textbf{5.1} & \cellcolor[RGB]{255,246,226}3.8 & \cellcolor[RGB]{255,250,239}2.1 & \cellcolor[RGB]{255,251,243}1.6 & \cellcolor[RGB]{255,248,232}3.1 & \cellcolor[RGB]{255,249,234}2.8 & \cellcolor[RGB]{255,254,252}0.5 & \cellcolor[RGB]{255,247,230}3.4 & \cellcolor[RGB]{255,255,254}0.2 & \cellcolor[RGB]{255,255,254}0.2 \\
Remove Node & \cellcolor[RGB]{255,234,187}\textbf{9.1} & \cellcolor[RGB]{255,235,191}8.6 & \cellcolor[RGB]{255,240,207}6.3 & \cellcolor[RGB]{255,247,229}3.5 & \cellcolor[RGB]{255,244,221}4.6 & \cellcolor[RGB]{255,246,225}4.0 & \cellcolor[RGB]{255,244,220}4.6 & \cellcolor[RGB]{255,249,236}2.5 & \cellcolor[RGB]{255,242,213}5.6 & \cellcolor[RGB]{255,255,254}0.1 & \cellcolor[RGB]{255,254,253}0.3 \\
Recolor Node & \cellcolor[RGB]{255,245,222}4.4 & \cellcolor[RGB]{255,245,221}\textbf{4.5} & \cellcolor[RGB]{255,254,253}0.2 & \cellcolor[RGB]{255,255,254}0.2 & \cellcolor[RGB]{255,255,254}0.1 & \cellcolor[RGB]{255,255,255}0.1 & \cellcolor[RGB]{255,255,255}0.0 & \cellcolor[RGB]{255,255,255}0.0 & \cellcolor[RGB]{255,255,254}0.1 & \cellcolor[RGB]{255,255,255}0.0 & \cellcolor[RGB]{255,255,255}0.0 \\
\midrule
\rowcolor{benchbg}
 & \cellcolor[RGB]{254,218,135}\textbf{15.9} & \cellcolor[RGB]{254,218,138}15.6 & \cellcolor[RGB]{255,242,215}5.3 & \cellcolor[RGB]{255,247,230}3.4 & \cellcolor[RGB]{255,249,235}2.7 & \cellcolor[RGB]{255,248,232}3.1 & \cellcolor[RGB]{255,248,232}3.1 & \cellcolor[RGB]{255,248,231}3.2 & \cellcolor[RGB]{255,247,229}3.4 & \cellcolor[RGB]{255,254,253}0.3 & \cellcolor[RGB]{255,255,254}0.2 \\[-2pt]
\rowcolor{benchbg}
\multirow{-2}{*}{\textbf{\tinygrafixbench\ Avg.}} & \cellcolor[RGB]{254,218,135}{\color{black}\scriptsize $\pm0.8$} & \cellcolor[RGB]{254,218,138}{\color{black}\scriptsize $\pm0.8$} & \cellcolor[RGB]{255,242,215}{\color{black}\scriptsize $\pm0.3$} & \cellcolor[RGB]{255,247,230}{\color{black}\scriptsize $\pm0.4$} & \cellcolor[RGB]{255,249,235}{\color{black}\scriptsize $\pm0.3$} & \cellcolor[RGB]{255,248,232}{\color{black}\scriptsize $\pm0.3$} & \cellcolor[RGB]{255,248,232}{\color{black}\scriptsize $\pm0.2$} & \cellcolor[RGB]{255,248,231}{\color{black}\scriptsize $\pm0.3$} & \cellcolor[RGB]{255,247,229}{\color{black}\scriptsize $\pm0.3$} & \cellcolor[RGB]{255,254,253}{\color{black}\scriptsize $\pm0.1$} & \cellcolor[RGB]{255,255,254}{\color{black}\scriptsize $\pm0.1$} \\
\bottomrule
\end{tabular}%
}
\end{table}

\begin{figure}[t]
    \centering
    \includegraphics[width=0.85\linewidth]{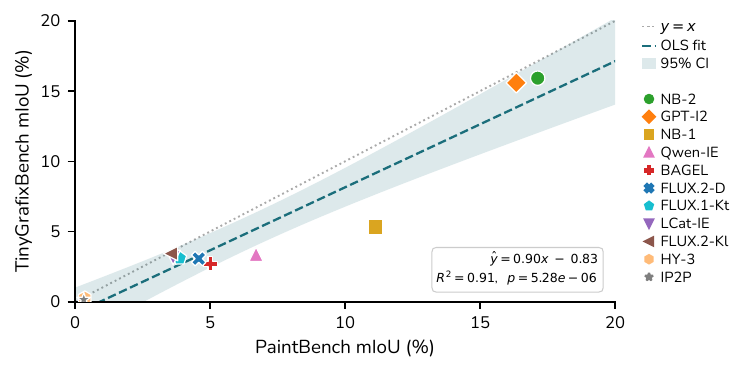}
    \caption{
        \textbf{\bench\ and \tinygrafixbench\ \miou scores are strongly linearly correlated across models.}
        Each point is one model. The OLS fit (\textcolor{olsteal}{dashed teal}, with \colorbox{olsteal!18}{95\% CI}) closely tracks $y = x$ (\textcolor{cigrey}{dotted gray}); regression yields $R^2 = 0.91$, $p < 0.001$.
    }
    \label{fig:pb-tgb-scatter}
\end{figure}

\paragraph{Model-specific patterns.}
Per-task results in \cref{tab:tinygrafixbench_metrics} reveal distinct model-specific patterns.
A clear capability gap separates \nanobanana from \nanobananaone on several tasks where \nanobanana performs well: add bar (32.3\% vs.\ 0.3\%), recolor bar (29.1\% vs.\ 0.6\%), shade interval (28.4\% vs.\ 0.3\%), and mask cells (24.1\% vs.\ 1.2\%).
Among open-weights models, task differentiation is notable: \bagel leads on filter series (10.2\%, above any closed-weights model), \fluxkontext on swap axes (15.7\%), \longcat on shift heatmap (18.0\%), and \fluxklein on sort bars (11.6\%).
Meanwhile, tasks requiring precise placement of small elements (\eg, draw best-fit line, add node, draw segments) are especially hard for all models.

\finding{}{Model scores on \tinygrafixbench are strongly correlated with those on \bench ($R^2 = 0.91$, $p < 0.001$), suggesting that \bench captures capabilities that generalize to applied visual editing tasks.}

\newpage
\section{Discussion}
\label{sec:conclusion}

We introduce \bench and \tinygrafixbench on the premise that a meaningful class of visual
editing tasks that contain unique correct outputs has been underserved by
evaluation frameworks built for subjective, open-ended generation.
By constructing problems procedurally from random seeds, both benchmarks support pixel-level
evaluation without bias-prone judge models or perceptual proxies, while protecting against contamination.
Because \bench enables controllable difficulty and scene variation,
practitioners can generate custom fine-grained task sets,
turning the benchmark from a static snapshot into a configurable diagnostic instrument.

Our experiments reveal that current image editing models cannot reliably execute basic raster
operations despite strong open-ended generation performance.
Geometric transformation, formula-based color change, and most structural manipulation tasks are effectively unsolved
across all models; even the most tractable removal and recoloring tasks are far
from reliably executed.
Yet amidst the generally low scores, distinct model specializations in different tasks and task categories emerge.
Our analysis shows that models are brittle to scene variation in various forms: high shape counts, striped backgrounds, nonstandard colors, and small edit-regions.
Promisingly, the strong correlation between \bench and \tinygrafixbench scores suggests that \bench captures capabilities that generalize to applied visual editing tasks built on similar primitives.

\paragraph{Limitations and future directions.}
Decomposing evaluation into edit- and preservation-regions provides an assumption-free assessment of model behavior.
However, depending on real-world application, this framework may warrant adjustment.
When the edit-region is small or thin, for instance, small spatial errors can cause a model to miss the edit-region entirely, resulting in stricter grading than
the magnitude of error may necessitate.
For certain applications, continuously weighted or
task-specific scoring functions may be more appropriate. Furthermore, precise editing tasks with non-unique answers (such as those involving text rendering or non-unique ways to fill the background exposed by a transformed shape) may require more complex deterministic evaluation metrics. Nevertheless, we encourage practitioners to carefully consider the feasibility of adopting a deterministic metric for the task at hand, and choose one that best captures the most important considerations.

This work focuses on 2D raster editing, but the procedurally generated, deterministically evaluated approach we advocate for can extend to a broader range of tasks with unique correct edits: scientific visualization, engineering drawing, physics and game simulation, 3D scene manipulation, and more. Our framework delivers an infinitude of \bench problems; the infinitude of tasks awaiting this broader paradigm is a torch for future researchers to carry.

\section*{Acknowledgements}

We are grateful to Vaibhavi Singh, Michael Hu, Valerie Chen, Jihan Yang, Xichen Pan, Peter Tong, and Chris Hoang for helpful feedback and discussions throughout this project.
K.X. and H.H. are supported by grants from Amazon and Cisco.
E.B. is supported by the NDSEG Fellowship.
S.X. acknowledges support from the MSIT IITP grant (RS-2024-00457882) and NSF Award IIS-2443404.
This work uses computing resources provided by NYU Torch and High Performance Computing.

\clearpage
\phantomsection
\addcontentsline{toc}{section}{References}
\bibliographystyle{unsrt}
\bibliography{references}

\clearpage
\appendix
\renewcommand{\theHsection}{appx.\Alph{section}}

\addtocontents{toc}{\protect\setcounter{tocdepth}{1}}

\phantomsection
\addcontentsline{toc}{section}{Appendix}

\section*{Appendix}\label{sec:appx}

\noindent This appendix provides benchmark construction details, experimental details,
qualitative examples, and extended results supporting the main paper:

\begin{itemize}[noitemsep,topsep=0pt,parsep=0pt,partopsep=0pt,leftmargin=1.5em]
    \item (\S\ref{sec:appx:construction}) \textbf{Benchmark Construction}:
        shape library, color system, scene generation, seed design,
        and evaluation pipeline.
    \item (\S\ref{sec:appx:details}) \textbf{Experimental Details}:
        model specifications, inference parameters, and benchmark generation
        pipeline configuration.
    \item (\S\ref{sec:appx:additional-experiments}) \textbf{Additional Experiments}:
        prompt augmentation case study using multimodal reasoning traces from \gemini to augment \nanobanana.
    \item (\S\ref{sec:appx:extended-results}) \textbf{Extended Results}:
        full per-task and per-mode results for \bench and \tinygrafixbench.
    \item (\S\ref{sec:appx:qualitative}) \textbf{Model Output Galleries}:
        per-problem galleries for \bench and \tinygrafixbench,
        showing model outputs alongside ground-truth answers.
\end{itemize}

\section{Benchmark Construction}
\label{sec:appx:construction}

\bench and \tinygrafixbench are generated entirely procedurally.
\cref{tab:appx:taxonomy} lists the 20 \bench tasks (grouped into four categories,
with task-mode breakdowns);
\cref{tab:appx:tgb-tasks} in \cref{sec:appx:tgb} lists the 20 \tinygrafixbench tasks.
The remainder of this section documents the design choices behind both benchmarks and
the shared evaluation pipeline; high-level configuration (problem counts, palette
definitions, visual condition parameters) lives in \cref{sec:appx:details}.

\begin{table}[ht]
    \centering
    \caption{\textbf{\bench\ task taxonomy.} 4 categories, 20 task types, 35 task-modes. Single-mode tasks are listed without a mode qualifier.}
    \label{tab:appx:taxonomy}
    \footnotesize
    \setlength{\tabcolsep}{4pt}
    \begin{tabular}{@{}l l l@{}}
        \toprule
        \textbf{Category} & \textbf{Task} & \textbf{Modes} \\
        \midrule
        \multirow{5}{*}{\shortstack[l]{\textit{Geometric}\\[0pt]\textit{Transformation}}}
            & Translation     & amount, align \\
            & Rotation        & local, external \\
            & Reflection      & local, external \\
            & Scaling         & amount, match \\
            & Shearing        & --- \\
        \midrule
        \multirow{5}{*}{\shortstack[l]{\textit{Structural}\\[0pt]\textit{Manipulation}}}
            & Construction    & circle, line, polygon \\
            & Removal         & attribute, location \\
            & Copying         & --- \\
            & Border          & --- \\
            & Cropping        & straight, tilted \\
        \midrule
        \multirow{5}{*}{\textit{Color Change}}
            & Recolor         & color\_code, dropper \\
            & Flood Fill      & background, foreground \\
            & Blending        & --- \\
            & Gradient        & background, foreground \\
            & Point Ops       & brightness, grayscale, invert \\
        \midrule
        \multirow{5}{*}{\shortstack[l]{\textit{Symbolic}\\[0pt]\textit{Reasoning}}}
            & Comparison      & --- \\
            & Ordering        & --- \\
            & Pattern         & grid, circular \\
            & Counting        & shape, color \\
            & Legend          & --- \\
        \bottomrule
    \end{tabular}
\end{table}

\subsection{\bench: Per-Task Descriptions}
\label{sec:appx:tasks}

The four \bench task categories are described in detail below; \cref{tab:appx:taxonomy}
above lists the 35 task-modes.

\paragraph{Geometric Transformation}
tests affine transformations of shapes.
\textit{Translation} moves a shape by a specified displacement (\texttt{amount} mode) or so that one of its named control points aligns with a named reference point elsewhere in the scene (\texttt{align} mode); the reference may be on another shape or on the canvas.
\textit{Rotation} rotates a shape by a specified angle, about a named control point on the shape itself (\texttt{local} mode) or about an external pivot (\texttt{external} mode); the external pivot is either a named reference point elsewhere in the scene or a random canvas coordinate.
\textit{Reflection} mirrors a shape across one of eight axes of its bounding box (four edges, two center lines, two diagonals; \texttt{local} mode) or across a line defined by two external points (\texttt{external} mode).
\textit{Scaling} resizes a shape by a specified multiplicative factor (\texttt{amount} mode) or to match a bounding-box dimension of another shape (\texttt{match} mode).
\textit{Shearing} applies a horizontal or vertical shear relative to a fixed bounding-box edge or center line.

\paragraph{Structural Manipulation}
tests addition and removal of elements, and modification
of scene composition.
\textit{Construction} places a new shape of specified type (line, circle, or polygon), color,
size, and position.
\textit{Removal} deletes a shape identified by an attribute such as color or shape type
(\texttt{attribute} mode) or by its canvas position (\texttt{location} mode).
\textit{Copying} duplicates a specified shape to a new canvas location.
\textit{Border} adds a colored border around a specified shape.
\textit{Cropping} extracts and upsamples a subregion of the canvas centered at a control point,
either aligned to canvas axes (\texttt{straight} mode) or at an arbitrary angle (\texttt{tilted} mode).

\paragraph{Color Change}
tests manipulation of pixel color values.
\textit{Recolor} requires changing all shapes matching a description to a target color, specified
either by hex code (\texttt{color\_code} mode) or by naming an existing scene color to match
(\texttt{dropper} mode).
\textit{Flood Fill} fills a connected region bounded by a shape with a new color; modes select
whether the edit-region is the background (\texttt{background}) or foreground (\texttt{foreground}).
\textit{Blending} alpha-blends a specified color at a given opacity over a target shape region.
\textit{Gradient} applies a linear gradient inside a parallelogram region; modes select
background or foreground edit-region.
\textit{Point Operations} applies a per-pixel intensity transformation (brightness adjustment,
grayscale conversion, or color inversion) to selected shapes.

\paragraph{Symbolic Reasoning}
tests edits that require spatial or numerical inference before execution.
\textit{Comparison} requires identifying and removing the shape at a specified size rank
(\eg, the second-largest).
\textit{Ordering} rearranges all shapes of a given type along an axis in ascending or descending
size order.
\textit{Pattern} completes a missing cell in a two-dimensional repeating grid, inferred from
the surrounding context; modes vary between rectangular (\texttt{grid}) and circular (\texttt{circular})
arrangements.
\textit{Counting} adjusts a tally strip to match the number of shapes meeting a criterion; the
model must determine the count by inspection, as the instruction does not state it;
modes select whether to count by shape type (\texttt{shape}) or color (\texttt{color}).
\textit{Legend} applies a set of color substitutions and shape removals specified by a key already present in the
image; the model must read the key and recolor or remove the corresponding shapes.

\subsection{\bench: Shape Library}
\label{sec:appx:shapes}

\bench scenes are composed from a vocabulary of 12 shape types,
each tagged with two constraints that govern its appearance.
\textit{Rotatable} shapes may be randomly oriented; non-rotatable shapes either look
identical under rotation (circle) or are semantically axis-aligned (rectangle, cross,
cloud).
\textit{Width/height-free} shapes allow independent width and height scaling, so their
aspect ratio varies across problems; fixed-ratio shapes always use their canonical
proportions.

\begin{table}[ht]
    \centering
    \caption{\textbf{Shape library properties.} ``R'' = rotatable; ``AR-free'' = width and height may be set independently.}
    \label{tab:appx:shapes}
    \footnotesize
    \setlength{\tabcolsep}{5pt}
    \begin{tabular}{@{}lcc@{\quad}lcc@{}}
        \toprule
        \textbf{Shape} & \textbf{R} & \textbf{AR-free} & \textbf{Shape} & \textbf{R} & \textbf{AR-free} \\
        \midrule
        Circle     & --           & --           & Arrow      & \checkmark & \checkmark \\
        Rectangle  & --           & \checkmark   & Heart      & \checkmark & --         \\
        Cloud      & --           & --           & Star       & \checkmark & --         \\
        Hexagon    & \checkmark   & --           & Semicircle & \checkmark & --         \\
        Triangle   & \checkmark   & --           & Cross      & --         & \checkmark \\
        Ring       & \checkmark   & \checkmark   & Diamond    & \checkmark & \checkmark \\
        \bottomrule
    \end{tabular}
\end{table}

Each shape type also exposes a set of named control points (e.g.\ ``center,''
``tip,'' ``30-degree vertex,'' ``arc midpoint'').
Task generators use these to express instructions in terms of named, semantically
meaningful locations.
Every placed shape additionally inherits nine axis-aligned bounding-box control points
(four corners, four edge midpoints, and the bounding-box center), which the geometric tasks
draw on alongside the intrinsic control points when constructing instructions. The canvas
itself contributes nine analogous reference points for instructions that target
absolute image locations.

\subsection{\bench: Color System}
\label{sec:appx:colors}

Two 11-color palettes are defined.
The \textit{standard} palette uses the exact web-color hex codes for common color names
(\eg, \textcolor[HTML]{FF0000}{\texttt{\#FF0000}} for red, \textcolor[HTML]{0000FF}{\texttt{\#0000FF}} for blue).
The \textit{nonstandard} palette uses perceptually distinct variants with uncommon hex codes
(\eg, gold at \textcolor[HTML]{E4BA18}{\texttt{\#E4BA18}} rather than \textcolor[HTML]{FFD700}{\texttt{\#FFD700}}).

At generation time, the active palette is shuffled deterministically using the problem seed.
The first shuffled color becomes the background fill; the second is a ``holdout'' color
reserved for two-color striped backgrounds.
The remaining nine colors populate the object color pool.
The background and holdout colors are excluded from the object pool,
ensuring that no shape visually merges with the background.

\begin{table}[ht]
\centering
\small
\renewcommand{\arraystretch}{1.25}
\setlength{\tabcolsep}{4pt}
\caption{\textbf{PaintBench color palettes.}
The standard palette uses common web hex codes; the nonstandard palette uses perceptually distinct variants with uncommon codes.}
\label{tab:appx:palettes}
\begin{tabular}{p{1.5em}ll@{\hspace{2em}}p{1.5em}ll}
\toprule
\multicolumn{3}{l}{\textbf{Standard palette}} & \multicolumn{3}{l}{\textbf{Nonstandard palette}} \\
\midrule
\cellcolor[RGB]{255,0,0}\phantom{~~}    & red    & \textcolor[HTML]{FF0000}{\texttt{\#FF0000}} & \cellcolor[RGB]{195,27,55}\phantom{~~}   & crimson           & \textcolor[HTML]{C31B37}{\texttt{\#C31B37}} \\
\cellcolor[RGB]{255,165,0}\phantom{~~}  & orange & \textcolor[HTML]{FFA500}{\texttt{\#FFA500}} & \cellcolor[RGB]{244,123,22}\phantom{~~}  & tangerine-colored & \textcolor[HTML]{F47B16}{\texttt{\#F47B16}} \\
\cellcolor[RGB]{255,255,0}\phantom{~~}  & yellow & \colorbox{black}{\textcolor[HTML]{FFFF00}{\texttt{\#FFFF00}}} & \cellcolor[RGB]{228,186,24}\phantom{~~}  & gold              & \textcolor[HTML]{E4BA18}{\texttt{\#E4BA18}} \\
\cellcolor[RGB]{0,255,0}\phantom{~~}    & green  & \textcolor[HTML]{00FF00}{\texttt{\#00FF00}} & \cellcolor[RGB]{113,122,30}\phantom{~~}  & olive-colored     & \textcolor[HTML]{717A1E}{\texttt{\#717A1E}} \\
\cellcolor[RGB]{0,0,255}\phantom{~~}    & blue   & \textcolor[HTML]{0000FF}{\texttt{\#0000FF}} & \cellcolor[RGB]{15,225,223}\phantom{~~}  & cyan              & \textcolor[HTML]{0FE1DF}{\texttt{\#0FE1DF}} \\
\cellcolor[RGB]{128,0,128}\phantom{~~}  & purple & \textcolor[HTML]{800080}{\texttt{\#800080}} & \cellcolor[RGB]{217,210,233}\phantom{~~} & lavender          & \colorbox{black}{\textcolor[HTML]{D9D2E9}{\texttt{\#D9D2E9}}} \\
\cellcolor[RGB]{255,192,203}\phantom{~~}& pink   & \colorbox{black}{\textcolor[HTML]{FFC0CB}{\texttt{\#FFC0CB}}} & \cellcolor[RGB]{242,13,216}\phantom{~~}  & magenta           & \textcolor[HTML]{F20DD8}{\texttt{\#F20DD8}} \\
\cellcolor[RGB]{139,69,19}\phantom{~~}  & brown  & \textcolor[HTML]{8B4513}{\texttt{\#8B4513}} & \cellcolor[RGB]{203,170,133}\phantom{~~} & tan-colored       & \textcolor[HTML]{CBAA85}{\texttt{\#CBAA85}} \\
\cellcolor[RGB]{0,0,0}\phantom{~~}      & black  & \textcolor[HTML]{000000}{\texttt{\#000000}} & \cellcolor[RGB]{16,18,17}\phantom{~~}    & jet black         & \textcolor[HTML]{101211}{\texttt{\#101211}} \\
\cellcolor[RGB]{128,128,128}\phantom{~~}& gray   & \textcolor[HTML]{808080}{\texttt{\#808080}} & \cellcolor[RGB]{187,188,186}\phantom{~~} & silver            & \textcolor[HTML]{BBBCBA}{\texttt{\#BBBCBA}} \\
\cellcolor[RGB]{255,255,255}\phantom{~~}& white  & \colorbox{black}{\textcolor[HTML]{FFFFFF}{\texttt{\#FFFFFF}}} & \cellcolor[RGB]{248,246,232}\phantom{~~} & ivory white       & \colorbox{black}{\textcolor[HTML]{F8F6E8}{\texttt{\#F8F6E8}}} \\
\bottomrule
\end{tabular}
\end{table}

\subsection{\bench: Background Rendering}
\label{sec:appx:background}

Main \bench problems use solid single-color backgrounds.
The \emph{striped} visual condition replaces the solid fill with a two-color striped
pattern alternating the background and holdout colors;
parameters (orientation $\in \{0^\circ, 45^\circ, 90^\circ\}$, band width $\in \{6\%, 8\%, 10\%\}$
of canvas width, and waveform sampled uniformly from \texttt{line}, \texttt{sine},
\texttt{square}, \texttt{triangle}, \texttt{sawtooth}) are randomized per problem,
producing a diverse but fully deterministic set of striped patterns.

\subsection{\bench: Scene Generation}
\label{sec:appx:scene-gen}

A scene consists of multiple shapes drawn on a background.
The same scene-generation routine is used across all \bench tasks: shapes are placed sequentially, and for each shape the generator samples a type, color, size, aspect ratio (for AR-free
shapes), rotation (for rotatable shapes), and canvas position, then rejects any
candidate whose axis-aligned bounding box overlaps an already-placed shape;
a 4-pixel gap margin is enforced so shape boundaries are always separated by at least
a few pixels.

Shape size is sampled from a density-dependent range
$[\ell_{\min}, \ell_{\max}]$ expressed as fractions of the shorter canvas dimension:
\[
    \ell_{\min} = \max\!\left(0.02,\, \frac{0.18}{\sqrt{n_{\text{mid}}}}\right),
    \qquad
    \ell_{\max} = \min\!\left(0.40,\, \frac{0.55}{\sqrt{n_{\text{mid}}}}\right),
\]
where $n_{\text{mid}}$ is the midpoint of the $[n_{\min}, n_{\max}]$ count range.
This $1/\sqrt{n}$ scaling keeps individual shapes readable as scene density increases.
For AR-free shapes, the aspect ratio is sampled log-uniformly from $[0.4, 2.5]$;
rectangle, ring, and diamond shapes additionally exclude the near-square band $[0.8, 1.25]$
to prevent visual ambiguity with tilted squares.

To enforce visual diversity, no two shapes in the same scene may share the same
(type, color) combination, and at most $\lceil n/3 \rceil$ shapes may share the same color.

\subsection{\bench: Seed Design}
\label{sec:appx:seeds}

Seeds are derived deterministically via SHA-256 from the string 
\begin{center}
    \small\texttt{"paintbench|\{task\}|\{cond\}|\{mode\}|\{slot\}|\{attempt\}"},
\end{center}
making them immune to Python hash randomization and stable across machines.
Every problem is uniquely identified by its (task, visual condition, mode, slot) tuple,
and each combination is assigned its own independent seed.
For each (task, condition, mode, slot), the generator searches sequentially over attempt
indices until a valid scene is produced; an attempt that fails validation (for instance,
due to shape placement collisions) is discarded.
\tinygrafixbench uses its own seed namespace (\cref{sec:appx:tgb}),
so no seed-space interactions exist between the two benchmarks.

\subsection{\tinygrafixbench: Benchmark Design}
\label{sec:appx:tgb}

\tinygrafixbench applies the same deterministic-edit framework to a different visual
domain: Matplotlib-rendered analytical charts instead of synthetic shape scenes.
Five chart types (\texttt{bar\_chart}, \texttt{heatmap}, \texttt{line\_chart},
\texttt{network}, \texttt{scatter\_plot}) each expose four editing tasks, giving
$5 \times 4 = 20$ task-modes; \cref{tab:appx:tgb-tasks} summarizes the task catalog and
\cref{fig:tinygrafixbench-examples} in \cref{sec:tinygrafixbench} shows three illustrative chart types (full per-chart galleries in \cref{sec:appx:tgb-gallery}).
All figures are rendered at $1024 \times 768$ pixels (160\,dpi, $6.4 \times 4.8$ inches)
using a bundled DejaVuSans font, so chart pixels are byte-identical across machines.
The on-disk layout (input PNG, answer PNG, instruction JSON per problem) is identical
to \bench, so the same inference and evaluation pipeline applies without modification.

\begin{table}[ht]
    \centering
    \caption{\textbf{\tinygrafixbench task catalog.} Each chart type's four tasks span the four editing operations: construction, transformation, removal, and recoloring. Brief descriptions in {\color{cigrey}gray}.}
    \label{tab:appx:tgb-tasks}
    \footnotesize
    \setlength{\tabcolsep}{4pt}
    \renewcommand{\arraystretch}{1.15}
    \begin{tabular}{@{}l p{0.20\linewidth} p{0.20\linewidth} p{0.19\linewidth} p{0.19\linewidth}@{}}
        \toprule
        \textbf{Chart type} & \textbf{Construction} & \textbf{Transformation} & \textbf{Removal} & \textbf{Recoloring} \\
        \midrule
        Bar chart    &
            Add Bar \newline {\scriptsize\color{cigrey}add missing bar} &
            Sort Bars \newline {\scriptsize\color{cigrey}sort bars ascending or descending} &
            Remove Bar \newline {\scriptsize\color{cigrey}remove bar and label} &
            Recolor Bar \newline {\scriptsize\color{cigrey}recolor one bar} \\
        Scatter plot &
            Draw Best Fit Line \newline {\scriptsize\color{cigrey}draw OLS best-fit line} &
            Swap Axes \newline {\scriptsize\color{cigrey}swap $x$ and $y$ coordinates} &
            Remove Outlier \newline {\scriptsize\color{cigrey}remove max-residual point} &
            Recolor Class \newline {\scriptsize\color{cigrey}recolor points and best-fit line} \\
        Line chart   &
            Draw Segments \newline {\scriptsize\color{cigrey}connect gaps using line segments} &
            Normalize Series \newline {\scriptsize\color{cigrey}stretch and shift series vertically to fit in range} &
            Filter Series \newline {\scriptsize\color{cigrey}clip to $y$ values above or below threshold} &
            Shade Interval \newline {\scriptsize\color{cigrey}shade area under curve within interval} \\
        Heatmap      &
            Add Cell \newline {\scriptsize\color{cigrey}fill empty cell with value-corresponding color} &
            Shift Heatmap \newline {\scriptsize\color{cigrey}shift cells in a direction} &
            Mask Cells \newline {\scriptsize\color{cigrey}mask cells above or below threshold} &
            Change Colormap \newline {\scriptsize\color{cigrey}change colormap gradient} \\
        Network      &
            Add Node \newline {\scriptsize\color{cigrey}add node and incident edges} &
            Swap Nodes \newline {\scriptsize\color{cigrey}swap two node positions} &
            Remove Node \newline {\scriptsize\color{cigrey}remove node and incident edges} &
            Recolor Node \newline {\scriptsize\color{cigrey}recolor node in graph and legend} \\
        \bottomrule
    \end{tabular}
\end{table}

\paragraph{State-then-render architecture.}
Each chart module factors generation into three stages.
A seeded state-builder produces a single base chart description (data,
colors, labels, title, ranges) for a given slot.
Each task function then takes that base state, mutates a copy to produce the desired
edit, and returns \emph{two} states (input and answer) plus a natural-language
instruction.
A single deterministic renderer per chart type emits both the input
and answer figures.
This factorization is essential for evaluation: any pixel difference between input
and answer comes entirely from the state edit, not from rendering nondeterminism, so
small ground-truth deltas (e.g.\ removing one outlier point) are recoverable to the
exact pixel.

\paragraph{Visual style sampling.}
Bg/text color pairs are sampled to guarantee luminance contrast (30\% dark-bg/light-text,
70\% light-bg/dark-text).
Object colors (bars, classes, nodes, colormap endpoints) are sampled uniformly in CIE~L*a*b* space,
rejecting any draw within $\Delta E^*_{76} \le 20$ of an \texttt{avoid} list that always
includes the background; multi-color palettes are built one color at a time with the
running set added to the avoid list, so all colors in a scene are mutually
perceptually distinct.
Titles, axis labels, and bar/node labels are random gibberish strings of
1--3 letter sequences each, so models cannot lean on real-world label semantics
(``height of building'' would prime a different distribution than ``\texttt{Qkx Lpvm}'').
A per-problem magnitude factor sampled from $\{10^{-3}, 10^{-2}, \ldots, 10^{3}\}$ scales
all numerical ranges, so axis values span seven orders of magnitude across the
benchmark.

\paragraph{Numerical-textual consistency.}
Every numerical value that appears in an instruction is rounded to three significant
figures \emph{before} being used both in the instruction string and in the answer
state's render parameters.
Without this step, the value displayed by Matplotlib (e.g.\ \texttt{0.123}) and the
value the model is told to draw (e.g.\ \texttt{0.12345}) could drift, producing a
small, persistent edit-region error invisible to the dataset author.

\paragraph{Unambiguity by construction.}
Each chart-edit task is constructed so a unique correct answer exists.
Bar values are resampled until every pair differs by at least 5\% of the $y$-axis
maximum, so the sorted order is visually unambiguous.
Line-chart gaps are placed with $\ge 2$ visible vertices between them, so every input
segment renders as a line (not a degenerate dot).
Heatmap base states always contain at least one empty cell (NaN), so
\texttt{add\_cell} has a target.
For scatter plots, the class with a best-fit line is constructed by drawing points along a line, then pushing one randomly chosen point by 3.5 noise standard deviations along the sign of its natural residual and clipping it back to the axis;
this guarantees the maximum-residual point is the same one in both input and answer,
so the ``remove outlier'' task has a unique solution.

\paragraph{Seed namespace.}
A SHA-256 seed derived from \texttt{``tinygrafixbench|\{graph\}|\{task\}|\{slot\}''}
determines all chart parameters and the target transformation;
no seed search is needed (every \texttt{generate\_task} is constructed to always
succeed), so the slot index alone identifies a problem.

\subsection{Evaluation Pipeline}
\label{sec:appx:eval-pipeline}

The evaluation pipeline is shared across \bench and \tinygrafixbench.
Given input $I$, answer $A$, and model output $O$ (all $W \times H$ RGB images):

\paragraph{Output normalization.}
$O$ is rescaled (preserving aspect ratio) and center-cropped to $A$'s resolution.
We use nearest-neighbor interpolation rather than bilinear or bicubic: smooth
interpolation would synthesize intermediate colors at edit boundaries that do not exist in $I$ or $A$.

\paragraph{Change mask.}
$M_{\mathrm{edit}}[p] = \mathbf{1}[I[p] \neq A[p]]$ identifies the $E$ pixels that changed
from input to answer.
The remaining $P = WH - E$ pixels are \emph{preservation} pixels (background and unchanged
shapes that the model should leave untouched).

\paragraph{CIE76 distance.}
$O$ and $A$ are converted from sRGB to CIE~L*a*b* (D65 illuminant, standard IEC~61966-2-1
piecewise linearization), and the per-pixel CIE76 distance is
$\Delta E^*_{76}[p] = \|O_{\mathrm{lab}}[p] - A_{\mathrm{lab}}[p]\|_2$.

\paragraph{Per-tolerance metrics.}
For each tolerance $t \in \{0, 1, \ldots, 10\}$, pixel $p$ is declared correct iff
$\Delta E^*_{76}[p] \le t$.
Three metrics are computed (consistent with the CE / IE / CP / IP partition used in \cref{sec:iou}):
\begin{align*}
    \text{Edit accuracy} &= \frac{|\{p : M_{\mathrm{edit}}[p] = 1 \wedge \Delta E^*_{76}[p] \le t\}|}{E}, \\
    \text{Preservation accuracy} &= \frac{|\{p : M_{\mathrm{edit}}[p] = 0 \wedge \Delta E^*_{76}[p] \le t\}|}{P}, \\
    \text{IoU} &= \frac{|\{p : M_{\mathrm{edit}}[p] = 1 \wedge \Delta E^*_{76}[p] \le t\}|}{E + |\{p : M_{\mathrm{edit}}[p] = 0 \wedge \Delta E^*_{76}[p] > t\}|}.
\end{align*}
IoU penalizes both missed edits and erroneous modifications to preservation-regions,
analogous to intersection over union for segmentation masks.

\paragraph{Mean-tolerance IoU.}
The primary reported metric is the mean IoU over all 11 tolerances
$t \in \{0,\ldots,10\}$, averaging over a range of color tolerance levels rather
than committing to a single tolerance.
Tolerances 0--10 span from exact pixel match ($t = 0$) to a lenient tolerance
($t = 10$).

\clearpage

\section{Experimental Details}
\label{sec:appx:details}

\subsection{Models}
\label{sec:appx:model-specs}

\Cref{tab:model_specs} summarizes the eleven models evaluated, spanning
proprietary native multimodal generators (\nanobanana, \nanobananaone, \gptitwo)
and open-weights flow-matching and diffusion editors (\fluxkontext, \fluxdev, \fluxklein,
\qwenedit, \longcat, \bagel, \iptop, \hunyuanimage).
Open-weights models are accessed under their respective published licenses:
\begin{center}
\footnotesize
\begin{tabular}{@{}l@{\quad}l@{}}
\qwenedit, \longcat, \bagel & Apache 2.0 \\
\iptop                      & MIT \\
\hunyuanimage               & Tencent Hunyuan Community License \\
\fluxkontext                & FLUX.1\,[dev] Non-Commercial License v1.1.1 \\
\fluxdev                    & FLUX\,[dev] Non-Commercial License v2.0 \\
\fluxklein                  & FLUX Non-Commercial License v2.1 \\
\end{tabular}
\end{center}
Proprietary models (\nanobanana, \nanobananaone, \gptitwo) are accessed via official commercial APIs under the providers' terms of service.

\begin{table}[t]
\centering
\small
\caption{\textbf{Model specifications.} Eleven models evaluated on \bench\ and \tinygrafixbench, spanning proprietary native multimodal generators and open-weights flow-matching and diffusion editors. All models perform image editing conditioned on a natural-language instruction.}
\label{tab:model_specs}
\resizebox{\linewidth}{!}{%
\begin{tabular}{lllcl}
\toprule
\textbf{Model} & \textbf{Architecture} & \textbf{Params} & \textbf{Open weights} & \textbf{Reference} \\
\midrule
\nanobanana    & Native multimodal generator        & ---             & ---        & \cite{gemini31flashimage2026} \\
\nanobananaone & Native multimodal generator        & ---             & ---        & \cite{google2025gemini25flashimage} \\
\gptitwo       & Native multimodal generator        & ---             & ---        & \cite{openai2026gptimage2} \\
\qwenedit      & Diffusion-based editor             & 20B             & \checkmark & \cite{wu2025qwen} \\
\longcat       & Diffusion-based editor             & 6B               & \checkmark & \cite{team2025longcat} \\
\bagel         & Mixture-of-Transformers generator  & 7B              & \checkmark & \cite{deng2025bagel} \\
\fluxkontext   & Rectified flow editor              & 12B             & \checkmark & \cite{flux2025kontext} \\
\fluxdev       & Flow-matching generator            & 32B             & \checkmark & \cite{bfl2025flux2} \\
\fluxklein     & Flow-matching generator (distilled)   & 9B              & \checkmark & \cite{bfl2025flux2} \\
\hunyuanimage  & MoE instruction-following generator & 80B (13B active) & \checkmark & \cite{cao2025hunyuanimage} \\
\iptop         & Diffusion-based editor (DDPM)         & 1B              & \checkmark & \cite{brooks2023instructpix2pix} \\
\bottomrule
\end{tabular}%
}
\end{table}

\subsection{Inference Parameters}
\label{sec:appx:inference}

\Cref{tab:inference_params} lists the generation parameters for the eight
locally-run models.
The proprietary API-only models (\nanobanana, \nanobananaone, \gptitwo) use the
provider's default API sampling and have no locally-controllable inference parameters.
All locally-run models are evaluated with a fixed random seed.
Inference was run on NVIDIA H200 GPUs via a Slurm cluster.

\begin{table}[t]
\centering
\small
\caption{\textbf{Inference parameters.} Generation settings for the eight locally-run models, taken from the per-run \texttt{inference\_metrics} sidecar JSONs. The proprietary API-only models (\nanobanana, \nanobananaone, \gptitwo) use the provider's default API sampling and are omitted.}
\label{tab:inference_params}
\begin{tabular}{lll}
\toprule
\textbf{Model} & \textbf{Steps} & \textbf{CFG Scale} \\
\midrule
\qwenedit      & 50                     & 4.0$^{\dagger}$ \\
\longcat       & 50                     & 4.0$^{\dagger}$ \\
\bagel         & 50                     & 4.0$^{\ddagger}$ \\
\fluxkontext   & 28                     & 3.5 \\
\fluxdev       & 50                     & 4.0 \\
\fluxklein     & 50                     & 4.0 \\
\hunyuanimage  & 8$^{\P}$               & --- \\
\iptop         & 100                    & 7.5 / 1.5$^{\S}$ \\
\bottomrule
\end{tabular}
\\[4pt]
\begin{flushleft}
\footnotesize{
$^{\dagger}$Pipeline default for the \qweneditS\ / \longcatS\ family
(\texttt{true\_cfg\_scale}); not overridden in our runs. \\
$^{\ddagger}$\bagelS\ uses dual-CFG: \texttt{cfg\_text\_scale}$=4.0$,
\texttt{cfg\_img\_scale}$=2.0$. \\
$^{\P}$\hunyuanimageS\ uses 8-step distilled sampling via its own
\texttt{generate\_image()} pipeline; classifier-free guidance is not
separately configurable. \\
$^{\S}$\iptop\ uses two guidance scales: text CFG $=$ 7.5, image CFG $=$ 1.5.
}
\end{flushleft}
\end{table}

\subsection{Benchmark Configuration}
\label{sec:appx:bench-config}

Construction-side details (shape vocabulary, palette definitions, scene-generation rules,
seed scheme) appear in \cref{sec:appx:construction};
this subsection records the configuration values used in our experiments.

\paragraph{Problem counts.}
The \bench\ test set contains 1{,}920 problems
(20 tasks $\times$ 8 visual conditions $\times$ 12 problems per task-condition cell),
rendered at either $1024 \times 1024$ (baseline and most conditions) or $1024 \times 576$ / $576 \times 1024$ (the \textit{horizontal} and \textit{vertical} aspect-ratio conditions).
The baseline condition uses a single-color background drawn from the standard palette and $n=3$ shapes for most tasks (see \cref{tab:appx:n-values} for task-specific values).
Visual conditions each change exactly one parameter; see \cref{tab:appx:ablation-variants} for the full enumeration.
\tinygrafixbench contributes 600 problems (20 tasks $\times$ 30 problems)
at $1024 \times 768$ resolution.

\paragraph{Visual conditions.}
Eight visual conditions are baked into every task of \bench, each changing exactly one parameter relative to the baseline.
Each condition receives its own independent set of seeds, so problems across conditions are drawn from distinct random scenes.

\begin{table}[h]
    \centering
    \caption{\textbf{Visual conditions.} Each condition changes exactly one scene parameter relative to the baseline. All 20 tasks are rendered for all 8 conditions at 12 problems each, for $20 \times 8 \times 12 = 1{,}920$ problems total.}
    \label{tab:appx:ablation-variants}
    \footnotesize
    \setlength{\tabcolsep}{6pt}
    \begin{tabular}{@{}llrlll@{}}
        \toprule
        \textbf{Condition} & \textbf{Axis} & \textbf{Canvas} & \textbf{$n$} & \textbf{Palette} & \textbf{Background} \\
        \midrule
        baseline    & ---             & $1024 \times 1024$ &  3 & standard    & solid   \\
        horizontal  & aspect ratio    & $1024 \times \phantom{0}576$  &  3 & standard    & solid   \\
        vertical    & aspect ratio    & $\phantom{0}576 \times 1024$  &  3 & standard    & solid   \\
        nonstandard & palette         & $1024 \times 1024$ &  3 & nonstandard & solid   \\
        striped     & background type & $1024 \times 1024$ &  3 & standard    & striped \\
        $n_{\text{med}}$    & object count    & $1024 \times 1024$ & 10 & standard    & solid   \\
        $n_{\text{high}}$   & object count    & $1024 \times 1024$ & 25 & standard    & solid   \\
        $n_{\text{xhigh}}$  & object count    & $1024 \times 1024$ & 60 & standard    & solid   \\
        \bottomrule
    \end{tabular}
\end{table}

\paragraph{Object counts by task.}
Most tasks use $n=3$ at baseline ($n_\text{low}=3$); ablation modes use $n_{\text{med}}=10$, $n_{\text{high}}=25$, and $n_{\text{xhigh}}=60$.
Three task groups use adjusted ranges suited to the structure of their task; \cref{tab:appx:n-values} lists the exact values and justifications.

\begin{table}[h]
    \centering
    \caption{\textbf{Object counts for different tasks ($n$).} Most tasks use default levels; four use adjusted ranges. Object count is the only variable changed by the $n_{\text{med}}$, $n_{\text{high}}$, and $n_{\text{xhigh}}$ conditions.}
    \label{tab:appx:n-values}
    \footnotesize
    \setlength{\tabcolsep}{5pt}
    \begin{tabular}{@{}llcccc p{0.30\linewidth}@{}}
        \toprule
        \textbf{Group} & \textbf{Tasks} & \textbf{baseline} & $n_{\text{med}}$ & $n_{\text{high}}$ & $n_{\text{xhigh}}$ & \textbf{Justification} \\
        \midrule
        Default & all other 17 tasks & 3 & 10 & 25 & 60 & Standard range covering sparse to very dense scenes. \\[3pt]
        Comparison, Ordering & comparison, ordering & 3 & 5 & 7 & 9 & Ranking $n$ shapes requires precise size discrimination; upper cap is reduced to keep the spatial-reasoning task visually tractable. \\[3pt]
        Pattern & pattern & 1 & 3 & 6 & 10 & $n$ indexes grid cells; starts at 1 (a single populated cell to infer from) and caps at 10 to preserve legible grid structure. \\[3pt]
        Counting & counting & 5 & 10 & 25 & 60 & Baseline raised to $n=5$ to ensure a non-trivial count; upper range matches the default. \\
        \bottomrule
    \end{tabular}
\end{table}

\paragraph{Natural-language instructions.}
Instructions are generated automatically from each problem's transformation parameters
and fully specify the target shape(s), the operation, and all parameters needed to
produce the unique answer.
Each problem is saved as an input PNG, an answer PNG, and an instruction JSON sidecar
that also records the seed and any task-specific metadata, enabling downstream analysis of problem characteristics beyond those reported in this paper.

\clearpage

\section{Additional Experiments}
\label{sec:appx:additional-experiments}

This section presents a case study exploring how prompt augmentation via reasoning traces from a multimodal language model affects image editing performance.

\subsection{Prompt Augmentation via Reasoning Traces}
\label{sec:appx:prompt-augmentation}

Can reasoning traces and structured solutions generated by a multimodal language model improve the performance of image editing models?
Standard pipelines pass the instruction and input image directly to an image editing model. We explore a two-stage augmentation approach in which a separate multimodal LLM first reasons over the input, producing a detailed reasoning trace that is then provided alongside the original instruction and image to the image editing model.
\footnote{This case study uses \nanobanana as the image editor and was conducted on an earlier version of \bench; we expect the qualitative findings to continue to hold on the current benchmark, and leave a comprehensive re-evaluation across the full model lineup to future work.}

\begin{figure}[H]
    \centering
    \includegraphics[width=\linewidth]{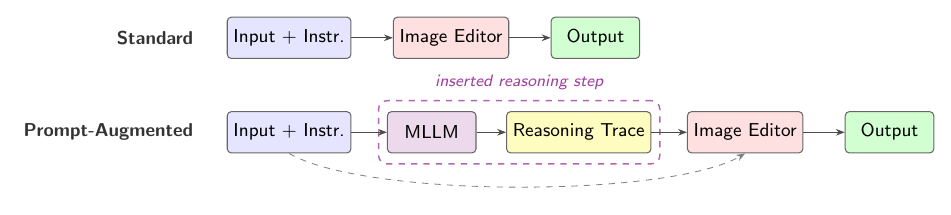}
    \caption{\textbf{Standard vs.\ prompt-augmented pipeline.} The standard pipeline (top) passes the input image and instruction directly to the image editing model. Prompt augmentation (bottom) inserts a reasoning step in which a multimodal LLM (\gemini) generates a reasoning trace with a structured solution from the same inputs; the solution plus the original inputs (dashed) are then passed to the image editing model (\nanobanana).}
    \label{fig:prompt-augmentation-pipeline}
\end{figure}

\Cref{fig:prompt-augmentation-pipeline} illustrates the two pipelines. In Stage 1, \gemini receives the input image and instruction and generates a reasoning trace elaborating on the editing task. In Stage 2, \nanobanana receives the original input image and instruction together with this reasoning trace to produce the output. The standard condition omits Stage 1 and passes the instruction and image directly to \nanobanana.

\begin{table}[ht!]
\centering
\small
\caption{\textbf{Prompt augmentation: edit-region and preservation-region accuracy (\%)} (\nanobanana, $\Delta E \leq 5$). Mean pixel accuracy in the edit-region and preservation-region, comparing the original and prompt-augmented conditions.}
\space
\label{tab:prompt-aug-region-acc}
\begin{tabular}{lrrrrrr}
\toprule
\textbf{Task} & \multicolumn{2}{c}{\textbf{Edit Region}} & \textbf{$\Delta$} & \multicolumn{2}{c}{\textbf{Preservation Region}} & \textbf{$\Delta$} \\
\cmidrule(lr){2-3} \cmidrule(lr){5-6}
             & Orig & Aug  &       & Orig & Aug  &       \\
\midrule
Recolor      & 85.6 & 88.2 & {\color{pos}+2.6}  & 99.6 & 99.7 & {\color{pos}+0.1} \\
Flood Fill   & 95.4 & 90.3 & {\color{neg}-5.1}  & 98.3 & 97.4 & {\color{neg}-0.9} \\
Blending     & 33.4 & 33.5 & {\color{pos}+0.1}  & 98.3 & 98.4 & {\color{pos}+0.0} \\
Gradient     & 39.2 & 35.3 & {\color{neg}-3.9}  & 89.8 & 90.4 & {\color{pos}+0.6} \\
Translation  & 45.6 & 51.8 & {\color{pos}+6.2}  & 96.0 & 97.4 & {\color{pos}+1.4} \\
Reflection   & 21.5 & 21.3 & {\color{neg}-0.2}  & 95.0 & 95.4 & {\color{pos}+0.3} \\
Rotation     & 49.6 & 58.2 & {\color{pos}+8.5}  & 93.2 & 96.6 & {\color{pos}+3.4} \\
Scaling      & 38.1 & 34.9 & {\color{neg}-3.2}  & 89.6 & 90.6 & {\color{pos}+1.0} \\
Shearing     & 37.1 & 42.6 & {\color{pos}+5.4}  & 95.1 & 90.0 & {\color{neg}-5.1} \\
Cropping     & 63.8 & 64.2 & {\color{pos}+0.3}  & 68.8 & 75.4 & {\color{pos}+6.6} \\
Construction & 43.8 & 51.9 & {\color{pos}+8.1}  & 92.4 & 90.8 & {\color{neg}-1.6} \\
Removal      & 89.7 & 99.9 & {\color{pos}+10.1} & 99.2 & 99.6 & {\color{pos}+0.4} \\
Comparison   & 61.1 & 61.1 & 0.0                & 99.0 & 98.9 & {\color{neg}-0.1} \\
Ordering     & 61.1 & 61.3 & {\color{pos}+0.3}  & 95.7 & 96.9 & {\color{pos}+1.2} \\
Pattern      & 80.5 & 81.6 & {\color{pos}+1.1}  & 92.1 & 92.5 & {\color{pos}+0.4} \\
Counting     & 93.5 & 99.3 & {\color{pos}+5.8}  & 99.4 & 99.5 & {\color{pos}+0.1} \\
\bottomrule
\end{tabular}
\end{table}

\FloatBarrier

\paragraph{When does augmentation help?}
A pattern emerges across tasks: augmentation consistently improves edit-region accuracy when the bottleneck is identifying the correct target or planning the transformation.
Tasks with the largest gains (Removal $+$10.1\%, Rotation $+$8.5\%, Construction $+$8.1\%, Translation $+$6.2\%, Counting $+$5.8\%, and Shearing $+$5.4\%) all require either locating a specific scene element, determining a spatial goal, or enumerating shapes before acting; a detailed reasoning trace can resolve these before the editing model generates its output.
In contrast, Flood Fill ($-$5.1\%), Gradient ($-$3.9\%), and Scaling ($-$3.2\%) regress, suggesting that for operations the model already executes via a rote pattern, the additional context may introduce distraction rather than guidance.

\clearpage

\clearpage

\section{Extended Results}
\label{sec:appx:extended-results}

\subsection{Full Bootstrap Confidence Intervals}
\label{sec:appx:bootstrap-ci}

\Cref{tab:appx:bootstrap_ci,tab:appx:bootstrap_ci_tgb,tab:appx:bootstrap_ci_ablations}
provide the complete asymmetric percentile bootstrap confidence intervals for the
Category, Plot Type, Visual Condition, and Benchmark macro-averages shown in
\cref{tab:paintbench_metrics,tab:tinygrafixbench_metrics,tab:ablations}.

\paragraph{Bootstrap procedure.}
CIs are computed by resampling per-problem IoU values with replacement
($B = 10{,}000$ iterations, seed 0, percentile method;
\cite{efron1993bootstrap}). All CIs in this paper share a single
hierarchical methodology that mirrors each main-table average's
aggregation hierarchy:
per-task-mode resample $\rightarrow$ per-task pooled mean.
\textit{Category} / \textit{Plot Type} / \textit{Visual Condition} rows
then macro-average across the inner units within that group: 5 tasks
for \bench categories, 4 subtasks for \tinygrafixbench plot types, and
the 20 tasks contributing to a given condition for \bench visual
conditions. \textit{Benchmark Avg.} rows perform one further macro
step across the outer units: 4 categories for \bench and 5 chart types
for \tinygrafixbench, so each benchmark mean is a doubly-macro average
in which every category / chart type contributes equally.
The benchmark task lists are treated as fixed (definitional) rather than
a random sample, so each interval should be read as ``how much would
this macro-average shift under a different draw of per-problem
instances,'' not ``how much would it shift under a different choice of
tasks.''
Intervals are only slightly asymmetric (max $0.13$\% deviation from symmetry in
our data), so the main paper's half-width $\pm$X.X averages the two true
endpoints; full {\color{cigrey}\textit{Lo}} and {\color{cigrey}\textit{Hi}}
columns are reported here for readers who want the exact bracket bounds.
To conserve space, the visual conditions table (\cref{tab:appx:bootstrap_ci_ablations}) uses a narrow
layout with three sub-rows per condition (\textbf{Mean}, {\color{cigrey}\textit{Lo}},
{\color{cigrey}\textit{Hi}}).

\definecolor{cigrey}{RGB}{136,136,136}

\begin{table}[t]
\centering
\small
\caption{\textbf{Full 95\% bootstrap CIs for \bench\ Category and Benchmark macro-averages.} Companion to \cref{tab:paintbench_metrics}; bootstrap procedure described in \cref{sec:appx:bootstrap-ci}.}
\label{tab:appx:bootstrap_ci}
\setlength{\tabcolsep}{4pt}
\resizebox{\linewidth}{!}{%
\begin{tabular}{l|rrr|rrr|rrr|rrr|rrr}
\toprule
{} & \multicolumn{3}{c|}{\textbf{Geom.\ Trans.}} & \multicolumn{3}{c|}{\textbf{Struct.\ Manip.}} & \multicolumn{3}{c|}{\textbf{Color Change}} & \multicolumn{3}{c|}{\textbf{Symbolic Reas.}} & \multicolumn{3}{c}{\textbf{Benchmark}} \\
\cmidrule(lr){2-4}\cmidrule(lr){5-7}\cmidrule(lr){8-10}\cmidrule(lr){11-13}\cmidrule(lr){14-16}
\textbf{Model} & Avg & {\color{cigrey}Lo} & {\color{cigrey}Hi} & Avg & {\color{cigrey}Lo} & {\color{cigrey}Hi} & Avg & {\color{cigrey}Lo} & {\color{cigrey}Hi} & Avg & {\color{cigrey}Lo} & {\color{cigrey}Hi} & Avg & {\color{cigrey}Lo} & {\color{cigrey}Hi} \\
\midrule
\textbf{NB-2} & 6.1 & {\color{cigrey}5.4} & {\color{cigrey}6.8} & 22.7 & {\color{cigrey}21.1} & {\color{cigrey}24.4} & 17.2 & {\color{cigrey}15.6} & {\color{cigrey}18.8} & 22.6 & {\color{cigrey}21.0} & {\color{cigrey}24.2} & 17.1 & {\color{cigrey}16.4} & {\color{cigrey}17.8} \\
\textbf{GPT-I2} & 11.1 & {\color{cigrey}10.2} & {\color{cigrey}12.0} & 24.5 & {\color{cigrey}23.0} & {\color{cigrey}26.0} & 13.8 & {\color{cigrey}12.4} & {\color{cigrey}15.3} & 15.9 & {\color{cigrey}14.5} & {\color{cigrey}17.4} & 16.3 & {\color{cigrey}15.7} & {\color{cigrey}17.0} \\
\textbf{NB-1} & 6.2 & {\color{cigrey}5.4} & {\color{cigrey}6.9} & 14.0 & {\color{cigrey}12.5} & {\color{cigrey}15.4} & 6.4 & {\color{cigrey}5.2} & {\color{cigrey}7.6} & 18.0 & {\color{cigrey}16.7} & {\color{cigrey}19.5} & 11.1 & {\color{cigrey}10.5} & {\color{cigrey}11.7} \\
\textbf{Qwen-IE} & 3.4 & {\color{cigrey}2.9} & {\color{cigrey}3.9} & 10.3 & {\color{cigrey}9.1} & {\color{cigrey}11.6} & 5.4 & {\color{cigrey}4.4} & {\color{cigrey}6.5} & 7.7 & {\color{cigrey}6.7} & {\color{cigrey}8.8} & 6.7 & {\color{cigrey}6.2} & {\color{cigrey}7.2} \\
\textbf{BAGEL} & 2.4 & {\color{cigrey}2.0} & {\color{cigrey}2.8} & 10.0 & {\color{cigrey}8.6} & {\color{cigrey}11.5} & 2.6 & {\color{cigrey}1.9} & {\color{cigrey}3.5} & 5.1 & {\color{cigrey}4.2} & {\color{cigrey}6.0} & 5.0 & {\color{cigrey}4.5} & {\color{cigrey}5.5} \\
\textbf{FLUX.2-D} & 3.3 & {\color{cigrey}2.8} & {\color{cigrey}3.9} & 8.2 & {\color{cigrey}7.1} & {\color{cigrey}9.4} & 2.7 & {\color{cigrey}2.0} & {\color{cigrey}3.5} & 4.1 & {\color{cigrey}3.5} & {\color{cigrey}4.7} & 4.6 & {\color{cigrey}4.2} & {\color{cigrey}5.0} \\
\textbf{FLUX.1-Kt} & 2.4 & {\color{cigrey}2.1} & {\color{cigrey}2.7} & 7.9 & {\color{cigrey}6.8} & {\color{cigrey}9.1} & 1.6 & {\color{cigrey}1.1} & {\color{cigrey}2.1} & 3.6 & {\color{cigrey}3.1} & {\color{cigrey}4.2} & 3.9 & {\color{cigrey}3.5} & {\color{cigrey}4.2} \\
\textbf{LCat-IE} & 2.2 & {\color{cigrey}1.9} & {\color{cigrey}2.6} & 7.1 & {\color{cigrey}5.9} & {\color{cigrey}8.3} & 1.8 & {\color{cigrey}1.2} & {\color{cigrey}2.4} & 3.5 & {\color{cigrey}2.9} & {\color{cigrey}4.2} & 3.6 & {\color{cigrey}3.3} & {\color{cigrey}4.0} \\
\textbf{FLUX.2-Kl} & 1.4 & {\color{cigrey}1.2} & {\color{cigrey}1.7} & 7.2 & {\color{cigrey}6.1} & {\color{cigrey}8.2} & 2.4 & {\color{cigrey}1.7} & {\color{cigrey}3.1} & 3.2 & {\color{cigrey}2.6} & {\color{cigrey}3.7} & 3.5 & {\color{cigrey}3.2} & {\color{cigrey}3.9} \\
\textbf{HY-3} & 0.1 & {\color{cigrey}0.0} & {\color{cigrey}0.1} & 0.8 & {\color{cigrey}0.5} & {\color{cigrey}1.2} & 0.2 & {\color{cigrey}0.2} & {\color{cigrey}0.3} & 0.3 & {\color{cigrey}0.2} & {\color{cigrey}0.4} & 0.4 & {\color{cigrey}0.3} & {\color{cigrey}0.5} \\
\textbf{IP2P} & 0.0 & {\color{cigrey}0.0} & {\color{cigrey}0.0} & 0.9 & {\color{cigrey}0.5} & {\color{cigrey}1.4} & 0.2 & {\color{cigrey}0.1} & {\color{cigrey}0.4} & 0.1 & {\color{cigrey}0.0} & {\color{cigrey}0.3} & 0.3 & {\color{cigrey}0.2} & {\color{cigrey}0.5} \\
\bottomrule
\end{tabular}%
}
\end{table}

\definecolor{cigrey}{RGB}{136,136,136}

\begin{table}[t]
\centering
\small
\caption{\textbf{Full 95\% bootstrap CIs for \tinygrafixbench\ Plot Type and Benchmark macro-averages.} Companion to \cref{tab:tinygrafixbench_metrics}; bootstrap procedure described in \cref{sec:appx:bootstrap-ci}.}
\label{tab:appx:bootstrap_ci_tgb}
\setlength{\tabcolsep}{4pt}
\resizebox{\linewidth}{!}{%
\begin{tabular}{l|rrr|rrr|rrr|rrr|rrr|rrr}
\toprule
{} & \multicolumn{3}{c|}{\textbf{Bar Chart}} & \multicolumn{3}{c|}{\textbf{Scatter Plot}} & \multicolumn{3}{c|}{\textbf{Line Chart}} & \multicolumn{3}{c|}{\textbf{Heatmap}} & \multicolumn{3}{c|}{\textbf{Network}} & \multicolumn{3}{c}{\textbf{Benchmark}} \\
\cmidrule(lr){2-4}\cmidrule(lr){5-7}\cmidrule(lr){8-10}\cmidrule(lr){11-13}\cmidrule(lr){14-16}\cmidrule(lr){17-19}
\textbf{Model} & Avg & {\color{cigrey}Lo} & {\color{cigrey}Hi} & Avg & {\color{cigrey}Lo} & {\color{cigrey}Hi} & Avg & {\color{cigrey}Lo} & {\color{cigrey}Hi} & Avg & {\color{cigrey}Lo} & {\color{cigrey}Hi} & Avg & {\color{cigrey}Lo} & {\color{cigrey}Hi} & Avg & {\color{cigrey}Lo} & {\color{cigrey}Hi} \\
\midrule
\textbf{NB-2} & 38.9 & {\color{cigrey}36.6} & {\color{cigrey}41.2} & 4.2 & {\color{cigrey}3.6} & {\color{cigrey}4.8} & 11.5 & {\color{cigrey}9.9} & {\color{cigrey}13.1} & 20.2 & {\color{cigrey}17.4} & {\color{cigrey}23.1} & 4.8 & {\color{cigrey}4.3} & {\color{cigrey}5.4} & 15.9 & {\color{cigrey}15.1} & {\color{cigrey}16.7} \\
\textbf{GPT-I2} & 34.8 & {\color{cigrey}32.5} & {\color{cigrey}37.1} & 6.8 & {\color{cigrey}6.1} & {\color{cigrey}7.4} & 16.0 & {\color{cigrey}14.5} & {\color{cigrey}17.5} & 15.5 & {\color{cigrey}13.1} & {\color{cigrey}18.1} & 4.8 & {\color{cigrey}4.0} & {\color{cigrey}5.7} & 15.6 & {\color{cigrey}14.8} & {\color{cigrey}16.4} \\
\textbf{NB-1} & 11.5 & {\color{cigrey}10.1} & {\color{cigrey}12.9} & 3.7 & {\color{cigrey}3.5} & {\color{cigrey}4.0} & 4.7 & {\color{cigrey}4.3} & {\color{cigrey}5.1} & 4.0 & {\color{cigrey}3.3} & {\color{cigrey}4.8} & 2.7 & {\color{cigrey}2.4} & {\color{cigrey}3.1} & 5.3 & {\color{cigrey}5.0} & {\color{cigrey}5.7} \\
\textbf{Qwen-IE} & 4.8 & {\color{cigrey}3.3} & {\color{cigrey}6.4} & 1.3 & {\color{cigrey}1.1} & {\color{cigrey}1.5} & 4.4 & {\color{cigrey}3.7} & {\color{cigrey}5.2} & 4.9 & {\color{cigrey}3.7} & {\color{cigrey}6.2} & 1.6 & {\color{cigrey}1.2} & {\color{cigrey}2.0} & 3.4 & {\color{cigrey}3.0} & {\color{cigrey}3.8} \\
\textbf{BAGEL} & 2.7 & {\color{cigrey}1.7} & {\color{cigrey}3.9} & 2.1 & {\color{cigrey}1.5} & {\color{cigrey}2.7} & 4.9 & {\color{cigrey}4.1} & {\color{cigrey}5.7} & 2.0 & {\color{cigrey}1.5} & {\color{cigrey}2.5} & 1.8 & {\color{cigrey}1.3} & {\color{cigrey}2.4} & 2.7 & {\color{cigrey}2.4} & {\color{cigrey}3.0} \\
\textbf{FLUX.2-D} & 2.3 & {\color{cigrey}1.6} & {\color{cigrey}3.0} & 4.2 & {\color{cigrey}3.9} & {\color{cigrey}4.6} & 3.8 & {\color{cigrey}3.1} & {\color{cigrey}4.4} & 3.2 & {\color{cigrey}2.1} & {\color{cigrey}4.5} & 1.9 & {\color{cigrey}1.5} & {\color{cigrey}2.3} & 3.1 & {\color{cigrey}2.8} & {\color{cigrey}3.4} \\
\textbf{FLUX.1-Kt} & 1.3 & {\color{cigrey}0.9} & {\color{cigrey}1.7} & 4.3 & {\color{cigrey}4.1} & {\color{cigrey}4.6} & 4.5 & {\color{cigrey}4.2} & {\color{cigrey}4.9} & 3.5 & {\color{cigrey}2.9} & {\color{cigrey}4.1} & 1.9 & {\color{cigrey}1.7} & {\color{cigrey}2.2} & 3.1 & {\color{cigrey}2.9} & {\color{cigrey}3.3} \\
\textbf{LCat-IE} & 2.9 & {\color{cigrey}2.1} & {\color{cigrey}3.9} & 1.0 & {\color{cigrey}0.7} & {\color{cigrey}1.3} & 3.2 & {\color{cigrey}2.7} & {\color{cigrey}3.6} & 7.9 & {\color{cigrey}6.5} & {\color{cigrey}9.3} & 0.8 & {\color{cigrey}0.5} & {\color{cigrey}1.1} & 3.2 & {\color{cigrey}2.8} & {\color{cigrey}3.5} \\
\textbf{FLUX.2-Kl} & 2.9 & {\color{cigrey}2.3} & {\color{cigrey}3.6} & 2.1 & {\color{cigrey}1.7} & {\color{cigrey}2.5} & 4.9 & {\color{cigrey}4.1} & {\color{cigrey}5.9} & 4.8 & {\color{cigrey}4.0} & {\color{cigrey}5.7} & 2.4 & {\color{cigrey}1.9} & {\color{cigrey}3.0} & 3.4 & {\color{cigrey}3.1} & {\color{cigrey}3.8} \\
\textbf{HY-3} & 0.2 & {\color{cigrey}0.1} & {\color{cigrey}0.4} & 0.1 & {\color{cigrey}0.1} & {\color{cigrey}0.1} & 0.0 & {\color{cigrey}0.0} & {\color{cigrey}0.1} & 0.9 & {\color{cigrey}0.6} & {\color{cigrey}1.2} & 0.1 & {\color{cigrey}0.0} & {\color{cigrey}0.1} & 0.3 & {\color{cigrey}0.2} & {\color{cigrey}0.3} \\
\textbf{IP2P} & 0.1 & {\color{cigrey}0.1} & {\color{cigrey}0.2} & 0.3 & {\color{cigrey}0.2} & {\color{cigrey}0.6} & 0.1 & {\color{cigrey}0.1} & {\color{cigrey}0.1} & 0.0 & {\color{cigrey}0.0} & {\color{cigrey}0.1} & 0.1 & {\color{cigrey}0.1} & {\color{cigrey}0.3} & 0.2 & {\color{cigrey}0.1} & {\color{cigrey}0.2} \\
\bottomrule
\end{tabular}%
}
\end{table}

\definecolor{cigrey}{RGB}{136,136,136}

\begin{table}[t]
\centering
\small
\caption{\textbf{Full 95\% bootstrap CIs for visual conditions.} Companion to \cref{tab:ablations}; bootstrap procedure described in \cref{sec:appx:bootstrap-ci}. Object-count conditions ($n_{\text{med}}$, $n_{\text{high}}$, $n_{\text{xhigh}}$) use the task-group-specific levels listed in \cref{tab:appx:n-values}.}
\label{tab:appx:bootstrap_ci_ablations}
\setlength{\tabcolsep}{4pt}
\resizebox{\linewidth}{!}{%
\begin{tabular}{llrrrrrrrrrrr}
\toprule
\textbf{Condition} & \textbf{Stat} & \textbf{NB-2} & \textbf{GPT-I2} & \textbf{NB-1} & \textbf{Qwen-IE} & \textbf{BAGEL} & \textbf{FLUX.2-D} & \textbf{FLUX.1-Kt} & \textbf{LCat-IE} & \textbf{FLUX.2-Kl} & \textbf{HY-3} & \textbf{IP2P} \\
\midrule
Baseline & Mean & \textbf{21.9} & 20.9 & 13.4 & 7.2 & 6.4 & 5.2 & 4.3 & 3.7 & 3.9 & 0.4 & 0.4 \\
           & {\color{cigrey}Lo} & {\color{cigrey}19.7} & {\color{cigrey}19.0} & {\color{cigrey}11.6} & {\color{cigrey}5.8} & {\color{cigrey}5.0} & {\color{cigrey}4.1} & {\color{cigrey}3.4} & {\color{cigrey}2.8} & {\color{cigrey}3.0} & {\color{cigrey}0.2} & {\color{cigrey}0.2} \\
           & {\color{cigrey}Hi} & {\color{cigrey}24.1} & {\color{cigrey}22.8} & {\color{cigrey}15.3} & {\color{cigrey}8.6} & {\color{cigrey}8.0} & {\color{cigrey}6.4} & {\color{cigrey}5.3} & {\color{cigrey}4.6} & {\color{cigrey}4.8} & {\color{cigrey}0.6} & {\color{cigrey}0.7} \\
\cmidrule(l){1-13}
Horizontal & Mean & 19.2 & \textbf{19.3} & 11.5 & 8.7 & 6.7 & 5.7 & 5.5 & 5.1 & 5.4 & 0.6 & 0.3 \\
           & {\color{cigrey}Lo} & {\color{cigrey}17.2} & {\color{cigrey}17.5} & {\color{cigrey}9.9} & {\color{cigrey}7.3} & {\color{cigrey}5.4} & {\color{cigrey}4.7} & {\color{cigrey}4.6} & {\color{cigrey}4.0} & {\color{cigrey}4.3} & {\color{cigrey}0.3} & {\color{cigrey}0.1} \\
           & {\color{cigrey}Hi} & {\color{cigrey}21.2} & {\color{cigrey}21.3} & {\color{cigrey}13.2} & {\color{cigrey}10.1} & {\color{cigrey}8.0} & {\color{cigrey}6.9} & {\color{cigrey}6.5} & {\color{cigrey}6.3} & {\color{cigrey}6.5} & {\color{cigrey}0.9} & {\color{cigrey}0.6} \\
\cmidrule(l){1-13}
Vertical & Mean & \textbf{21.0} & 17.4 & 13.3 & 8.0 & 5.0 & 6.4 & 3.8 & 3.5 & 5.7 & 0.5 & 0.2 \\
           & {\color{cigrey}Lo} & {\color{cigrey}19.1} & {\color{cigrey}15.7} & {\color{cigrey}11.7} & {\color{cigrey}6.7} & {\color{cigrey}3.8} & {\color{cigrey}5.2} & {\color{cigrey}3.2} & {\color{cigrey}2.6} & {\color{cigrey}4.5} & {\color{cigrey}0.3} & {\color{cigrey}0.1} \\
           & {\color{cigrey}Hi} & {\color{cigrey}22.9} & {\color{cigrey}19.1} & {\color{cigrey}15.1} & {\color{cigrey}9.2} & {\color{cigrey}6.4} & {\color{cigrey}7.8} & {\color{cigrey}4.4} & {\color{cigrey}4.5} & {\color{cigrey}6.9} & {\color{cigrey}0.7} & {\color{cigrey}0.4} \\
\cmidrule(l){1-13}
Nonstandard & Mean & \textbf{23.2} & 22.9 & 17.5 & 8.6 & 4.5 & 4.7 & 5.3 & 6.7 & 3.7 & 0.3 & 0.2 \\
           & {\color{cigrey}Lo} & {\color{cigrey}21.2} & {\color{cigrey}21.0} & {\color{cigrey}15.6} & {\color{cigrey}7.1} & {\color{cigrey}3.5} & {\color{cigrey}3.7} & {\color{cigrey}4.3} & {\color{cigrey}5.5} & {\color{cigrey}2.8} & {\color{cigrey}0.1} & {\color{cigrey}0.0} \\
           & {\color{cigrey}Hi} & {\color{cigrey}25.1} & {\color{cigrey}24.7} & {\color{cigrey}19.3} & {\color{cigrey}10.2} & {\color{cigrey}5.6} & {\color{cigrey}5.7} & {\color{cigrey}6.4} & {\color{cigrey}7.9} & {\color{cigrey}4.6} & {\color{cigrey}0.6} & {\color{cigrey}0.5} \\
\cmidrule(l){1-13}
Striped & Mean & 10.8 & \textbf{12.2} & 7.4 & 4.5 & 4.0 & 4.7 & 2.4 & 1.7 & 2.0 & 0.4 & 0.3 \\
           & {\color{cigrey}Lo} & {\color{cigrey}9.7} & {\color{cigrey}11.0} & {\color{cigrey}6.3} & {\color{cigrey}3.8} & {\color{cigrey}3.2} & {\color{cigrey}3.9} & {\color{cigrey}1.9} & {\color{cigrey}1.3} & {\color{cigrey}1.5} & {\color{cigrey}0.2} & {\color{cigrey}0.1} \\
           & {\color{cigrey}Hi} & {\color{cigrey}11.9} & {\color{cigrey}13.5} & {\color{cigrey}8.6} & {\color{cigrey}5.4} & {\color{cigrey}4.9} & {\color{cigrey}5.7} & {\color{cigrey}3.0} & {\color{cigrey}2.2} & {\color{cigrey}2.5} & {\color{cigrey}0.6} & {\color{cigrey}0.5} \\
\cmidrule(l){1-13}
$n_{\text{med}}$ & Mean & \textbf{17.9} & 17.0 & 11.9 & 6.8 & 5.0 & 3.6 & 3.9 & 3.4 & 3.1 & 0.4 & 0.6 \\
           & {\color{cigrey}Lo} & {\color{cigrey}16.1} & {\color{cigrey}15.2} & {\color{cigrey}10.1} & {\color{cigrey}5.4} & {\color{cigrey}3.8} & {\color{cigrey}2.8} & {\color{cigrey}2.9} & {\color{cigrey}2.4} & {\color{cigrey}2.3} & {\color{cigrey}0.1} & {\color{cigrey}0.1} \\
           & {\color{cigrey}Hi} & {\color{cigrey}19.8} & {\color{cigrey}18.8} & {\color{cigrey}13.7} & {\color{cigrey}8.3} & {\color{cigrey}6.4} & {\color{cigrey}4.5} & {\color{cigrey}5.0} & {\color{cigrey}4.5} & {\color{cigrey}4.0} & {\color{cigrey}0.8} & {\color{cigrey}1.1} \\
\cmidrule(l){1-13}
$n_{\text{high}}$ & Mean & \textbf{13.1} & 11.5 & 7.7 & 5.3 & 4.2 & 2.8 & 3.3 & 2.4 & 2.2 & 0.2 & 0.5 \\
           & {\color{cigrey}Lo} & {\color{cigrey}11.6} & {\color{cigrey}10.0} & {\color{cigrey}6.5} & {\color{cigrey}4.1} & {\color{cigrey}3.0} & {\color{cigrey}2.2} & {\color{cigrey}2.4} & {\color{cigrey}1.5} & {\color{cigrey}1.5} & {\color{cigrey}0.1} & {\color{cigrey}0.1} \\
           & {\color{cigrey}Hi} & {\color{cigrey}14.7} & {\color{cigrey}12.9} & {\color{cigrey}8.9} & {\color{cigrey}6.6} & {\color{cigrey}5.5} & {\color{cigrey}3.6} & {\color{cigrey}4.3} & {\color{cigrey}3.5} & {\color{cigrey}3.0} & {\color{cigrey}0.4} & {\color{cigrey}1.0} \\
\cmidrule(l){1-13}
$n_{\text{xhigh}}$ & Mean & \textbf{10.0} & 9.6 & 6.3 & 4.6 & 4.2 & 3.4 & 2.5 & 2.6 & 2.3 & 0.2 & 0.1 \\
           & {\color{cigrey}Lo} & {\color{cigrey}8.5} & {\color{cigrey}8.1} & {\color{cigrey}5.3} & {\color{cigrey}3.4} & {\color{cigrey}2.8} & {\color{cigrey}2.3} & {\color{cigrey}1.5} & {\color{cigrey}1.5} & {\color{cigrey}1.4} & {\color{cigrey}0.1} & {\color{cigrey}0.0} \\
           & {\color{cigrey}Hi} & {\color{cigrey}11.6} & {\color{cigrey}11.2} & {\color{cigrey}7.5} & {\color{cigrey}5.8} & {\color{cigrey}5.6} & {\color{cigrey}4.7} & {\color{cigrey}3.5} & {\color{cigrey}3.8} & {\color{cigrey}3.4} & {\color{cigrey}0.4} & {\color{cigrey}0.1} \\
\bottomrule
\end{tabular}%
}
\end{table}

\subsection{Full Results Tables}
\label{sec:appx:full-results}

The following tables provide complete per-task, and per-mode breakdowns.
\Cref{tab:paintbench_full_iou} covers all 35 \bench task-modes (three sub-tables: mean IoU,
edit-region accuracy, and preservation-region accuracy);
\Cref{tab:ablations_full_iou} covers all 8 visual conditions across all 20 \bench tasks;
\Cref{tab:tinygrafixbench_full_iou} covers all 20 \tinygrafixbench task-modes.
Per-cell CIs are not shown to keep these full-results tables compact;
aggregate-level bootstrap CIs (categories, chart types, visual conditions, benchmark averages)
are reported in \cref{sec:appx:bootstrap-ci}.

\definecolor{cigrey}{RGB}{136,136,136}
\definecolor{benchbg}{RGB}{240,240,240}
\definecolor{catgeometrictransformationlight}{RGB}{235,242,255}
\definecolor{catgeometrictransformationdark}{RGB}{43,94,167}
\definecolor{catstructuralmanipulationlight}{RGB}{255,243,235}
\definecolor{catstructuralmanipulationdark}{RGB}{192,86,33}
\definecolor{catcolorchangelight}{RGB}{234,250,240}
\definecolor{catcolorchangedark}{RGB}{39,103,73}
\definecolor{catsymbolicreasoninglight}{RGB}{245,238,255}
\definecolor{catsymbolicreasoningdark}{RGB}{107,45,139}

\begin{table}[p]
\centering
\scriptsize
\caption{\textbf{\bench\ Mean IoU (\%) per task-mode.} Best per row in bold. Aggregate-level bootstrap CIs are reported in \cref{sec:appx:bootstrap-ci}.}
\label{tab:paintbench_full_iou}
\resizebox{\linewidth}{!}{%
\begin{tabular}{lllrrrrrrrrrrr}
\toprule
\textbf{Category} & \textbf{Task} & \textbf{Mode} & \textbf{NB-2} & \textbf{GPT-I2} & \textbf{NB-1} & \textbf{Qwen-IE} & \textbf{BAGEL} & \textbf{FLUX.2-D} & \textbf{FLUX.1-Kt} & \textbf{LCat-IE} & \textbf{FLUX.2-Kl} & \textbf{HY-3} & \textbf{IP2P} \\
\midrule
\rowcolor{catgeometrictransformationdark}
\multicolumn{14}{l}{\textbf{\color{white}Geometric Transformation}\strut} \\
\rowcolor{catgeometrictransformationlight!50!white}
\multicolumn{3}{l}{} & \cellcolor[RGB]{255,238,199}6.1 & \cellcolor[RGB]{254,223,154}\textbf{11.1} & \cellcolor[RGB]{255,237,199}6.2 & \cellcolor[RGB]{255,245,224}3.4 & \cellcolor[RGB]{255,248,233}2.4 & \cellcolor[RGB]{255,246,225}3.3 & \cellcolor[RGB]{255,248,233}2.4 & \cellcolor[RGB]{255,249,235}2.2 & \cellcolor[RGB]{255,251,242}1.4 & \cellcolor[RGB]{255,255,254}0.1 & \cellcolor[RGB]{255,255,255}0.0 \\[-2pt]
\rowcolor{catgeometrictransformationlight!50!white}
\multicolumn{3}{l}{\multirow{-2}{*}{\textit{\quad Category Avg.}}} & \cellcolor[RGB]{255,238,199}{\color{black}\scriptsize $\pm0.7$} & \cellcolor[RGB]{254,223,154}{\color{black}\scriptsize $\pm0.9$} & \cellcolor[RGB]{255,237,199}{\color{black}\scriptsize $\pm0.8$} & \cellcolor[RGB]{255,245,224}{\color{black}\scriptsize $\pm0.5$} & \cellcolor[RGB]{255,248,233}{\color{black}\scriptsize $\pm0.4$} & \cellcolor[RGB]{255,246,225}{\color{black}\scriptsize $\pm0.5$} & \cellcolor[RGB]{255,248,233}{\color{black}\scriptsize $\pm0.3$} & \cellcolor[RGB]{255,249,235}{\color{black}\scriptsize $\pm0.4$} & \cellcolor[RGB]{255,251,242}{\color{black}\scriptsize $\pm0.3$} & \cellcolor[RGB]{255,255,254}{\color{black}\scriptsize $\pm0.1$} & \cellcolor[RGB]{255,255,255}{\color{black}\scriptsize $\pm0.0$} \\
\bottomrule
\end{tabular}%
}
\end{table}

\begin{table}[p]
\centering
\scriptsize
\caption{\textbf{\bench\ Edit-Region Accuracy (\%) per task-mode.} Best per row in bold. Aggregate-level bootstrap CIs are reported in \cref{sec:appx:bootstrap-ci}.}
\label{tab:paintbench_full_edit_acc}
\resizebox{\linewidth}{!}{%
\begin{tabular}{lllrrrrrrrrrrr}
\toprule
\textbf{Category} & \textbf{Task} & \textbf{Mode} & \textbf{NB-2} & \textbf{GPT-I2} & \textbf{NB-1} & \textbf{Qwen-IE} & \textbf{BAGEL} & \textbf{FLUX.2-D} & \textbf{FLUX.1-Kt} & \textbf{LCat-IE} & \textbf{FLUX.2-Kl} & \textbf{HY-3} & \textbf{IP2P} \\
\midrule
\rowcolor{catgeometrictransformationdark}
\multicolumn{14}{l}{\textbf{\color{white}Geometric Transformation}\strut} \\
\rowcolor{catgeometrictransformationlight!50!white}
\multicolumn{3}{l}{\textit{\quad Category Avg.}} & \cellcolor[RGB]{254,199,89}35.9 & \cellcolor[RGB]{254,196,88}\textbf{37.4} & \cellcolor[RGB]{254,219,139}23.5 & \cellcolor[RGB]{254,222,148}21.6 & \cellcolor[RGB]{254,228,169}17.3 & \cellcolor[RGB]{255,230,175}16.3 & \cellcolor[RGB]{254,225,160}19.2 & \cellcolor[RGB]{254,222,150}21.2 & \cellcolor[RGB]{255,235,192}12.7 & \cellcolor[RGB]{255,250,240}3.0 & \cellcolor[RGB]{255,251,243}2.4 \\
 & \multirow{2}{*}{Translation} & align & \cellcolor[RGB]{254,188,84}\textbf{41.7} & \cellcolor[RGB]{254,190,85}40.8 & \cellcolor[RGB]{254,221,147}21.9 & \cellcolor[RGB]{254,217,133}24.7 & \cellcolor[RGB]{254,226,162}18.9 & \cellcolor[RGB]{254,229,171}17.0 & \cellcolor[RGB]{255,232,183}14.6 & \cellcolor[RGB]{254,222,149}21.4 & \cellcolor[RGB]{255,231,178}15.6 & \cellcolor[RGB]{255,252,245}2.0 & \cellcolor[RGB]{255,254,252}0.6 \\
 &  & amount & \cellcolor[RGB]{254,185,82}\textbf{43.4} & \cellcolor[RGB]{254,189,84}41.1 & \cellcolor[RGB]{254,225,158}19.6 & \cellcolor[RGB]{254,223,152}20.9 & \cellcolor[RGB]{255,241,210}9.1 & \cellcolor[RGB]{255,242,213}8.5 & \cellcolor[RGB]{255,234,189}13.4 & \cellcolor[RGB]{254,224,157}19.8 & \cellcolor[RGB]{255,238,202}10.7 & \cellcolor[RGB]{255,252,244}2.2 & \cellcolor[RGB]{255,254,253}0.5 \\
\bottomrule
\end{tabular}%
}
\end{table}

\begin{table}[p]
\centering
\scriptsize
\caption{\textbf{\bench\ Preservation-Region Accuracy (\%) per task-mode.} Best per row in bold. Aggregate-level bootstrap CIs are reported in \cref{sec:appx:bootstrap-ci}.}
\label{tab:paintbench_full_pres_acc}
\resizebox{\linewidth}{!}{%
\begin{tabular}{lllrrrrrrrrrrr}
\toprule
\textbf{Category} & \textbf{Task} & \textbf{Mode} & \textbf{NB-2} & \textbf{GPT-I2} & \textbf{NB-1} & \textbf{Qwen-IE} & \textbf{BAGEL} & \textbf{FLUX.2-D} & \textbf{FLUX.1-Kt} & \textbf{LCat-IE} & \textbf{FLUX.2-Kl} & \textbf{HY-3} & \textbf{IP2P} \\
\midrule
\rowcolor{catgeometrictransformationdark}
\multicolumn{14}{l}{\textbf{\color{white}Geometric Transformation}\strut} \\
\rowcolor{catgeometrictransformationlight!50!white}
\multicolumn{3}{l}{\textit{\quad Category Avg.}} & \cellcolor[RGB]{203,30,43}{\color{white}77.7} & \cellcolor[RGB]{215,58,47}{\color{white}72.2} & \cellcolor[RGB]{209,44,45}{\color{white}74.9} & \cellcolor[RGB]{202,28,42}{\color{white}\textbf{78.2}} & \cellcolor[RGB]{214,55,47}{\color{white}72.9} & \cellcolor[RGB]{234,99,53}{\color{white}64.3} & \cellcolor[RGB]{219,66,48}{\color{white}70.7} & \cellcolor[RGB]{253,153,66}50.4 & \cellcolor[RGB]{253,146,62}53.9 & \cellcolor[RGB]{255,232,181}12.6 & \cellcolor[RGB]{255,246,227}4.7 \\
 & \multirow{2}{*}{Translation} & align & \cellcolor[RGB]{207,40,44}{\color{white}75.8} & \cellcolor[RGB]{225,79,50}{\color{white}68.1} & \cellcolor[RGB]{232,95,53}{\color{white}64.9} & \cellcolor[RGB]{205,34,43}{\color{white}\textbf{76.9}} & \cellcolor[RGB]{219,66,48}{\color{white}70.6} & \cellcolor[RGB]{235,101,54}{\color{white}63.9} & \cellcolor[RGB]{225,80,51}{\color{white}67.9} & \cellcolor[RGB]{253,161,70}46.8 & \cellcolor[RGB]{251,136,59}56.9 & \cellcolor[RGB]{255,242,213}7.1 & \cellcolor[RGB]{255,249,235}3.3 \\
\bottomrule
\end{tabular}%
}
\end{table}

\clearpage

\begin{table}[p]
\centering
\small
\caption{\textbf{Visual conditions: Mean IoU (\%).} All values averaged over all 20 \bench tasks. Best per row in bold. Object-count conditions ($n_{\text{med}}$, $n_{\text{high}}$, $n_{\text{xhigh}}$) use the task-group-specific levels listed in \cref{tab:appx:n-values}.}
\label{tab:ablations_full_iou}
\resizebox{\linewidth}{!}{%
\begin{tabular}{lrrrrrrrrrrr}
\toprule
\textbf{Condition} & \textbf{NB-2} & \textbf{GPT-I2} & \textbf{NB-1} & \textbf{Qwen-IE} & \textbf{BAGEL} & \textbf{FLUX.2-D} & \textbf{FLUX.1-Kt} & \textbf{LCat-IE} & \textbf{FLUX.2-Kl} & \textbf{HY-3} & \textbf{IP2P} \\
\midrule
\rowcolor{ours}
Baseline & \textbf{21.9} & 20.9 & 13.4 & 7.2 & 6.4 & 5.2 & 4.3 & 3.7 & 3.9 & 0.4 & 0.4 \\
Horizontal & 19.2 & \textbf{19.3} & 11.5 & 8.7 & 6.7 & 5.7 & 5.5 & 5.1 & 5.4 & 0.6 & 0.3 \\
Vertical & \textbf{21.0} & 17.4 & 13.3 & 8.0 & 5.0 & 6.4 & 3.8 & 3.5 & 5.7 & 0.5 & 0.2 \\
Nonstandard & \textbf{23.2} & 22.9 & 17.5 & 8.6 & 4.5 & 4.7 & 5.3 & 6.7 & 3.7 & 0.3 & 0.2 \\
Striped & 10.8 & \textbf{12.2} & 7.4 & 4.5 & 4.0 & 4.7 & 2.4 & 1.7 & 2.0 & 0.4 & 0.3 \\
$n_{\text{med}}$ & \textbf{17.9} & 17.0 & 11.9 & 6.8 & 5.0 & 3.6 & 3.9 & 3.4 & 3.1 & 0.4 & 0.6 \\
$n_{\text{high}}$ & \textbf{13.1} & 11.5 & 7.7 & 5.3 & 4.2 & 2.8 & 3.3 & 2.4 & 2.2 & 0.2 & 0.5 \\
$n_{\text{xhigh}}$ & \textbf{10.0} & 9.6 & 6.3 & 4.6 & 4.2 & 3.4 & 2.5 & 2.6 & 2.3 & 0.2 & 0.1 \\
\bottomrule
\end{tabular}%
}
\end{table}

\begin{table}[p]
\centering
\small
\caption{\textbf{Visual conditions: Edit-Region Accuracy (\%).} All values averaged over all 20 \bench tasks. Best per row in bold. Object-count conditions ($n_{\text{med}}$, $n_{\text{high}}$, $n_{\text{xhigh}}$) use the task-group-specific levels listed in \cref{tab:appx:n-values}.}
\label{tab:ablations_full_edit_acc}
\resizebox{\linewidth}{!}{%
\begin{tabular}{lrrrrrrrrrrr}
\toprule
\textbf{Condition} & \textbf{NB-2} & \textbf{GPT-I2} & \textbf{NB-1} & \textbf{Qwen-IE} & \textbf{BAGEL} & \textbf{FLUX.2-D} & \textbf{FLUX.1-Kt} & \textbf{LCat-IE} & \textbf{FLUX.2-Kl} & \textbf{HY-3} & \textbf{IP2P} \\
\midrule
\rowcolor{ours}
Baseline & \textbf{48.1} & 42.8 & 28.1 & 21.6 & 17.8 & 18.5 & 13.3 & 17.1 & 15.2 & 4.9 & 4.2 \\
Horizontal & \textbf{47.7} & 44.1 & 30.0 & 27.6 & 19.2 & 21.1 & 28.6 & 22.3 & 20.3 & 6.5 & 4.2 \\
Vertical & \textbf{49.7} & 42.5 & 32.8 & 23.3 & 15.8 & 20.9 & 26.4 & 17.7 & 19.6 & 5.5 & 3.0 \\
Nonstandard & \textbf{45.6} & 42.1 & 29.8 & 21.0 & 14.4 & 13.6 & 14.5 & 20.9 & 13.0 & 3.3 & 1.5 \\
Striped & \textbf{42.6} & 40.7 & 25.6 & 25.4 & 23.0 & 23.4 & 18.2 & 15.9 & 15.0 & 5.0 & 3.2 \\
$n_{\text{med}}$ & \textbf{44.8} & 37.5 & 27.5 & 21.5 & 17.0 & 13.6 & 13.0 & 16.5 & 11.8 & 4.2 & 4.4 \\
$n_{\text{high}}$ & \textbf{41.7} & 34.9 & 24.2 & 18.4 & 15.0 & 13.1 & 12.1 & 9.9 & 10.3 & 3.3 & 2.7 \\
$n_{\text{xhigh}}$ & \textbf{36.4} & 31.9 & 19.2 & 17.1 & 12.5 & 13.0 & 8.8 & 11.4 & 8.5 & 3.4 & 3.9 \\
\bottomrule
\end{tabular}%
}
\end{table}

\begin{table}[p]
\centering
\small
\caption{\textbf{Visual conditions: Preservation-Region Accuracy (\%).} All values averaged over all 20 \bench tasks. Best per row in bold. Object-count conditions ($n_{\text{med}}$, $n_{\text{high}}$, $n_{\text{xhigh}}$) use the task-group-specific levels listed in \cref{tab:appx:n-values}.}
\label{tab:ablations_full_pres_acc}
\resizebox{\linewidth}{!}{%
\begin{tabular}{lrrrrrrrrrrr}
\toprule
\textbf{Condition} & \textbf{NB-2} & \textbf{GPT-I2} & \textbf{NB-1} & \textbf{Qwen-IE} & \textbf{BAGEL} & \textbf{FLUX.2-D} & \textbf{FLUX.1-Kt} & \textbf{LCat-IE} & \textbf{FLUX.2-Kl} & \textbf{HY-3} & \textbf{IP2P} \\
\midrule
\rowcolor{ours}
Baseline & \textbf{77.4} & 68.2 & 72.1 & 75.5 & 69.0 & 61.1 & 70.6 & 43.5 & 49.1 & 8.8 & 5.9 \\
Horizontal & \textbf{78.1} & 73.4 & 71.7 & 77.9 & 74.1 & 58.1 & 64.2 & 50.1 & 58.4 & 11.5 & 5.5 \\
Vertical & \textbf{79.8} & 70.4 & 73.8 & 78.3 & 75.3 & 59.3 & 63.5 & 45.2 & 64.1 & 11.9 & 5.7 \\
Nonstandard & 76.1 & 68.9 & 76.4 & \textbf{79.3} & 52.7 & 54.3 & 70.3 & 57.4 & 49.2 & 6.1 & 3.1 \\
Striped & \textbf{75.6} & 63.9 & 74.0 & 70.4 & 67.6 & 62.7 & 59.1 & 37.7 & 40.5 & 12.7 & 9.7 \\
$n_{\text{med}}$ & \textbf{79.5} & 68.5 & 74.6 & 76.6 & 67.0 & 60.0 & 71.6 & 43.5 & 53.5 & 10.2 & 6.9 \\
$n_{\text{high}}$ & \textbf{79.7} & 64.4 & 76.6 & 77.0 & 68.9 & 60.9 & 74.2 & 42.1 & 49.0 & 10.5 & 4.2 \\
$n_{\text{xhigh}}$ & \textbf{78.2} & 62.1 & 77.7 & 74.2 & 68.0 & 58.3 & 72.0 & 41.1 & 49.7 & 11.0 & 6.4 \\
\bottomrule
\end{tabular}%
}
\end{table}

\clearpage

\definecolor{cigrey}{RGB}{136,136,136}
\definecolor{benchbg}{RGB}{240,240,240}
\definecolor{chartbarchartlight}{RGB}{235,242,255}
\definecolor{chartbarchartdark}{RGB}{43,94,167}
\definecolor{chartscatterplotlight}{RGB}{234,250,240}
\definecolor{chartscatterplotdark}{RGB}{39,103,73}
\definecolor{chartlinechartlight}{RGB}{255,243,235}
\definecolor{chartlinechartdark}{RGB}{192,86,33}
\definecolor{chartheatmaplight}{RGB}{245,238,255}
\definecolor{chartheatmapdark}{RGB}{107,45,139}
\definecolor{chartnetworklight}{RGB}{230,244,247}
\definecolor{chartnetworkdark}{RGB}{27,110,123}

\begin{table}[p]
\centering
\scriptsize
\caption{\textbf{\tinygrafixbench\ Mean IoU (\%) per task-mode.} Best per row in bold. Aggregate-level bootstrap CIs are reported in \cref{sec:appx:bootstrap-ci}.}
\label{tab:tinygrafixbench_full_iou}
\resizebox{\linewidth}{!}{%
\begin{tabular}{llrrrrrrrrrrr}
\toprule
\textbf{Chart Type} & \textbf{Task} & \textbf{NB-2} & \textbf{GPT-I2} & \textbf{NB-1} & \textbf{Qwen-IE} & \textbf{BAGEL} & \textbf{FLUX.2-D} & \textbf{FLUX.1-Kt} & \textbf{LCat-IE} & \textbf{FLUX.2-Kl} & \textbf{HY-3} & \textbf{IP2P} \\
\midrule
\rowcolor{chartbarchartdark}
\multicolumn{13}{l}{\textbf{\color{white}Bar Chart}\strut} \\
\rowcolor{chartbarchartlight!50!white}
\multicolumn{2}{l}{} & \cellcolor[RGB]{253,156,67}\textbf{38.9} & \cellcolor[RGB]{253,167,73}34.8 & \cellcolor[RGB]{254,228,169}11.5 & \cellcolor[RGB]{255,244,219}4.8 & \cellcolor[RGB]{255,249,234}2.7 & \cellcolor[RGB]{255,250,238}2.3 & \cellcolor[RGB]{255,252,245}1.3 & \cellcolor[RGB]{255,248,233}2.9 & \cellcolor[RGB]{255,248,233}2.9 & \cellcolor[RGB]{255,255,254}0.2 & \cellcolor[RGB]{255,255,254}0.1 \\[-2pt]
\rowcolor{chartbarchartlight!50!white}
\multicolumn{2}{l}{\multirow{-2}{*}{\textit{\quad Plot Type Avg.}}} & \cellcolor[RGB]{253,156,67}{\color{black}\scriptsize $\pm2.3$} & \cellcolor[RGB]{253,167,73}{\color{black}\scriptsize $\pm2.3$} & \cellcolor[RGB]{254,228,169}{\color{black}\scriptsize $\pm1.4$} & \cellcolor[RGB]{255,244,219}{\color{black}\scriptsize $\pm1.5$} & \cellcolor[RGB]{255,249,234}{\color{black}\scriptsize $\pm1.1$} & \cellcolor[RGB]{255,250,238}{\color{black}\scriptsize $\pm0.7$} & \cellcolor[RGB]{255,252,245}{\color{black}\scriptsize $\pm0.4$} & \cellcolor[RGB]{255,248,233}{\color{black}\scriptsize $\pm0.9$} & \cellcolor[RGB]{255,248,233}{\color{black}\scriptsize $\pm0.6$} & \cellcolor[RGB]{255,255,254}{\color{black}\scriptsize $\pm0.2$} & \cellcolor[RGB]{255,255,254}{\color{black}\scriptsize $\pm0.1$} \\
 & Add Bar & \cellcolor[RGB]{254,174,77}\textbf{32.3} & \cellcolor[RGB]{254,175,77}31.9 & \cellcolor[RGB]{255,254,252}0.3 & \cellcolor[RGB]{255,254,250}0.6 & \cellcolor[RGB]{255,255,255}0.0 & \cellcolor[RGB]{255,255,255}0.0 & \cellcolor[RGB]{255,255,255}0.0 & \cellcolor[RGB]{255,254,253}0.2 & \cellcolor[RGB]{255,255,255}0.0 & \cellcolor[RGB]{255,255,255}0.0 & \cellcolor[RGB]{255,255,255}0.0 \\
 & Sort Bars & \cellcolor[RGB]{193,9,39}{\color{white}64.4} & \cellcolor[RGB]{189,0,38}{\color{white}\textbf{65.8}} & \cellcolor[RGB]{254,177,78}31.3 & \cellcolor[RGB]{255,231,177}10.4 & \cellcolor[RGB]{255,245,224}4.1 & \cellcolor[RGB]{255,234,189}8.8 & \cellcolor[RGB]{255,247,228}3.5 & \cellcolor[RGB]{255,234,188}8.9 & \cellcolor[RGB]{254,228,168}11.6 & \cellcolor[RGB]{255,255,254}0.1 & \cellcolor[RGB]{255,254,252}0.4 \\
 & Remove Bar & \cellcolor[RGB]{254,181,81}\textbf{29.8} & \cellcolor[RGB]{254,224,155}13.3 & \cellcolor[RGB]{254,223,151}13.8 & \cellcolor[RGB]{255,236,195}8.0 & \cellcolor[RGB]{255,239,204}6.8 & \cellcolor[RGB]{255,255,255}0.0 & \cellcolor[RGB]{255,251,243}1.6 & \cellcolor[RGB]{255,250,240}2.1 & \cellcolor[RGB]{255,255,254}0.1 & \cellcolor[RGB]{255,253,250}0.7 & \cellcolor[RGB]{255,255,254}0.1 \\
 & Recolor Bar & \cellcolor[RGB]{254,183,81}\textbf{29.1} & \cellcolor[RGB]{254,186,83}28.2 & \cellcolor[RGB]{255,253,250}0.6 & \cellcolor[RGB]{255,255,255}0.1 & \cellcolor[RGB]{255,255,255}0.0 & \cellcolor[RGB]{255,254,253}0.3 & \cellcolor[RGB]{255,255,255}0.0 & \cellcolor[RGB]{255,254,251}0.5 & \cellcolor[RGB]{255,255,254}0.1 & \cellcolor[RGB]{255,255,255}0.0 & \cellcolor[RGB]{255,255,255}0.0 \\
\rowcolor{chartscatterplotdark}
\multicolumn{13}{l}{\textbf{\color{white}Scatter Plot}\strut} \\
\rowcolor{chartscatterplotlight!50!white}
\multicolumn{2}{l}{} & \cellcolor[RGB]{255,245,224}4.2 & \cellcolor[RGB]{255,239,204}\textbf{6.8} & \cellcolor[RGB]{255,246,227}3.7 & \cellcolor[RGB]{255,252,245}1.3 & \cellcolor[RGB]{255,250,239}2.1 & \cellcolor[RGB]{255,245,223}4.2 & \cellcolor[RGB]{255,245,222}4.3 & \cellcolor[RGB]{255,253,247}1.0 & \cellcolor[RGB]{255,250,240}2.1 & \cellcolor[RGB]{255,255,254}0.1 & \cellcolor[RGB]{255,254,252}0.3 \\[-2pt]
\rowcolor{chartscatterplotlight!50!white}
\multicolumn{2}{l}{\multirow{-2}{*}{\textit{\quad Plot Type Avg.}}} & \cellcolor[RGB]{255,245,224}{\color{black}\scriptsize $\pm0.6$} & \cellcolor[RGB]{255,239,204}{\color{black}\scriptsize $\pm0.7$} & \cellcolor[RGB]{255,246,227}{\color{black}\scriptsize $\pm0.2$} & \cellcolor[RGB]{255,252,245}{\color{black}\scriptsize $\pm0.2$} & \cellcolor[RGB]{255,250,239}{\color{black}\scriptsize $\pm0.6$} & \cellcolor[RGB]{255,245,223}{\color{black}\scriptsize $\pm0.4$} & \cellcolor[RGB]{255,245,222}{\color{black}\scriptsize $\pm0.3$} & \cellcolor[RGB]{255,253,247}{\color{black}\scriptsize $\pm0.3$} & \cellcolor[RGB]{255,250,240}{\color{black}\scriptsize $\pm0.4$} & \cellcolor[RGB]{255,255,254}{\color{black}\scriptsize $\pm0.0$} & \cellcolor[RGB]{255,254,252}{\color{black}\scriptsize $\pm0.2$} \\
 & Draw Best Fit Line & \cellcolor[RGB]{255,253,249}0.8 & \cellcolor[RGB]{255,252,247}\textbf{1.1} & \cellcolor[RGB]{255,254,251}0.5 & \cellcolor[RGB]{255,253,250}0.7 & \cellcolor[RGB]{255,253,249}0.8 & \cellcolor[RGB]{255,253,248}0.9 & \cellcolor[RGB]{255,254,252}0.5 & \cellcolor[RGB]{255,255,253}0.2 & \cellcolor[RGB]{255,253,248}0.9 & \cellcolor[RGB]{255,255,255}0.0 & \cellcolor[RGB]{255,255,254}0.1 \\
 & Swap Axes & \cellcolor[RGB]{255,234,187}9.0 & \cellcolor[RGB]{254,217,135}\textbf{16.0} & \cellcolor[RGB]{254,224,156}13.1 & \cellcolor[RGB]{255,248,231}3.1 & \cellcolor[RGB]{255,239,203}6.9 & \cellcolor[RGB]{254,220,145}14.7 & \cellcolor[RGB]{254,218,137}15.7 & \cellcolor[RGB]{255,247,229}3.4 & \cellcolor[RGB]{255,242,215}5.4 & \cellcolor[RGB]{255,254,253}0.3 & \cellcolor[RGB]{255,253,248}0.9 \\
 & Remove Outlier & \cellcolor[RGB]{255,254,252}0.4 & \cellcolor[RGB]{255,254,253}0.3 & \cellcolor[RGB]{255,254,253}0.3 & \cellcolor[RGB]{255,254,251}0.5 & \cellcolor[RGB]{255,254,251}0.5 & \cellcolor[RGB]{255,253,248}\textbf{0.9} & \cellcolor[RGB]{255,253,250}0.7 & \cellcolor[RGB]{255,254,253}0.3 & \cellcolor[RGB]{255,253,249}0.8 & \cellcolor[RGB]{255,255,255}0.0 & \cellcolor[RGB]{255,254,253}0.2 \\
 & Recolor Class & \cellcolor[RGB]{255,239,205}6.6 & \cellcolor[RGB]{255,232,183}\textbf{9.6} & \cellcolor[RGB]{255,253,248}0.9 & \cellcolor[RGB]{255,253,250}0.7 & \cellcolor[RGB]{255,255,254}0.1 & \cellcolor[RGB]{255,254,252}0.4 & \cellcolor[RGB]{255,254,251}0.5 & \cellcolor[RGB]{255,255,254}0.1 & \cellcolor[RGB]{255,252,246}1.2 & \cellcolor[RGB]{255,255,255}0.0 & \cellcolor[RGB]{255,255,254}0.1 \\
\rowcolor{chartlinechartdark}
\multicolumn{13}{l}{\textbf{\color{white}Line Chart}\strut} \\
\rowcolor{chartlinechartlight!50!white}
\multicolumn{2}{l}{} & \cellcolor[RGB]{254,228,169}11.5 & \cellcolor[RGB]{254,217,135}\textbf{16.0} & \cellcolor[RGB]{255,244,220}4.7 & \cellcolor[RGB]{255,245,222}4.4 & \cellcolor[RGB]{255,243,218}4.9 & \cellcolor[RGB]{255,246,227}3.8 & \cellcolor[RGB]{255,244,221}4.5 & \cellcolor[RGB]{255,248,231}3.2 & \cellcolor[RGB]{255,243,218}4.9 & \cellcolor[RGB]{255,255,255}0.0 & \cellcolor[RGB]{255,255,254}0.1 \\[-2pt]
\rowcolor{chartlinechartlight!50!white}
\multicolumn{2}{l}{\multirow{-2}{*}{\textit{\quad Plot Type Avg.}}} & \cellcolor[RGB]{254,228,169}{\color{black}\scriptsize $\pm1.6$} & \cellcolor[RGB]{254,217,135}{\color{black}\scriptsize $\pm1.5$} & \cellcolor[RGB]{255,244,220}{\color{black}\scriptsize $\pm0.4$} & \cellcolor[RGB]{255,245,222}{\color{black}\scriptsize $\pm0.7$} & \cellcolor[RGB]{255,243,218}{\color{black}\scriptsize $\pm0.8$} & \cellcolor[RGB]{255,246,227}{\color{black}\scriptsize $\pm0.7$} & \cellcolor[RGB]{255,244,221}{\color{black}\scriptsize $\pm0.3$} & \cellcolor[RGB]{255,248,231}{\color{black}\scriptsize $\pm0.4$} & \cellcolor[RGB]{255,243,218}{\color{black}\scriptsize $\pm0.9$} & \cellcolor[RGB]{255,255,255}{\color{black}\scriptsize $\pm0.0$} & \cellcolor[RGB]{255,255,254}{\color{black}\scriptsize $\pm0.0$} \\
 & Draw Segments & \cellcolor[RGB]{255,251,243}\textbf{1.6} & \cellcolor[RGB]{255,253,247}1.0 & \cellcolor[RGB]{255,254,252}0.4 & \cellcolor[RGB]{255,253,247}1.0 & \cellcolor[RGB]{255,253,250}0.6 & \cellcolor[RGB]{255,253,249}0.9 & \cellcolor[RGB]{255,254,252}0.4 & \cellcolor[RGB]{255,254,253}0.3 & \cellcolor[RGB]{255,254,250}0.6 & \cellcolor[RGB]{255,255,255}0.0 & \cellcolor[RGB]{255,255,254}0.1 \\
 & Normalize Series & \cellcolor[RGB]{255,241,209}6.2 & \cellcolor[RGB]{254,219,139}\textbf{15.5} & \cellcolor[RGB]{254,229,172}11.0 & \cellcolor[RGB]{254,228,168}11.5 & \cellcolor[RGB]{255,236,193}8.3 & \cellcolor[RGB]{255,234,188}8.9 & \cellcolor[RGB]{255,230,174}10.8 & \cellcolor[RGB]{255,235,190}8.6 & \cellcolor[RGB]{255,232,180}10.0 & \cellcolor[RGB]{255,255,255}0.0 & \cellcolor[RGB]{255,255,255}0.1 \\
 & Filter Series & \cellcolor[RGB]{255,232,181}9.8 & \cellcolor[RGB]{255,232,182}9.7 & \cellcolor[RGB]{255,238,202}7.1 & \cellcolor[RGB]{255,251,243}1.6 & \cellcolor[RGB]{255,231,178}\textbf{10.2} & \cellcolor[RGB]{255,244,220}4.7 & \cellcolor[RGB]{255,239,203}6.9 & \cellcolor[RGB]{255,247,231}3.2 & \cellcolor[RGB]{255,238,201}7.2 & \cellcolor[RGB]{255,255,254}0.1 & \cellcolor[RGB]{255,254,253}0.2 \\
 & Shade Interval & \cellcolor[RGB]{254,185,82}28.4 & \cellcolor[RGB]{253,158,69}\textbf{37.9} & \cellcolor[RGB]{255,254,253}0.3 & \cellcolor[RGB]{255,247,229}3.4 & \cellcolor[RGB]{255,254,251}0.5 & \cellcolor[RGB]{255,254,251}0.6 & \cellcolor[RGB]{255,255,255}0.0 & \cellcolor[RGB]{255,254,252}0.5 & \cellcolor[RGB]{255,251,241}1.9 & \cellcolor[RGB]{255,255,255}0.0 & \cellcolor[RGB]{255,255,255}0.0 \\
\rowcolor{chartheatmapdark}
\multicolumn{13}{l}{\textbf{\color{white}Heatmap}\strut} \\
\rowcolor{chartheatmaplight!50!white}
\multicolumn{2}{l}{} & \cellcolor[RGB]{254,207,103}\textbf{20.2} & \cellcolor[RGB]{254,219,138}15.5 & \cellcolor[RGB]{255,246,225}4.0 & \cellcolor[RGB]{255,243,218}4.9 & \cellcolor[RGB]{255,250,240}2.0 & \cellcolor[RGB]{255,247,231}3.2 & \cellcolor[RGB]{255,247,229}3.5 & \cellcolor[RGB]{255,236,196}7.9 & \cellcolor[RGB]{255,244,219}4.8 & \cellcolor[RGB]{255,253,248}0.9 & \cellcolor[RGB]{255,255,255}0.0 \\[-2pt]
\rowcolor{chartheatmaplight!50!white}
\multicolumn{2}{l}{\multirow{-2}{*}{\textit{\quad Plot Type Avg.}}} & \cellcolor[RGB]{254,207,103}{\color{black}\scriptsize $\pm2.8$} & \cellcolor[RGB]{254,219,138}{\color{black}\scriptsize $\pm2.5$} & \cellcolor[RGB]{255,246,225}{\color{black}\scriptsize $\pm0.7$} & \cellcolor[RGB]{255,243,218}{\color{black}\scriptsize $\pm1.3$} & \cellcolor[RGB]{255,250,240}{\color{black}\scriptsize $\pm0.5$} & \cellcolor[RGB]{255,247,231}{\color{black}\scriptsize $\pm1.2$} & \cellcolor[RGB]{255,247,229}{\color{black}\scriptsize $\pm0.6$} & \cellcolor[RGB]{255,236,196}{\color{black}\scriptsize $\pm1.4$} & \cellcolor[RGB]{255,244,219}{\color{black}\scriptsize $\pm0.8$} & \cellcolor[RGB]{255,253,248}{\color{black}\scriptsize $\pm0.3$} & \cellcolor[RGB]{255,255,255}{\color{black}\scriptsize $\pm0.0$} \\
 & Add Cell & \cellcolor[RGB]{255,249,235}2.6 & \cellcolor[RGB]{255,237,198}\textbf{7.6} & \cellcolor[RGB]{255,254,252}0.4 & \cellcolor[RGB]{255,254,250}0.6 & \cellcolor[RGB]{255,255,255}0.0 & \cellcolor[RGB]{255,253,248}1.0 & \cellcolor[RGB]{255,254,252}0.4 & \cellcolor[RGB]{255,254,253}0.2 & \cellcolor[RGB]{255,252,247}1.1 & \cellcolor[RGB]{255,255,255}0.0 & \cellcolor[RGB]{255,255,255}0.0 \\
 & Shift Heatmap & \cellcolor[RGB]{253,149,64}\textbf{41.1} & \cellcolor[RGB]{254,179,79}30.5 & \cellcolor[RGB]{254,223,153}13.5 & \cellcolor[RGB]{254,222,150}14.0 & \cellcolor[RGB]{255,238,200}7.4 & \cellcolor[RGB]{255,239,204}6.8 & \cellcolor[RGB]{255,231,179}10.1 & \cellcolor[RGB]{254,213,120}18.0 & \cellcolor[RGB]{254,216,131}16.5 & \cellcolor[RGB]{255,250,239}2.1 & \cellcolor[RGB]{255,255,254}0.1 \\
 & Mask Cells & \cellcolor[RGB]{254,197,89}\textbf{24.1} & \cellcolor[RGB]{254,213,122}17.8 & \cellcolor[RGB]{255,252,246}1.2 & \cellcolor[RGB]{255,244,220}4.6 & \cellcolor[RGB]{255,255,254}0.1 & \cellcolor[RGB]{255,244,219}4.7 & \cellcolor[RGB]{255,247,229}3.4 & \cellcolor[RGB]{254,224,155}13.3 & \cellcolor[RGB]{255,255,253}0.2 & \cellcolor[RGB]{255,252,246}1.2 & \cellcolor[RGB]{255,255,255}0.0 \\
 & Change Colormap & \cellcolor[RGB]{254,224,157}\textbf{13.1} & \cellcolor[RGB]{255,240,208}6.2 & \cellcolor[RGB]{255,253,249}0.8 & \cellcolor[RGB]{255,254,252}0.4 & \cellcolor[RGB]{255,254,252}0.4 & \cellcolor[RGB]{255,254,252}0.3 & \cellcolor[RGB]{255,255,255}0.0 & \cellcolor[RGB]{255,255,255}0.0 & \cellcolor[RGB]{255,252,244}1.5 & \cellcolor[RGB]{255,255,254}0.2 & \cellcolor[RGB]{255,255,255}0.0 \\
\rowcolor{chartnetworkdark}
\multicolumn{13}{l}{\textbf{\color{white}Network}\strut} \\
\rowcolor{chartnetworklight!50!white}
\multicolumn{2}{l}{} & \cellcolor[RGB]{255,244,219}\textbf{4.8} & \cellcolor[RGB]{255,244,219}\textbf{4.8} & \cellcolor[RGB]{255,249,235}2.7 & \cellcolor[RGB]{255,251,243}1.6 & \cellcolor[RGB]{255,251,242}1.8 & \cellcolor[RGB]{255,251,241}1.9 & \cellcolor[RGB]{255,250,241}1.9 & \cellcolor[RGB]{255,253,249}0.8 & \cellcolor[RGB]{255,249,237}2.4 & \cellcolor[RGB]{255,255,254}0.1 & \cellcolor[RGB]{255,255,254}0.1 \\[-2pt]
\rowcolor{chartnetworklight!50!white}
\multicolumn{2}{l}{\multirow{-2}{*}{\textit{\quad Plot Type Avg.}}} & \cellcolor[RGB]{255,244,219}{\color{black}\scriptsize $\pm0.6$} & \cellcolor[RGB]{255,244,219}{\color{black}\scriptsize $\pm0.8$} & \cellcolor[RGB]{255,249,235}{\color{black}\scriptsize $\pm0.4$} & \cellcolor[RGB]{255,251,243}{\color{black}\scriptsize $\pm0.4$} & \cellcolor[RGB]{255,251,242}{\color{black}\scriptsize $\pm0.5$} & \cellcolor[RGB]{255,251,241}{\color{black}\scriptsize $\pm0.4$} & \cellcolor[RGB]{255,250,241}{\color{black}\scriptsize $\pm0.2$} & \cellcolor[RGB]{255,253,249}{\color{black}\scriptsize $\pm0.3$} & \cellcolor[RGB]{255,249,237}{\color{black}\scriptsize $\pm0.6$} & \cellcolor[RGB]{255,255,254}{\color{black}\scriptsize $\pm0.0$} & \cellcolor[RGB]{255,255,254}{\color{black}\scriptsize $\pm0.1$} \\
 & Add Node & \cellcolor[RGB]{255,253,249}0.7 & \cellcolor[RGB]{255,252,246}\textbf{1.2} & \cellcolor[RGB]{255,254,251}0.5 & \cellcolor[RGB]{255,254,251}0.6 & \cellcolor[RGB]{255,253,248}1.0 & \cellcolor[RGB]{255,254,253}0.3 & \cellcolor[RGB]{255,254,253}0.3 & \cellcolor[RGB]{255,255,254}0.1 & \cellcolor[RGB]{255,254,251}0.6 & \cellcolor[RGB]{255,255,255}0.0 & \cellcolor[RGB]{255,255,254}0.1 \\
 & Swap Nodes & \cellcolor[RGB]{255,243,217}\textbf{5.0} & \cellcolor[RGB]{255,243,217}\textbf{5.1} & \cellcolor[RGB]{255,246,226}3.8 & \cellcolor[RGB]{255,250,239}2.1 & \cellcolor[RGB]{255,251,243}1.6 & \cellcolor[RGB]{255,248,232}3.1 & \cellcolor[RGB]{255,249,234}2.8 & \cellcolor[RGB]{255,254,252}0.5 & \cellcolor[RGB]{255,247,230}3.4 & \cellcolor[RGB]{255,255,254}0.2 & \cellcolor[RGB]{255,255,254}0.2 \\
 & Remove Node & \cellcolor[RGB]{255,234,187}\textbf{9.1} & \cellcolor[RGB]{255,235,191}8.6 & \cellcolor[RGB]{255,240,207}6.3 & \cellcolor[RGB]{255,247,229}3.5 & \cellcolor[RGB]{255,244,221}4.6 & \cellcolor[RGB]{255,246,225}4.0 & \cellcolor[RGB]{255,244,220}4.6 & \cellcolor[RGB]{255,249,236}2.5 & \cellcolor[RGB]{255,242,213}5.6 & \cellcolor[RGB]{255,255,254}0.1 & \cellcolor[RGB]{255,254,253}0.3 \\
 & Recolor Node & \cellcolor[RGB]{255,245,222}4.4 & \cellcolor[RGB]{255,245,221}\textbf{4.5} & \cellcolor[RGB]{255,254,253}0.2 & \cellcolor[RGB]{255,255,254}0.2 & \cellcolor[RGB]{255,255,254}0.1 & \cellcolor[RGB]{255,255,255}0.1 & \cellcolor[RGB]{255,255,255}0.0 & \cellcolor[RGB]{255,255,255}0.0 & \cellcolor[RGB]{255,255,254}0.1 & \cellcolor[RGB]{255,255,255}0.0 & \cellcolor[RGB]{255,255,255}0.0 \\
\midrule
\rowcolor{benchbg}
\multicolumn{2}{l}{} & \cellcolor[RGB]{254,218,135}\textbf{15.9} & \cellcolor[RGB]{254,218,138}15.6 & \cellcolor[RGB]{255,242,215}5.3 & \cellcolor[RGB]{255,247,230}3.4 & \cellcolor[RGB]{255,249,235}2.7 & \cellcolor[RGB]{255,248,232}3.1 & \cellcolor[RGB]{255,248,232}3.1 & \cellcolor[RGB]{255,248,231}3.2 & \cellcolor[RGB]{255,247,229}3.4 & \cellcolor[RGB]{255,254,253}0.3 & \cellcolor[RGB]{255,255,254}0.2 \\[-2pt]
\rowcolor{benchbg}
\multicolumn{2}{l}{\multirow{-2}{*}{\textbf{\tinygrafixbench\ Avg.}}} & \cellcolor[RGB]{254,218,135}{\color{black}\scriptsize $\pm0.8$} & \cellcolor[RGB]{254,218,138}{\color{black}\scriptsize $\pm0.8$} & \cellcolor[RGB]{255,242,215}{\color{black}\scriptsize $\pm0.3$} & \cellcolor[RGB]{255,247,230}{\color{black}\scriptsize $\pm0.4$} & \cellcolor[RGB]{255,249,235}{\color{black}\scriptsize $\pm0.3$} & \cellcolor[RGB]{255,248,232}{\color{black}\scriptsize $\pm0.3$} & \cellcolor[RGB]{255,248,232}{\color{black}\scriptsize $\pm0.2$} & \cellcolor[RGB]{255,248,231}{\color{black}\scriptsize $\pm0.3$} & \cellcolor[RGB]{255,247,229}{\color{black}\scriptsize $\pm0.3$} & \cellcolor[RGB]{255,254,253}{\color{black}\scriptsize $\pm0.1$} & \cellcolor[RGB]{255,255,254}{\color{black}\scriptsize $\pm0.1$} \\
\bottomrule
\end{tabular}%
}
\end{table}

\begin{table}[p]
\centering
\scriptsize
\caption{\textbf{\tinygrafixbench\ Edit-Region Accuracy (\%) per task-mode.} Best per row in bold. Aggregate-level bootstrap CIs are reported in \cref{sec:appx:bootstrap-ci}.}
\label{tab:tinygrafixbench_full_edit_acc}
\resizebox{\linewidth}{!}{%
\begin{tabular}{llrrrrrrrrrrr}
\toprule
\textbf{Chart Type} & \textbf{Task} & \textbf{NB-2} & \textbf{GPT-I2} & \textbf{NB-1} & \textbf{Qwen-IE} & \textbf{BAGEL} & \textbf{FLUX.2-D} & \textbf{FLUX.1-Kt} & \textbf{LCat-IE} & \textbf{FLUX.2-Kl} & \textbf{HY-3} & \textbf{IP2P} \\
\midrule
\rowcolor{chartbarchartdark}
\multicolumn{13}{l}{\textbf{\color{white}Bar Chart}\strut} \\
\rowcolor{chartbarchartlight!50!white}
\multicolumn{2}{l}{\textit{\quad Plot Type Avg.}} & \cellcolor[RGB]{224,77,50}{\color{white}\textbf{70.5}} & \cellcolor[RGB]{252,138,60}58.1 & \cellcolor[RGB]{254,204,93}28.1 & \cellcolor[RGB]{255,235,192}10.9 & \cellcolor[RGB]{255,243,218}6.4 & \cellcolor[RGB]{255,249,236}3.3 & \cellcolor[RGB]{255,243,216}6.8 & \cellcolor[RGB]{255,243,216}6.8 & \cellcolor[RGB]{255,248,232}4.0 & \cellcolor[RGB]{255,251,244}2.0 & \cellcolor[RGB]{255,254,253}0.4 \\
 & Add Bar & \cellcolor[RGB]{248,129,58}\textbf{59.9} & \cellcolor[RGB]{253,150,65}53.1 & \cellcolor[RGB]{255,253,248}1.2 & \cellcolor[RGB]{255,253,250}0.9 & \cellcolor[RGB]{255,255,255}0.0 & \cellcolor[RGB]{255,255,255}0.1 & \cellcolor[RGB]{255,255,255}0.0 & \cellcolor[RGB]{255,254,251}0.7 & \cellcolor[RGB]{255,255,255}0.0 & \cellcolor[RGB]{255,255,255}0.0 & \cellcolor[RGB]{255,255,255}0.0 \\
 & Sort Bars & \cellcolor[RGB]{216,59,47}{\color{white}74.0} & \cellcolor[RGB]{215,57,47}{\color{white}\textbf{74.4}} & \cellcolor[RGB]{254,177,78}40.7 & \cellcolor[RGB]{255,232,181}12.8 & \cellcolor[RGB]{255,246,226}5.1 & \cellcolor[RGB]{255,236,193}10.8 & \cellcolor[RGB]{255,245,222}5.8 & \cellcolor[RGB]{255,231,178}13.3 & \cellcolor[RGB]{254,228,170}14.8 & \cellcolor[RGB]{255,255,254}0.1 & \cellcolor[RGB]{255,254,251}0.7 \\
 & Remove Bar & \cellcolor[RGB]{198,20,41}{\color{white}\textbf{81.9}} & \cellcolor[RGB]{253,149,64}53.7 & \cellcolor[RGB]{230,90,52}{\color{white}67.8} & \cellcolor[RGB]{254,201,90}29.8 & \cellcolor[RGB]{254,218,137}20.4 & \cellcolor[RGB]{255,255,254}0.1 & \cellcolor[RGB]{254,216,131}21.5 & \cellcolor[RGB]{255,235,190}11.2 & \cellcolor[RGB]{255,254,253}0.4 & \cellcolor[RGB]{255,241,211}7.7 & \cellcolor[RGB]{255,253,250}0.9 \\
 & Recolor Bar & \cellcolor[RGB]{233,98,53}{\color{white}\textbf{66.2}} & \cellcolor[RGB]{253,155,67}51.3 & \cellcolor[RGB]{255,250,238}2.9 & \cellcolor[RGB]{255,255,254}0.2 & \cellcolor[RGB]{255,255,255}0.0 & \cellcolor[RGB]{255,251,243}2.0 & \cellcolor[RGB]{255,255,255}0.0 & \cellcolor[RGB]{255,252,244}1.9 & \cellcolor[RGB]{255,254,251}0.6 & \cellcolor[RGB]{255,255,255}0.0 & \cellcolor[RGB]{255,255,255}0.0 \\
\rowcolor{chartscatterplotdark}
\multicolumn{13}{l}{\textbf{\color{white}Scatter Plot}\strut} \\
\rowcolor{chartscatterplotlight!50!white}
\multicolumn{2}{l}{\textit{\quad Plot Type Avg.}} & \cellcolor[RGB]{254,215,128}22.1 & \cellcolor[RGB]{254,217,133}21.2 & \cellcolor[RGB]{254,213,120}23.5 & \cellcolor[RGB]{254,223,153}17.7 & \cellcolor[RGB]{255,231,177}13.5 & \cellcolor[RGB]{254,212,119}23.7 & \cellcolor[RGB]{254,191,86}\textbf{34.2} & \cellcolor[RGB]{255,230,174}14.0 & \cellcolor[RGB]{254,221,148}18.6 & \cellcolor[RGB]{255,254,250}0.8 & \cellcolor[RGB]{255,241,211}7.7 \\
 & Draw Best Fit Line & \cellcolor[RGB]{255,245,222}5.7 & \cellcolor[RGB]{255,244,221}\textbf{6.0} & \cellcolor[RGB]{255,245,223}5.6 & \cellcolor[RGB]{255,245,224}5.4 & \cellcolor[RGB]{255,245,224}5.3 & \cellcolor[RGB]{255,247,228}4.6 & \cellcolor[RGB]{255,246,225}5.2 & \cellcolor[RGB]{255,250,240}2.6 & \cellcolor[RGB]{255,246,226}5.1 & \cellcolor[RGB]{255,254,252}0.6 & \cellcolor[RGB]{255,254,251}0.7 \\
 & Swap Axes & \cellcolor[RGB]{254,214,123}23.0 & \cellcolor[RGB]{254,189,84}35.3 & \cellcolor[RGB]{254,176,78}41.3 & \cellcolor[RGB]{255,240,208}8.1 & \cellcolor[RGB]{255,231,179}13.3 & \cellcolor[RGB]{254,211,113}24.7 & \cellcolor[RGB]{254,175,77}\textbf{41.8} & \cellcolor[RGB]{255,233,184}12.3 & \cellcolor[RGB]{255,230,177}13.6 & \cellcolor[RGB]{255,253,249}1.0 & \cellcolor[RGB]{255,251,241}2.4 \\
 & Remove Outlier & \cellcolor[RGB]{254,202,91}29.0 & \cellcolor[RGB]{255,233,183}12.5 & \cellcolor[RGB]{254,178,79}40.5 & \cellcolor[RGB]{253,151,65}53.1 & \cellcolor[RGB]{254,189,84}35.2 & \cellcolor[RGB]{244,122,57}{\color{white}61.4} & \cellcolor[RGB]{189,0,38}{\color{white}\textbf{85.9}} & \cellcolor[RGB]{254,179,79}40.1 & \cellcolor[RGB]{253,153,66}52.1 & \cellcolor[RGB]{255,253,247}1.4 & \cellcolor[RGB]{254,206,100}27.0 \\
 & Recolor Class & \cellcolor[RGB]{254,199,89}30.6 & \cellcolor[RGB]{254,198,89}\textbf{30.9} & \cellcolor[RGB]{255,243,216}6.8 & \cellcolor[RGB]{255,248,231}4.1 & \cellcolor[RGB]{255,254,253}0.3 & \cellcolor[RGB]{255,248,232}4.1 & \cellcolor[RGB]{255,248,232}3.9 & \cellcolor[RGB]{255,253,249}1.1 & \cellcolor[RGB]{255,248,233}3.8 & \cellcolor[RGB]{255,254,253}0.3 & \cellcolor[RGB]{255,254,250}0.8 \\
\rowcolor{chartlinechartdark}
\multicolumn{13}{l}{\textbf{\color{white}Line Chart}\strut} \\
\rowcolor{chartlinechartlight!50!white}
\multicolumn{2}{l}{\textit{\quad Plot Type Avg.}} & \cellcolor[RGB]{254,178,79}40.2 & \cellcolor[RGB]{253,170,75}\textbf{44.2} & \cellcolor[RGB]{254,193,87}33.3 & \cellcolor[RGB]{254,215,129}22.0 & \cellcolor[RGB]{254,209,109}25.3 & \cellcolor[RGB]{254,225,159}16.6 & \cellcolor[RGB]{254,195,88}32.3 & \cellcolor[RGB]{254,216,130}21.7 & \cellcolor[RGB]{254,212,119}23.7 & \cellcolor[RGB]{255,253,249}1.0 & \cellcolor[RGB]{255,253,249}1.0 \\
 & Draw Segments & \cellcolor[RGB]{254,219,140}\textbf{20.0} & \cellcolor[RGB]{255,235,191}11.2 & \cellcolor[RGB]{255,238,201}9.4 & \cellcolor[RGB]{254,226,163}16.0 & \cellcolor[RGB]{255,243,215}6.9 & \cellcolor[RGB]{255,240,208}8.1 & \cellcolor[RGB]{255,240,207}8.3 & \cellcolor[RGB]{255,242,214}7.1 & \cellcolor[RGB]{255,238,202}9.3 & \cellcolor[RGB]{255,253,249}1.0 & \cellcolor[RGB]{255,252,245}1.7 \\
 & Normalize Series & \cellcolor[RGB]{254,222,150}18.3 & \cellcolor[RGB]{254,176,78}41.4 & \cellcolor[RGB]{253,166,73}45.7 & \cellcolor[RGB]{253,165,72}\textbf{46.3} & \cellcolor[RGB]{254,214,124}22.7 & \cellcolor[RGB]{254,203,91}29.0 & \cellcolor[RGB]{253,168,74}45.1 & \cellcolor[RGB]{254,189,84}35.4 & \cellcolor[RGB]{254,204,92}28.6 & \cellcolor[RGB]{255,254,252}0.6 & \cellcolor[RGB]{255,254,252}0.5 \\
 & Filter Series & \cellcolor[RGB]{223,74,50}{\color{white}70.9} & \cellcolor[RGB]{221,71,49}{\color{white}71.7} & \cellcolor[RGB]{209,45,45}{\color{white}\textbf{76.9}} & \cellcolor[RGB]{254,226,162}16.1 & \cellcolor[RGB]{223,75,50}{\color{white}70.8} & \cellcolor[RGB]{254,205,95}27.9 & \cellcolor[RGB]{212,50,46}{\color{white}75.8} & \cellcolor[RGB]{253,172,76}43.1 & \cellcolor[RGB]{253,151,65}52.7 & \cellcolor[RGB]{255,251,242}2.2 & \cellcolor[RGB]{255,252,245}1.7 \\
 & Shade Interval & \cellcolor[RGB]{253,154,67}51.6 & \cellcolor[RGB]{253,152,66}\textbf{52.5} & \cellcolor[RGB]{255,253,248}1.2 & \cellcolor[RGB]{255,238,200}9.5 & \cellcolor[RGB]{255,254,250}0.8 & \cellcolor[RGB]{255,252,246}1.5 & \cellcolor[RGB]{255,255,255}0.0 & \cellcolor[RGB]{255,253,248}1.2 & \cellcolor[RGB]{255,247,230}4.4 & \cellcolor[RGB]{255,255,255}0.0 & \cellcolor[RGB]{255,255,254}0.1 \\
\rowcolor{chartheatmapdark}
\multicolumn{13}{l}{\textbf{\color{white}Heatmap}\strut} \\
\rowcolor{chartheatmaplight!50!white}
\multicolumn{2}{l}{\textit{\quad Plot Type Avg.}} & \cellcolor[RGB]{254,206,98}\textbf{27.3} & \cellcolor[RGB]{254,217,134}21.0 & \cellcolor[RGB]{255,245,224}5.3 & \cellcolor[RGB]{255,242,213}7.3 & \cellcolor[RGB]{255,251,243}2.1 & \cellcolor[RGB]{255,246,227}4.9 & \cellcolor[RGB]{255,240,207}8.4 & \cellcolor[RGB]{255,230,174}14.1 & \cellcolor[RGB]{255,245,222}5.7 & \cellcolor[RGB]{255,252,246}1.6 & \cellcolor[RGB]{255,255,255}0.1 \\
 & Add Cell & \cellcolor[RGB]{255,241,210}7.7 & \cellcolor[RGB]{254,227,166}\textbf{15.4} & \cellcolor[RGB]{255,250,239}2.8 & \cellcolor[RGB]{255,246,225}5.3 & \cellcolor[RGB]{255,255,255}0.1 & \cellcolor[RGB]{255,250,240}2.6 & \cellcolor[RGB]{255,236,195}10.5 & \cellcolor[RGB]{255,248,232}3.9 & \cellcolor[RGB]{255,251,241}2.4 & \cellcolor[RGB]{255,254,252}0.5 & \cellcolor[RGB]{255,255,255}0.0 \\
 & Shift Heatmap & \cellcolor[RGB]{253,171,75}\textbf{43.6} & \cellcolor[RGB]{254,196,88}32.2 & \cellcolor[RGB]{254,227,166}15.5 & \cellcolor[RGB]{254,227,165}15.7 & \cellcolor[RGB]{255,241,210}7.8 & \cellcolor[RGB]{255,242,214}7.1 & \cellcolor[RGB]{255,230,175}13.9 & \cellcolor[RGB]{254,218,136}20.7 & \cellcolor[RGB]{254,223,151}18.0 & \cellcolor[RGB]{255,250,240}2.7 & \cellcolor[RGB]{255,255,254}0.1 \\
 & Mask Cells & \cellcolor[RGB]{254,175,77}\textbf{41.9} & \cellcolor[RGB]{254,204,92}28.6 & \cellcolor[RGB]{255,251,243}2.1 & \cellcolor[RGB]{255,241,210}7.8 & \cellcolor[RGB]{255,255,254}0.2 & \cellcolor[RGB]{255,238,200}9.5 & \cellcolor[RGB]{255,239,202}9.1 & \cellcolor[RGB]{254,197,88}31.6 & \cellcolor[RGB]{255,255,253}0.3 & \cellcolor[RGB]{255,250,239}2.9 & \cellcolor[RGB]{255,255,255}0.0 \\
 & Change Colormap & \cellcolor[RGB]{254,226,164}\textbf{15.9} & \cellcolor[RGB]{255,241,210}7.8 & \cellcolor[RGB]{255,253,250}0.9 & \cellcolor[RGB]{255,254,253}0.4 & \cellcolor[RGB]{255,254,252}0.5 & \cellcolor[RGB]{255,254,252}0.5 & \cellcolor[RGB]{255,255,255}0.0 & \cellcolor[RGB]{255,255,255}0.1 & \cellcolor[RGB]{255,251,242}2.2 & \cellcolor[RGB]{255,254,253}0.4 & \cellcolor[RGB]{255,255,255}0.1 \\
\rowcolor{chartnetworkdark}
\multicolumn{13}{l}{\textbf{\color{white}Network}\strut} \\
\rowcolor{chartnetworklight!50!white}
\multicolumn{2}{l}{\textit{\quad Plot Type Avg.}} & \cellcolor[RGB]{254,193,86}\textbf{33.6} & \cellcolor[RGB]{254,215,126}22.5 & \cellcolor[RGB]{254,214,124}22.7 & \cellcolor[RGB]{255,240,207}8.3 & \cellcolor[RGB]{255,241,211}7.6 & \cellcolor[RGB]{255,236,196}10.3 & \cellcolor[RGB]{254,208,105}26.0 & \cellcolor[RGB]{255,243,217}6.6 & \cellcolor[RGB]{255,233,186}12.1 & \cellcolor[RGB]{255,254,250}0.8 & \cellcolor[RGB]{255,252,246}1.5 \\
 & Add Node & \cellcolor[RGB]{255,244,219}6.3 & \cellcolor[RGB]{255,239,205}\textbf{8.6} & \cellcolor[RGB]{255,244,221}6.0 & \cellcolor[RGB]{255,247,229}4.5 & \cellcolor[RGB]{255,246,226}5.0 & \cellcolor[RGB]{255,250,238}2.9 & \cellcolor[RGB]{255,247,229}4.5 & \cellcolor[RGB]{255,251,242}2.2 & \cellcolor[RGB]{255,247,231}4.3 & \cellcolor[RGB]{255,254,253}0.4 & \cellcolor[RGB]{255,252,245}1.8 \\
 & Swap Nodes & \cellcolor[RGB]{254,218,135}20.8 & \cellcolor[RGB]{254,227,165}15.7 & \cellcolor[RGB]{254,212,117}24.0 & \cellcolor[RGB]{255,241,210}7.9 & \cellcolor[RGB]{255,248,233}3.8 & \cellcolor[RGB]{255,236,193}10.8 & \cellcolor[RGB]{254,208,103}\textbf{26.4} & \cellcolor[RGB]{255,249,237}3.1 & \cellcolor[RGB]{255,233,186}12.0 & \cellcolor[RGB]{255,252,246}1.6 & \cellcolor[RGB]{255,253,248}1.2 \\
 & Remove Node & \cellcolor[RGB]{247,129,58}60.0 & \cellcolor[RGB]{254,176,78}41.1 & \cellcolor[RGB]{253,141,60}57.6 & \cellcolor[RGB]{254,221,146}19.0 & \cellcolor[RGB]{254,217,134}21.0 & \cellcolor[RGB]{254,207,101}26.7 & \cellcolor[RGB]{221,69,49}{\color{white}\textbf{71.9}} & \cellcolor[RGB]{254,217,135}20.9 & \cellcolor[RGB]{254,199,90}30.5 & \cellcolor[RGB]{255,253,248}1.2 & \cellcolor[RGB]{255,250,238}2.9 \\
 & Recolor Node & \cellcolor[RGB]{253,163,71}\textbf{47.4} & \cellcolor[RGB]{254,211,114}24.5 & \cellcolor[RGB]{255,249,235}3.4 & \cellcolor[RGB]{255,252,245}1.7 & \cellcolor[RGB]{255,254,252}0.5 & \cellcolor[RGB]{255,253,250}0.8 & \cellcolor[RGB]{255,253,247}1.4 & \cellcolor[RGB]{255,255,254}0.2 & \cellcolor[RGB]{255,252,246}1.5 & \cellcolor[RGB]{255,255,254}0.1 & \cellcolor[RGB]{255,255,255}0.0 \\
\midrule
\rowcolor{benchbg}
\multicolumn{2}{l}{\textbf{\tinygrafixbench\ Avg.}} & \cellcolor[RGB]{254,182,81}\textbf{38.7} & \cellcolor[RGB]{254,193,86}33.4 & \cellcolor[RGB]{254,214,125}22.6 & \cellcolor[RGB]{255,231,179}13.2 & \cellcolor[RGB]{255,235,192}11.0 & \cellcolor[RGB]{255,234,187}11.8 & \cellcolor[RGB]{254,216,131}21.6 & \cellcolor[RGB]{255,232,182}12.6 & \cellcolor[RGB]{255,232,181}12.8 & \cellcolor[RGB]{255,253,248}1.2 & \cellcolor[RGB]{255,251,243}2.1 \\
\bottomrule
\end{tabular}%
}
\end{table}

\begin{table}[p]
\centering
\scriptsize
\caption{\textbf{\tinygrafixbench\ Preservation-Region Accuracy (\%) per task-mode.} Best per row in bold. Aggregate-level bootstrap CIs are reported in \cref{sec:appx:bootstrap-ci}.}
\label{tab:tinygrafixbench_full_pres_acc}
\resizebox{\linewidth}{!}{%
\begin{tabular}{llrrrrrrrrrrr}
\toprule
\textbf{Chart Type} & \textbf{Task} & \textbf{NB-2} & \textbf{GPT-I2} & \textbf{NB-1} & \textbf{Qwen-IE} & \textbf{BAGEL} & \textbf{FLUX.2-D} & \textbf{FLUX.1-Kt} & \textbf{LCat-IE} & \textbf{FLUX.2-Kl} & \textbf{HY-3} & \textbf{IP2P} \\
\midrule
\rowcolor{chartbarchartdark}
\multicolumn{13}{l}{\textbf{\color{white}Bar Chart}\strut} \\
\rowcolor{chartbarchartlight!50!white}
\multicolumn{2}{l}{\textit{\quad Plot Type Avg.}} & \cellcolor[RGB]{205,34,43}{\color{white}78.5} & \cellcolor[RGB]{222,73,49}{\color{white}70.8} & \cellcolor[RGB]{212,51,46}{\color{white}75.1} & \cellcolor[RGB]{202,30,43}{\color{white}\textbf{79.5}} & \cellcolor[RGB]{219,65,48}{\color{white}72.4} & \cellcolor[RGB]{249,133,59}58.8 & \cellcolor[RGB]{250,134,59}58.5 & \cellcolor[RGB]{253,170,75}43.6 & \cellcolor[RGB]{238,107,55}{\color{white}63.9} & \cellcolor[RGB]{254,212,119}23.5 & \cellcolor[RGB]{255,235,192}10.9 \\
 & Add Bar & \cellcolor[RGB]{203,32,43}{\color{white}\textbf{79.1}} & \cellcolor[RGB]{220,68,49}{\color{white}71.8} & \cellcolor[RGB]{210,46,45}{\color{white}76.2} & \cellcolor[RGB]{204,34,43}{\color{white}78.7} & \cellcolor[RGB]{218,64,48}{\color{white}72.6} & \cellcolor[RGB]{253,153,66}51.8 & \cellcolor[RGB]{240,112,55}{\color{white}63.1} & \cellcolor[RGB]{253,148,63}54.2 & \cellcolor[RGB]{229,89,52}{\color{white}67.6} & \cellcolor[RGB]{254,218,135}20.7 & \cellcolor[RGB]{255,243,217}6.5 \\
 & Sort Bars & \cellcolor[RGB]{205,35,43}{\color{white}\textbf{78.5}} & \cellcolor[RGB]{215,58,47}{\color{white}73.8} & \cellcolor[RGB]{212,52,46}{\color{white}75.1} & \cellcolor[RGB]{206,38,44}{\color{white}77.8} & \cellcolor[RGB]{219,65,48}{\color{white}72.3} & \cellcolor[RGB]{231,92,52}{\color{white}66.9} & \cellcolor[RGB]{252,139,60}57.7 & \cellcolor[RGB]{253,149,64}53.5 & \cellcolor[RGB]{237,106,54}{\color{white}64.3} & \cellcolor[RGB]{254,213,121}23.2 & \cellcolor[RGB]{255,244,220}6.1 \\
 & Remove Bar & \cellcolor[RGB]{205,36,44}{\color{white}78.2} & \cellcolor[RGB]{233,98,53}{\color{white}65.9} & \cellcolor[RGB]{216,59,47}{\color{white}73.6} & \cellcolor[RGB]{197,17,41}{\color{white}\textbf{82.0}} & \cellcolor[RGB]{223,75,50}{\color{white}70.3} & \cellcolor[RGB]{230,91,52}{\color{white}67.2} & \cellcolor[RGB]{253,145,62}55.6 & \cellcolor[RGB]{254,182,81}38.2 & \cellcolor[RGB]{223,76,50}{\color{white}70.2} & \cellcolor[RGB]{254,214,124}22.6 & \cellcolor[RGB]{254,219,141}19.7 \\
 & Recolor Bar & \cellcolor[RGB]{205,35,43}{\color{white}78.4} & \cellcolor[RGB]{220,69,49}{\color{white}71.5} & \cellcolor[RGB]{211,48,46}{\color{white}75.7} & \cellcolor[RGB]{203,30,43}{\color{white}\textbf{79.4}} & \cellcolor[RGB]{215,56,47}{\color{white}74.1} & \cellcolor[RGB]{253,158,69}49.4 & \cellcolor[RGB]{252,138,60}57.7 & \cellcolor[RGB]{254,203,91}28.8 & \cellcolor[RGB]{253,149,64}53.6 & \cellcolor[RGB]{254,205,96}27.6 & \cellcolor[RGB]{255,235,190}11.2 \\
\rowcolor{chartscatterplotdark}
\multicolumn{13}{l}{\textbf{\color{white}Scatter Plot}\strut} \\
\rowcolor{chartscatterplotlight!50!white}
\multicolumn{2}{l}{\textit{\quad Plot Type Avg.}} & \cellcolor[RGB]{202,29,43}{\color{white}79.6} & \cellcolor[RGB]{216,60,47}{\color{white}73.3} & \cellcolor[RGB]{199,22,41}{\color{white}81.0} & \cellcolor[RGB]{192,8,39}{\color{white}\textbf{83.9}} & \cellcolor[RGB]{211,49,46}{\color{white}75.5} & \cellcolor[RGB]{237,105,54}{\color{white}64.4} & \cellcolor[RGB]{201,27,42}{\color{white}80.0} & \cellcolor[RGB]{254,183,81}38.0 & \cellcolor[RGB]{215,57,47}{\color{white}74.0} & \cellcolor[RGB]{254,228,167}15.1 & \cellcolor[RGB]{254,223,151}17.9 \\
 & Draw Best Fit Line & \cellcolor[RGB]{200,25,42}{\color{white}80.4} & \cellcolor[RGB]{219,67,48}{\color{white}72.0} & \cellcolor[RGB]{203,30,43}{\color{white}79.4} & \cellcolor[RGB]{194,10,40}{\color{white}\textbf{83.3}} & \cellcolor[RGB]{206,38,44}{\color{white}77.8} & \cellcolor[RGB]{218,65,48}{\color{white}72.5} & \cellcolor[RGB]{210,46,45}{\color{white}76.2} & \cellcolor[RGB]{254,180,80}39.5 & \cellcolor[RGB]{211,49,46}{\color{white}75.6} & \cellcolor[RGB]{255,239,203}9.0 & \cellcolor[RGB]{255,235,193}10.8 \\
 & Swap Axes & \cellcolor[RGB]{207,40,44}{\color{white}77.4} & \cellcolor[RGB]{213,53,46}{\color{white}74.8} & \cellcolor[RGB]{196,15,40}{\color{white}82.3} & \cellcolor[RGB]{192,7,39}{\color{white}\textbf{84.0}} & \cellcolor[RGB]{198,20,41}{\color{white}81.4} & \cellcolor[RGB]{217,61,48}{\color{white}73.1} & \cellcolor[RGB]{199,22,41}{\color{white}81.0} & \cellcolor[RGB]{254,183,81}37.8 & \cellcolor[RGB]{211,49,46}{\color{white}75.5} & \cellcolor[RGB]{254,222,150}18.2 & \cellcolor[RGB]{255,238,200}9.5 \\
 & Remove Outlier & \cellcolor[RGB]{198,20,41}{\color{white}81.5} & \cellcolor[RGB]{212,52,46}{\color{white}75.1} & \cellcolor[RGB]{197,17,41}{\color{white}82.0} & \cellcolor[RGB]{191,4,39}{\color{white}\textbf{84.6}} & \cellcolor[RGB]{217,62,48}{\color{white}73.0} & \cellcolor[RGB]{216,59,47}{\color{white}73.6} & \cellcolor[RGB]{199,22,41}{\color{white}81.0} & \cellcolor[RGB]{253,168,74}44.7 & \cellcolor[RGB]{202,29,42}{\color{white}79.7} & \cellcolor[RGB]{254,222,149}18.4 & \cellcolor[RGB]{254,205,95}27.6 \\
 & Recolor Class & \cellcolor[RGB]{203,32,43}{\color{white}79.0} & \cellcolor[RGB]{221,70,49}{\color{white}71.5} & \cellcolor[RGB]{200,25,42}{\color{white}80.4} & \cellcolor[RGB]{193,9,39}{\color{white}\textbf{83.5}} & \cellcolor[RGB]{224,77,50}{\color{white}70.0} & \cellcolor[RGB]{254,182,81}38.3 & \cellcolor[RGB]{197,18,41}{\color{white}81.9} & \cellcolor[RGB]{254,200,90}30.2 & \cellcolor[RGB]{235,102,54}{\color{white}65.0} & \cellcolor[RGB]{254,228,168}15.0 & \cellcolor[RGB]{254,212,117}23.9 \\
\rowcolor{chartlinechartdark}
\multicolumn{13}{l}{\textbf{\color{white}Line Chart}\strut} \\
\rowcolor{chartlinechartlight!50!white}
\multicolumn{2}{l}{\textit{\quad Plot Type Avg.}} & \cellcolor[RGB]{207,39,44}{\color{white}\textbf{77.6}} & \cellcolor[RGB]{221,70,49}{\color{white}71.4} & \cellcolor[RGB]{214,55,47}{\color{white}74.5} & \cellcolor[RGB]{212,51,46}{\color{white}75.3} & \cellcolor[RGB]{212,51,46}{\color{white}75.1} & \cellcolor[RGB]{244,122,57}{\color{white}61.0} & \cellcolor[RGB]{221,70,49}{\color{white}71.5} & \cellcolor[RGB]{249,132,59}59.0 & \cellcolor[RGB]{228,85,51}{\color{white}68.3} & \cellcolor[RGB]{255,246,226}5.1 & \cellcolor[RGB]{255,231,178}13.4 \\
 & Draw Segments & \cellcolor[RGB]{202,29,42}{\color{white}79.6} & \cellcolor[RGB]{220,69,49}{\color{white}71.6} & \cellcolor[RGB]{202,29,43}{\color{white}79.5} & \cellcolor[RGB]{199,22,41}{\color{white}\textbf{81.0}} & \cellcolor[RGB]{204,32,43}{\color{white}78.9} & \cellcolor[RGB]{224,78,50}{\color{white}69.8} & \cellcolor[RGB]{207,41,44}{\color{white}77.3} & \cellcolor[RGB]{245,123,57}{\color{white}60.9} & \cellcolor[RGB]{220,69,49}{\color{white}71.5} & \cellcolor[RGB]{255,248,232}3.9 & \cellcolor[RGB]{255,234,189}11.3 \\
 & Normalize Series & \cellcolor[RGB]{199,22,41}{\color{white}\textbf{81.1}} & \cellcolor[RGB]{217,61,48}{\color{white}73.2} & \cellcolor[RGB]{201,26,42}{\color{white}80.1} & \cellcolor[RGB]{202,29,43}{\color{white}79.6} & \cellcolor[RGB]{211,48,45}{\color{white}75.9} & \cellcolor[RGB]{239,109,55}{\color{white}63.6} & \cellcolor[RGB]{202,28,42}{\color{white}79.8} & \cellcolor[RGB]{238,108,55}{\color{white}63.7} & \cellcolor[RGB]{209,45,45}{\color{white}76.4} & \cellcolor[RGB]{255,251,243}2.1 & \cellcolor[RGB]{255,246,227}4.9 \\
 & Filter Series & \cellcolor[RGB]{207,40,44}{\color{white}77.3} & \cellcolor[RGB]{223,75,50}{\color{white}70.3} & \cellcolor[RGB]{212,50,46}{\color{white}75.4} & \cellcolor[RGB]{211,47,45}{\color{white}75.9} & \cellcolor[RGB]{208,42,45}{\color{white}77.0} & \cellcolor[RGB]{229,88,52}{\color{white}67.8} & \cellcolor[RGB]{207,39,44}{\color{white}\textbf{77.6}} & \cellcolor[RGB]{243,118,56}{\color{white}61.8} & \cellcolor[RGB]{223,74,50}{\color{white}70.6} & \cellcolor[RGB]{255,238,202}9.1 & \cellcolor[RGB]{254,213,119}23.5 \\
 & Shade Interval & \cellcolor[RGB]{219,65,48}{\color{white}\textbf{72.3}} & \cellcolor[RGB]{223,75,50}{\color{white}70.4} & \cellcolor[RGB]{240,113,56}{\color{white}62.7} & \cellcolor[RGB]{236,104,54}{\color{white}64.6} & \cellcolor[RGB]{227,84,51}{\color{white}68.7} & \cellcolor[RGB]{253,172,76}42.8 & \cellcolor[RGB]{253,154,67}51.1 & \cellcolor[RGB]{253,158,69}49.4 & \cellcolor[RGB]{253,146,63}54.7 & \cellcolor[RGB]{255,246,225}5.2 & \cellcolor[RGB]{255,230,175}13.8 \\
\rowcolor{chartheatmapdark}
\multicolumn{13}{l}{\textbf{\color{white}Heatmap}\strut} \\
\rowcolor{chartheatmaplight!50!white}
\multicolumn{2}{l}{\textit{\quad Plot Type Avg.}} & \cellcolor[RGB]{228,85,51}{\color{white}68.3} & \cellcolor[RGB]{235,100,54}{\color{white}65.3} & \cellcolor[RGB]{223,75,50}{\color{white}70.4} & \cellcolor[RGB]{219,67,48}{\color{white}\textbf{72.0}} & \cellcolor[RGB]{243,119,57}{\color{white}61.6} & \cellcolor[RGB]{253,147,63}54.6 & \cellcolor[RGB]{254,203,92}28.5 & \cellcolor[RGB]{254,190,85}34.7 & \cellcolor[RGB]{253,149,64}53.5 & \cellcolor[RGB]{255,233,184}12.3 & \cellcolor[RGB]{255,232,180}12.9 \\
 & Add Cell & \cellcolor[RGB]{208,42,45}{\color{white}77.1} & \cellcolor[RGB]{215,56,47}{\color{white}74.2} & \cellcolor[RGB]{218,64,48}{\color{white}72.6} & \cellcolor[RGB]{206,37,44}{\color{white}\textbf{78.1}} & \cellcolor[RGB]{227,84,51}{\color{white}68.7} & \cellcolor[RGB]{242,117,56}{\color{white}62.0} & \cellcolor[RGB]{254,198,89}31.1 & \cellcolor[RGB]{254,198,89}31.1 & \cellcolor[RGB]{238,109,55}{\color{white}63.6} & \cellcolor[RGB]{255,232,183}12.4 & \cellcolor[RGB]{255,232,182}12.6 \\
 & Shift Heatmap & \cellcolor[RGB]{212,51,46}{\color{white}\textbf{75.2}} & \cellcolor[RGB]{217,61,48}{\color{white}73.2} & \cellcolor[RGB]{228,85,51}{\color{white}68.4} & \cellcolor[RGB]{221,71,49}{\color{white}71.2} & \cellcolor[RGB]{213,53,46}{\color{white}74.7} & \cellcolor[RGB]{221,71,49}{\color{white}71.3} & \cellcolor[RGB]{254,195,87}32.5 & \cellcolor[RGB]{253,164,72}46.5 & \cellcolor[RGB]{253,146,62}55.0 & \cellcolor[RGB]{254,217,133}21.1 & \cellcolor[RGB]{255,239,203}9.0 \\
 & Mask Cells & \cellcolor[RGB]{239,111,55}{\color{white}63.2} & \cellcolor[RGB]{249,133,59}58.9 & \cellcolor[RGB]{226,82,51}{\color{white}\textbf{69.0}} & \cellcolor[RGB]{229,89,52}{\color{white}67.7} & \cellcolor[RGB]{251,137,59}58.0 & \cellcolor[RGB]{253,148,64}53.8 & \cellcolor[RGB]{254,197,88}31.5 & \cellcolor[RGB]{254,208,104}26.1 & \cellcolor[RGB]{241,114,56}{\color{white}62.5} & \cellcolor[RGB]{255,237,197}10.0 & \cellcolor[RGB]{254,208,106}25.8 \\
 & Change Colormap & \cellcolor[RGB]{251,137,59}58.0 & \cellcolor[RGB]{253,146,62}55.1 & \cellcolor[RGB]{220,69,49}{\color{white}\textbf{71.7}} & \cellcolor[RGB]{221,71,49}{\color{white}71.2} & \cellcolor[RGB]{253,167,73}45.2 & \cellcolor[RGB]{254,197,89}31.2 & \cellcolor[RGB]{254,221,146}18.8 & \cellcolor[RGB]{254,189,84}35.0 & \cellcolor[RGB]{254,194,87}32.7 & \cellcolor[RGB]{255,245,223}5.6 & \cellcolor[RGB]{255,247,230}4.4 \\
\rowcolor{chartnetworkdark}
\multicolumn{13}{l}{\textbf{\color{white}Network}\strut} \\
\rowcolor{chartnetworklight!50!white}
\multicolumn{2}{l}{\textit{\quad Plot Type Avg.}} & \cellcolor[RGB]{201,26,42}{\color{white}80.1} & \cellcolor[RGB]{221,71,49}{\color{white}71.2} & \cellcolor[RGB]{197,17,41}{\color{white}81.9} & \cellcolor[RGB]{191,4,39}{\color{white}\textbf{84.5}} & \cellcolor[RGB]{202,30,43}{\color{white}79.5} & \cellcolor[RGB]{249,133,59}58.8 & \cellcolor[RGB]{213,52,46}{\color{white}75.0} & \cellcolor[RGB]{254,185,82}36.8 & \cellcolor[RGB]{211,48,46}{\color{white}75.7} & \cellcolor[RGB]{255,235,193}10.8 & \cellcolor[RGB]{254,228,169}14.8 \\
 & Add Node & \cellcolor[RGB]{209,45,45}{\color{white}76.5} & \cellcolor[RGB]{224,77,50}{\color{white}69.9} & \cellcolor[RGB]{198,19,41}{\color{white}81.6} & \cellcolor[RGB]{193,9,39}{\color{white}\textbf{83.6}} & \cellcolor[RGB]{205,36,44}{\color{white}78.2} & \cellcolor[RGB]{239,109,55}{\color{white}63.6} & \cellcolor[RGB]{213,53,46}{\color{white}74.8} & \cellcolor[RGB]{253,170,75}43.9 & \cellcolor[RGB]{215,56,47}{\color{white}74.1} & \cellcolor[RGB]{255,247,229}4.4 & \cellcolor[RGB]{254,217,133}21.0 \\
 & Swap Nodes & \cellcolor[RGB]{201,26,42}{\color{white}80.2} & \cellcolor[RGB]{217,62,48}{\color{white}73.1} & \cellcolor[RGB]{196,15,40}{\color{white}82.3} & \cellcolor[RGB]{189,0,38}{\color{white}\textbf{85.4}} & \cellcolor[RGB]{203,30,43}{\color{white}79.3} & \cellcolor[RGB]{221,71,49}{\color{white}71.3} & \cellcolor[RGB]{216,60,47}{\color{white}73.3} & \cellcolor[RGB]{254,198,89}31.0 & \cellcolor[RGB]{205,35,44}{\color{white}78.3} & \cellcolor[RGB]{254,228,169}14.8 & \cellcolor[RGB]{254,208,106}25.7 \\
 & Remove Node & \cellcolor[RGB]{198,19,41}{\color{white}81.6} & \cellcolor[RGB]{213,52,46}{\color{white}74.9} & \cellcolor[RGB]{196,15,40}{\color{white}82.3} & \cellcolor[RGB]{189,0,38}{\color{white}\textbf{85.3}} & \cellcolor[RGB]{199,21,41}{\color{white}81.1} & \cellcolor[RGB]{227,84,51}{\color{white}68.6} & \cellcolor[RGB]{214,55,47}{\color{white}74.3} & \cellcolor[RGB]{253,165,72}46.0 & \cellcolor[RGB]{211,47,45}{\color{white}75.9} & \cellcolor[RGB]{255,230,176}13.7 & \cellcolor[RGB]{255,242,212}7.4 \\
 & Recolor Node & \cellcolor[RGB]{196,16,41}{\color{white}82.1} & \cellcolor[RGB]{231,92,52}{\color{white}67.0} & \cellcolor[RGB]{198,19,41}{\color{white}81.6} & \cellcolor[RGB]{193,8,39}{\color{white}\textbf{83.7}} & \cellcolor[RGB]{203,31,43}{\color{white}79.2} & \cellcolor[RGB]{254,196,88}31.9 & \cellcolor[RGB]{207,39,44}{\color{white}77.6} & \cellcolor[RGB]{254,207,103}26.3 & \cellcolor[RGB]{214,54,46}{\color{white}74.6} & \cellcolor[RGB]{255,237,197}10.1 & \cellcolor[RGB]{255,246,226}5.0 \\
\midrule
\rowcolor{benchbg}
\multicolumn{2}{l}{\textbf{\tinygrafixbench\ Avg.}} & \cellcolor[RGB]{208,43,45}{\color{white}76.8} & \cellcolor[RGB]{223,75,50}{\color{white}70.4} & \cellcolor[RGB]{209,44,45}{\color{white}76.6} & \cellcolor[RGB]{203,32,43}{\color{white}\textbf{79.0}} & \cellcolor[RGB]{218,63,48}{\color{white}72.8} & \cellcolor[RGB]{248,129,58}59.5 & \cellcolor[RGB]{241,114,56}{\color{white}62.7} & \cellcolor[RGB]{254,173,76}42.4 & \cellcolor[RGB]{231,92,52}{\color{white}67.1} & \cellcolor[RGB]{255,231,178}13.4 & \cellcolor[RGB]{255,230,174}14.0 \\
\bottomrule
\end{tabular}%
}
\end{table}

\clearpage

\section{Model Output Galleries}
\label{sec:appx:qualitative}

This section presents model output galleries for \bench (\cref{sec:appx:pb-gallery})
and \tinygrafixbench (\cref{sec:appx:tgb-gallery}),
showing model outputs alongside ground-truth answers for representative problems.

\subsection{Per-Problem Galleries: \bench}
\label{sec:appx:pb-gallery}

\begin{figure*}[p]
    \centering
    \includegraphics[width=0.85\linewidth]{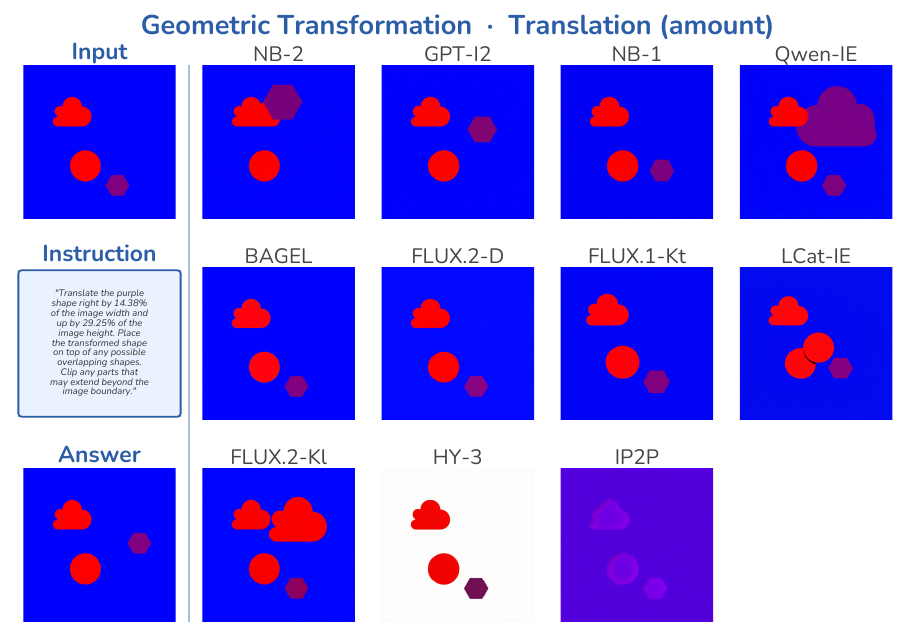}
    \caption{
        \textbf{\bench\ gallery: Geometric Transformation.}
    }
    \label{fig:appx:pb-gallery-geometric}
\end{figure*}

\begin{figure*}[p]
    \centering
    \includegraphics[width=0.85\linewidth]{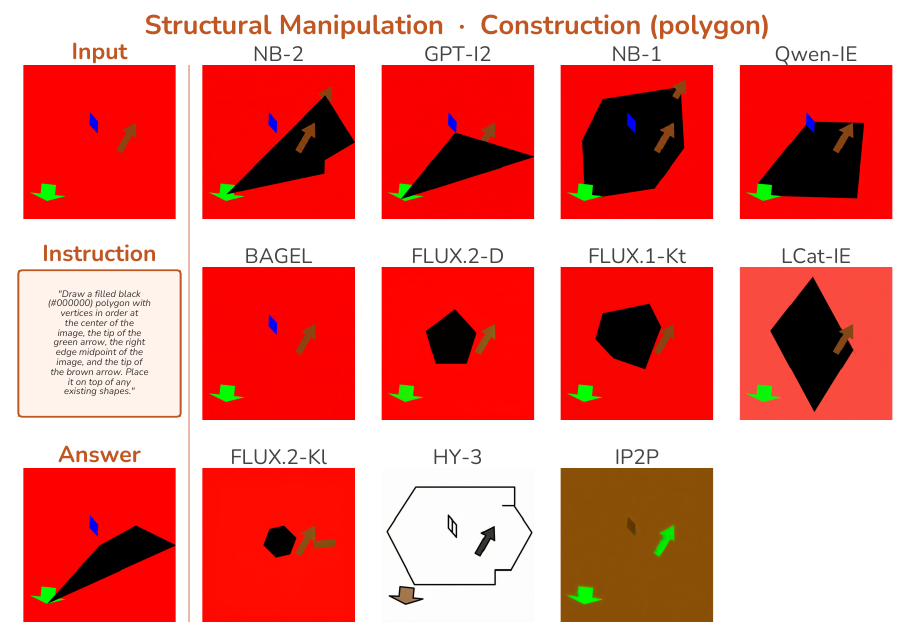}
    \caption{
        \textbf{\bench\ gallery: Structural Manipulation (construction task, polygon mode).}
        Input, ground-truth answer, and outputs from all eleven models.
    }
    \label{fig:appx:pb-gallery-structural}
\end{figure*}

\begin{figure*}[p]
    \centering
    \includegraphics[width=0.85\linewidth]{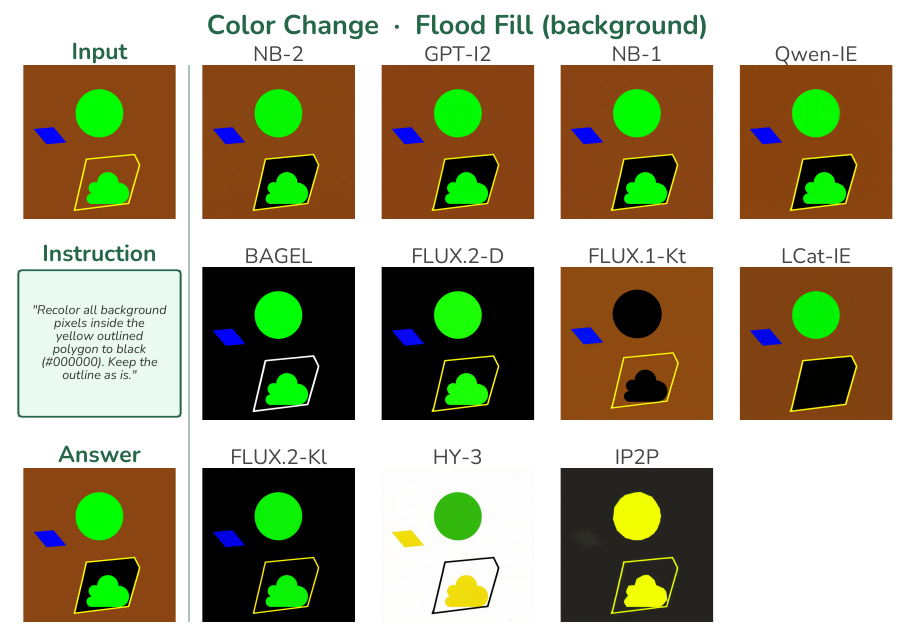}
    \caption{
        \textbf{\bench\ gallery: Color Change (flood fill task, background mode).}
        Input, ground-truth answer, and outputs from all eleven models.
    }
    \label{fig:appx:pb-gallery-color}
\end{figure*}

\begin{figure*}[p]
    \centering
    \includegraphics[width=0.85\linewidth]{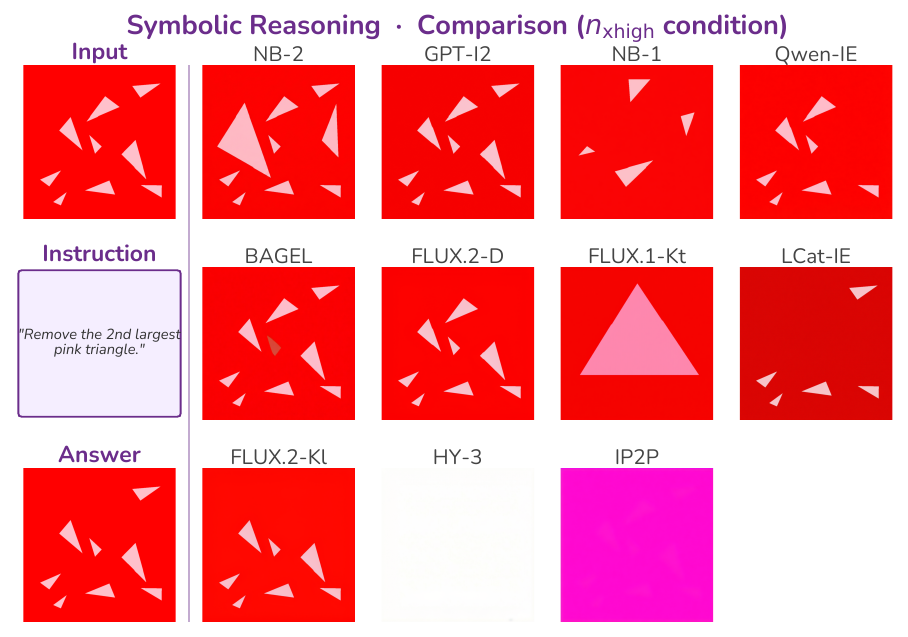}
    \caption{
        \textbf{\bench\ gallery: Symbolic Reasoning (comparison task, $n_{\text{xhigh}}$ condition).}
        Input, ground-truth answer, and outputs from all eleven models.
        \emph{N.B., \hunyuanimageS\ outputs a white image.}
        }
        \label{fig:appx:pb-gallery-symbolic}
    \end{figure*}

\Cref{fig:appx:pb-gallery-geometric,fig:appx:pb-gallery-structural,fig:appx:pb-gallery-color,fig:appx:pb-gallery-symbolic}
show one representative problem per \bench category alongside the outputs of all eleven models.
Each figure shows the input image, instruction, answer, and one output per model.

\subsection{Per-Problem Galleries: \tinygrafixbench}
\label{sec:appx:tgb-gallery}

\begin{figure*}[p]
    \centering
    \includegraphics[width=0.85\linewidth]{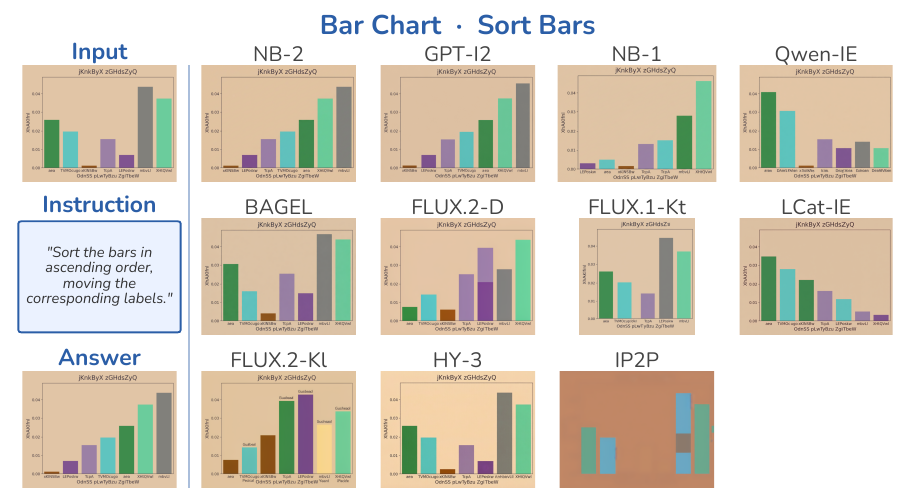}
    \caption{
        \textbf{\tinygrafixbench\ gallery: Bar Chart (sort bars task).}
    }
    \label{fig:appx:tgb-gallery-bar}
\end{figure*}

\begin{figure*}[p]
    \centering
    \includegraphics[width=0.85\linewidth]{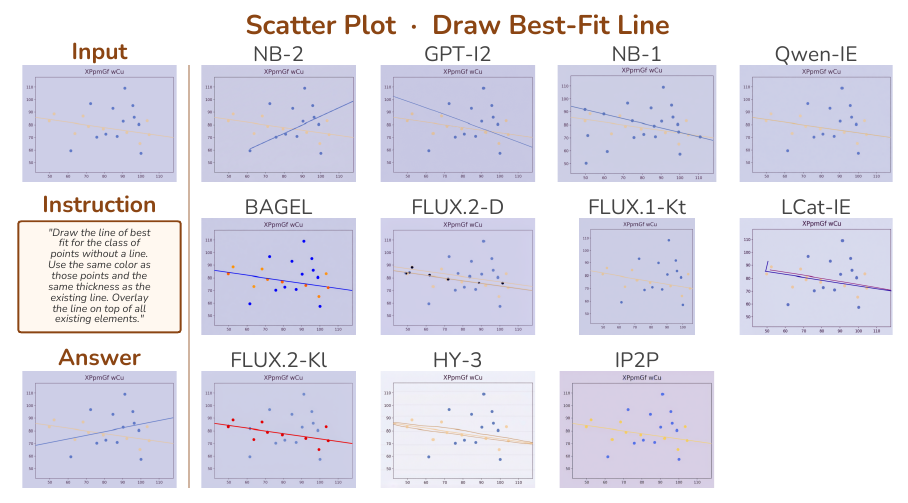}
    \caption{
        \textbf{\tinygrafixbench\ gallery: Scatter Plot (draw best-fit line task).}
    }
    \label{fig:appx:tgb-gallery-scatter}
\end{figure*}

\begin{figure*}[p]
    \centering
    \includegraphics[width=0.85\linewidth]{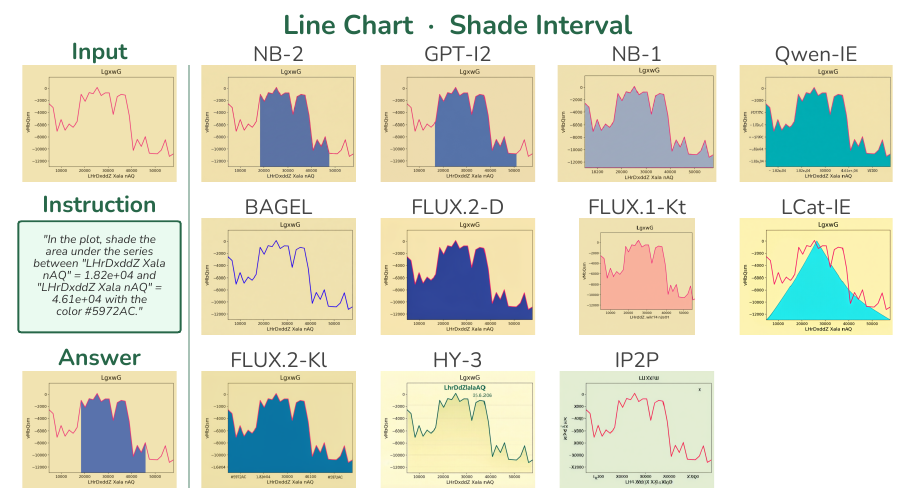}
    \caption{
        \textbf{\tinygrafixbench\ gallery: Line Chart (shade interval task).}
    }
    \label{fig:appx:tgb-gallery-line}
\end{figure*}

\begin{figure*}[p]
    \centering
    \includegraphics[width=0.85\linewidth]{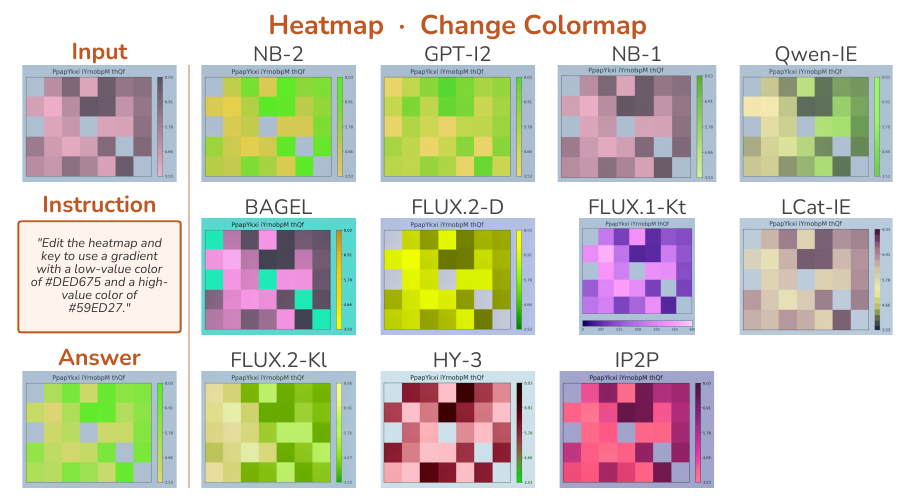}
    \caption{
        \textbf{\tinygrafixbench\ gallery: Heatmap (change colormap task).}
    }
    \label{fig:appx:tgb-gallery-heatmap}
\end{figure*}

\begin{figure*}[p]
    \centering
    \includegraphics[width=0.85\linewidth]{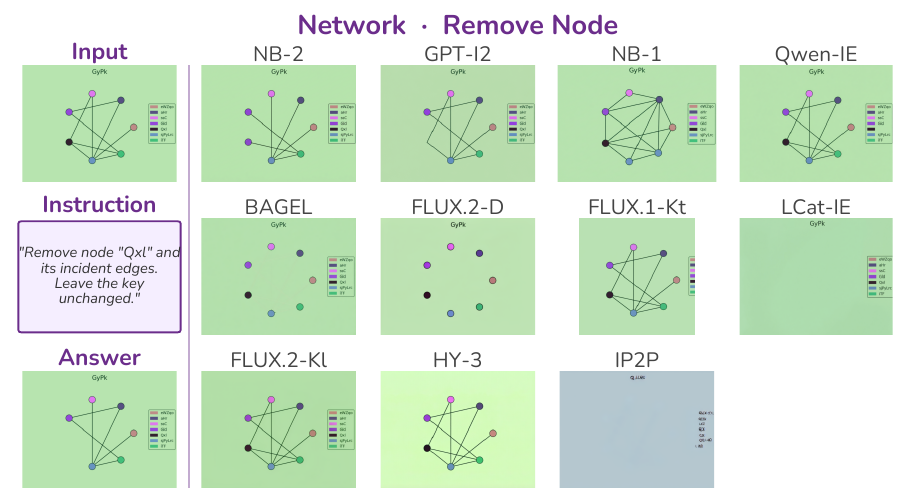}
    \caption{
        \textbf{\tinygrafixbench\ gallery: Network (remove node task).}
    }
    \label{fig:appx:tgb-gallery-network}
\end{figure*}

\Cref{fig:appx:tgb-gallery-bar,fig:appx:tgb-gallery-scatter,fig:appx:tgb-gallery-line,fig:appx:tgb-gallery-heatmap,fig:appx:tgb-gallery-network}
show one representative problem per chart type across the five \tinygrafixbench\ chart families.
Each figure shows the input image, instruction, answer, and one output per model.

\end{document}